\documentclass[11pt]{article}
\pdfoutput=1

\usepackage{euscript}
\usepackage{amssymb}
\usepackage{amsfonts}
\usepackage{amsbsy}
\usepackage{epsfig}
\usepackage{amsthm}
\usepackage{amscd}
\usepackage{amstext}
\usepackage{verbatim}
\usepackage{amsmath}
\usepackage{cancel}
\usepackage{capt-of}
\usepackage{empheq}
\usepackage{subfigure}
\usepackage{xcolor}

\usepackage{cite}
\usepackage{bm}
\usepackage{authblk}
\usepackage[T1]{fontenc}

\usepackage[pdftex,linktoc=all]{hyperref}
\hypersetup{ 
colorlinks=true, 
linkcolor=black, 
citecolor=red, 
}

\def\ben{\begin{equation}}
\def\een{\end{equation}}

\let\a=\alpha

\def\be{\begin{equation}}
\def\ee{\end{equation}}
\def\beq{\begin{equation}}
\def\eeq{\end{equation}}
\def\ba{\begin{array}}
\def\ea{\end{array}}

\def\wtd{\widetilde}

\def\dalemb#1#2{{\vbox{\hrule height .#2pt
       \hbox{\vrule width.#2pt height#1pt \kern#1pt
               \vrule width.#2pt}
       \hrule height.#2pt}}}

\newcommand{\bea}{\begin{eqnarray}}
\newcommand{\eea}{\end{eqnarray}}
\DeclareMathOperator{\tr}{tr}

\makeatletter
\newcommand*\bigcdot{\mathpalette\bigcdot@{.5}}
\newcommand*\bigcdot@[2]{\mathbin{\vcenter{\hbox{\scalebox{#2}{$\m@th#1\bullet$}}}}}
\makeatother

\renewcommand{\eqref}[1]{(\ref{#1})}

\def\Im{{{\frak{Im}}}}

\newcommand{\inv}{\text{inv}}

\newcommand{\E}{\mathbb{E}}

\newcommand{\fd}{\varphi}
\newcommand{\jm}{{j_{\max}}}
\newcommand{\modesum}{\sum_{0 \leq j \leq \jm}^{|m| \leq j}}

\newcommand{\cH}{\mathcal{H}}

\DeclareMathOperator{\im}{im}
\DeclareMathOperator{\adj}{adj}

\renewcommand{\Im}[0]{\operatorname{Im}}

\numberwithin{equation}{section}

\title{Deep Quantum Geometry of Matrices}
\author{Xizhi Han and Sean A. Hartnoll}
\affil{Department of Physics, Stanford University, \\
Stanford, CA 94305-4060, USA}
\date{}                     

\begin{document}
\frenchspacing

\maketitle

\begin{abstract}
We employ machine learning techniques to provide accurate variational wavefunctions for matrix quantum mechanics, with multiple bosonic and fermionic matrices. Variational quantum Monte Carlo is implemented with deep generative flows to search for gauge invariant low energy states. The ground state, and also long-lived metastable states, of an $\mathrm{SU}(N)$ matrix quantum mechanics with three bosonic matrices, as well as its supersymmetric  `mini-BMN' extension, are studied as a function of coupling and $N$. Known semiclassical fuzzy sphere states are recovered, and the collapse of these geometries in more strongly quantum regimes is probed using the variational wavefunction. We then describe a factorization of the quantum mechanical Hilbert space that corresponds to a spatial partition of the emergent geometry. Under this partition, the fuzzy sphere states show a boundary-law entanglement entropy in the large $N$ limit.
\end{abstract}

\newpage

\tableofcontents

\newpage

\section{Introduction}

A quantitative, first principles understanding of the emergence of spacetime from non-geometric microscopic degrees of freedom remains among the key challenges in quantum gravity. Holographic duality has provided a firm foundation for attacking this problem; we now know that supersymmetric large $N$ matrix theories can lead to emergent geometry \cite{Maldacena:1997re, Polchinski:2010hw}. What remains is the technical challenge of solving these strongly quantum mechanical systems and extracting the emergent spacetime dynamics from their quantum states. Recent years have seen significant progress in numerical studies of large $N$ matrix quantum mechanics at nonzero temperature. Using Monte Carlo simulations, quantitatively correct features of emergent black hole geometries have been obtained, e.g. \cite{Anagnostopoulos:2007fw, Catterall:2008yz, Berkowitz:2016jlq}. To grapple with questions such as the emergence of local spacetime physics, and its associated short distance entanglement \cite{Bombelli:1986rw,Srednicki:1993im}, new and inherently quantum mechanical tools are needed.

Variational wavefunctions can capture essential aspects of low energy physics. However, the design of accurate many-body wavefunction ansatze has typically required significant physical insight. For example, the power of tensor network states, such as Matrix Product States, hinges upon an understanding of entanglement in local systems \cite{MPS,ORUS2014117}. We are faced, in contrast, with models where there is an emergent locality that is not manifest in the microscopic interactions. This locality cannot be used {\it a priori}; it must be uncovered.
Facing a similar challenge of extracting the most relevant variables in high-dimensional data, deep learning has demonstrated remarkable success \cite{Hinton504, LeCun, Goodfellow-et-al-2016}, in tasks ranging from image classification \cite{Krizhevsky:2012:ICD:2999134.2999257} to game playing \cite{Go}. These successes, and others, have motivated tackling many-body physics problems with the machine learning toolbox \cite{doi:10.1063/PT.3.4164}. For example, there has been much interest and progress in applications of Restricted Boltzmann Machines to characterize states of spin systems \cite{Carleo602, PhysRevX.7.021021, Gao2017, PhysRevX.8.011006}. 

In this work we solve for low-energy states of quantum mechanical Hamiltonians with both bosons and fermions, using generative flows (normalizing flows \cite{dinh2014nice, 2015arXiv150505770J, DBLP:journals/corr/DinhSB16} and masked autoregressive flows \cite{DBLP:journals/corr/GermainGML15, DBLP:journals/corr/KingmaSW16, papamakarios2017masked} in particular) and variational quantum Monte Carlo. Compared with spin systems, the problem we are trying to solve contains continuous degrees of freedom and gauge symmetry, and there is no explicit spatial locality. Recent works have applied generative models to physics problems \cite{Carrasquilla2019, PhysRevX.8.031012, PhysRevLett.122.080602} and have aimed to understand holographic geometry, broadly conceived, with machine learning \cite{PhysRevB.97.045153, PhysRevD.98.046019, Hu:2019nea}. We will use generative flows to characterize emergent geometry in large $N$ multimatrix quantum mechanics. As we have noted above, such models form the microscopic basis of established holographic dualities.

We will focus on quantum mechanical models with three bosonic large $N$ matrices. These are among the simplest models with the core structure that is common to holographic theories. The bosonic part of the Hamiltonian takes the form
\be\label{eq:HamB}
H_B = \mathrm{tr}\,\left( \frac{1}{2} \Pi^i \Pi^i - \frac{1}{4} [X^i, X^j] [X^i, X^j] + \frac{1}{2} \nu^2 X^i X^i + i \nu \epsilon^{i j k} X^i X^j X^k \right) \,.
\ee
Here the $X^i$ are $N$ by $N$ traceless Hermitian matrices, with $i=1,2, 3$. The $\Pi^i$ are conjugate momenta and $\nu$ is a mass deformation parameter.
The potential energy in (\ref{eq:HamB}) is a total square: $V(X) = \frac{1}{4}\tr \left[\left(\nu \epsilon^{ijk} X^k + i [X^i,X^j] \right)^2\right]$.
The supersymmetric extension of this model \cite{Asplund:2015yda}, discussed
below, can be thought of as a simplified version of the BMN matrix quantum mechanics \cite{Berenstein:2002jq}. We refer to the supersymmetric model as `mini-BMN', following \cite{Anous:2017mwr}. For the low energy physics we will be exploring, the large $N$ planar diagram expansion in this model is controlled by the dimensionless coupling $\lambda \equiv N/\nu^3$. Here $\lambda$ can be understood as the usual dimensionful 't Hooft coupling of a large $N$ quantum mechanics at an energy scale set by the mass term (cf. \cite{Itzhaki:1998dd}).

The mass deformation in the Hamiltonian (\ref{eq:HamB}) inhibits the spatial spread of wavefunctions --- which will be helpful for numerics --- and leads to minima of the potential at
\be
[X^i,X^j] = i \nu \epsilon^{ijk} X^k \,.
\ee
In particular, one can have $X^i = \nu J^i$ with the $J^i$ being, for example, the $N$ dimensional irreducible representation of the $\mathfrak{su}(2)$ algebra. This set of matrices defines a `fuzzy sphere' \cite{Madore:1991bw}. There are two important features of this solution. Firstly, in the large $N$ limit the noncommutative algebra generated by the $X^i$ approaches the commutative algebra of functions on a smooth two dimensional sphere \cite{Hoppe:1988gk,DEWIT1988545}. Secondly, the large $\nu$ limit is a semiclassical limit in which the classical fuzzy sphere solution accurately describes the quantum state. In this semiclassical limit, the low energy excitations above the fuzzy sphere state are obtained from classical harmonic perturbations of the matrices about the fuzzy sphere \cite{Jatkar:2001uh}. See also \cite{Dasgupta:2002hx} for an analogous study of the large-mass BMN theory. At large $N$ and $\nu$, these excitations describe fields propagating on an emergent spatial geometry.

By using variational Monte Carlo with generative flows we will obtain a fully quantum mechanical description of this emergent space. This, in itself, is excessive given that the physics of the fuzzy sphere is accessible to semiclassical computations. Our variational wavefunctions will quantitatively reproduce the semiclassical results in the large $\nu$ limit, thereby providing a solid starting point for extending the variational method across the entire $N$ and $\nu$ phase diagram. Exploring the parameter space, we find that the fuzzy sphere collapses upon moving into the small $\nu$, quantum regime. We will consider two different `sectors' of the model, with different fermion number $R$. The first will be purely bosonic states, with $R=0$. The second will have a $R = N^2-N$. In this latter sector, the fuzzy sphere state is supersymmetric at large positive $\nu$, so we refer to this as the `supersymmetric sector'.
In the bosonic sector of the model the fuzzy sphere is a metastable state, and collapses in a first order large $N$ transition at $\nu \sim \nu_\text{c} \approx 4$. See Figs. \ref{fig:r} and \ref{fig:e} below. In the supersymmetric sector of the model, where the fuzzy sphere is stable, the collapse is found to be more gradual. See Figs. \ref{fig:susy_e} and \ref{fig:susy_r}. In Fig. \ref{fig:radius_small} we start to explore the small $\nu$ limit of the supersymmetric sector.

Beyond the energetics of the fuzzy sphere state, we will define a factorization of the microscopic quantum mechanical Hilbert space that leads to a boundary-law entanglement entropy at large $\nu$. See (\ref{eq:Nscaling}) below. This factorization at once captures the emergent local dynamics of fields on the fuzzy sphere and also reveals a microscopic cutoff to this dynamics at a scale set by $N$. The nature of the emergent fields and their cutoff can be usefully discussed in string theory realizations of the model. In string-theoretic constructions, fuzzy spheres arise from the polarization of D branes in background fields \cite{Myers:1999ps,Alekseev:2000fd,McGreevy:2000cw,Myers:2003bw}. A matrix quantum mechanics theory such as (\ref{eq:HamB}) describes $N$ `D0 branes' --- see \cite{Asplund:2015yda} and the discussion section below for a more precise characterization of the string theory embedding of mini-BMN theory --- and the maximal fuzzy sphere corresponds to a configuration in which the D0 branes polarize into a single spherical D2 brane. There is no gravity associated to this emergent space, the emergent fields describe the low energy worldvolume dynamics of the D2 brane. In this case, the emergent fields are a Maxwell field and a single scalar field corresponding to transverse fluctuations of the brane. In the final section of the paper we will discuss how richer, gravitating states may arise in the opposite small $\nu$ limit of the model.

\section{The mini-BMN model}
\label{sec:mini}

The mini-BMN Hamiltonian is \cite{Asplund:2015yda}
\be \label{eq:fermionic_potential}
    H = H_B + \text{tr} \left( \lambda^\dagger \sigma^k [X^k, \lambda] + \frac{3}{2} \nu \lambda^\dagger \lambda\right) - \frac{3}{2} \nu (N^2 - 1) \,.
\ee
The bosonic part $H_B$ is given in (\ref{eq:HamB}). The $\sigma^k$ are Pauli matrices. The $\lambda$ are matrices of two-component SO(3) spinors. It can be useful to write the matrices in terms of the $\mathfrak{su}(N)$ generators $T_A$, with $A = 1, 2, \ldots, N^2 - 1$, which obey $[T_A, T_B] = i f_{A B C} T_C$ and are Hermitian and orthonormal (with respect to the Killing form). That is, $X^i = X^i_A T^A$ and $\lambda^\alpha = \lambda^\alpha_A T^A$.\footnote{The $i j k$ and $A B C$ indices are freely raised and lowered. Lower $\alpha \beta$ indices are for spinors transforming in the $\mathbf{2}$ representation of SO(3), while
upper indices are for $\bar{\mathbf{2}}$. We will not raise or lower spinor indices.} The full Hamiltonian can then be written
\begin{align} \label{eq:hamitonian_full}
    H =& - \frac{1}{2} \frac{\partial^2}{(\partial X^i_A)^2}  + \frac{1}{4} \left(f_{A B C} X^i_B X^j_C\right)^2 + \frac{1}{2} \nu^2 \left(X^i_A\right)^2 - \frac{1}{2} \nu f_{A B C} \epsilon^{i j k} X^i_A X^j_B X^k_C \nonumber \\
    & + i f_{A B C} \lambda_A^{\alpha \dagger } X^k_B \sigma^{k\beta}_{\alpha} \lambda_{C \beta} + \frac{3}{2} \nu \lambda_A^{\alpha \dagger} \lambda_{A \alpha} - \frac{3}{2} \nu (N^2 - 1) ,
\end{align}
where $\lambda_A^{\alpha \dagger} \equiv (\lambda_{A \alpha})^\dagger$ and $\{\lambda_A^{\alpha \dagger}, \lambda_{B \beta}\} = \delta_{A B} \delta^\alpha_{\beta}$ are complex fermion creation and annihilation operators. This Hamiltonian is seen to have four supercharges 
\begin{equation}
    Q_\alpha = \left(- i \frac{\partial}{\partial X^i_A} + i \nu X^i_A - \frac{i}{2} f_{A B C} \epsilon_{i j k} X^j_B X^k_C\right) \sigma^{i\beta}_{\alpha} \lambda_{A\beta}, \quad \bar{Q}^\alpha = (Q_\alpha)^\dagger \,, \label{eq:supercharge}
\end{equation}
that obey
\begin{equation}
    \{Q_\alpha, \bar{Q}^{\alpha}\} = 4 H.
\end{equation}
States that are invariant under all supercharges therefore have vanishing energy.

Matrix quantum mechanics theories arising from microscopic string theory constructions are typically gauged. This means that physical states must be invariant under the $\mathrm{SU}(N)$ symmetry. In particular, physical state are annihilated by the generators
\be\label{eq:gens}
G_A = - i f_{ABC} \left(X^i_B \frac{\partial}{\partial X^i_C} + \lambda^{\a \dagger}_B \lambda_{C\a} \right) \,.
\ee

\subsection{Representation of the fermion wavefunction}

The mini-BMN wavefunction can be represented as a function from bosonic matrix coordinates to fermionic states $\psi(X) = f(X) | M(X) \rangle$. Here $X$ denotes the three bosonic traceless Hermitian matrices. The function $f(X) \geq 0$ is the norm of the wavefunction at $X$ while $|M(X) \rangle$ is a normalized state of matrix fermions. A fermionic state with definite fermion number $R$ is parametrized by a complex tensor $M^{r a}_{A \alpha}$ such that
\begin{equation}
    |M \rangle \equiv \sum_{r = 1}^D \prod_{a = 1}^R \Big(\sum_{\alpha = 1}^2  \sum_{A = 1}^{N^2 - 1} M_{A \alpha}^{r a} \lambda_A^{\alpha \dagger}\Big) |0\rangle, \label{eq:fermionic_states}
\end{equation}
where $|0\rangle$ is the state with all fermionic modes unoccupied. 

The definition (\ref{eq:fermionic_states}) is parsed as follows: for any fixed $r$ and $a$, $\eta^{r a \dagger} = \sum_{\alpha A} M^{r a}_{A \alpha} \lambda_A^{\alpha \dagger}$ is the creation operator for the matrix fermionic modes, where $A$ runs over some orthonormal basis of the $\mathfrak{su}(N)$ Lie algebra and $\alpha = 1, 2$ for two fermionic matrices. Then $\prod_a \eta^{r a \dagger} |0\rangle $ is a state of multiple free fermions created by $\eta^{ \dagger}$. The final summation over $r$ in (\ref{eq:fermionic_states}) is a decomposition of a general fermionic state into a sum of free fermion states. Such a representation is seen to be completely general (but not unique) if we have the number of free fermion states $D$ sufficiently large.  

For purely bosonic models, $|M(X) \rangle$ is simply the phase of the wavefunction.

\subsection{Gauge invariance and gauge fixing}
\label{sec:fix}

The generators (\ref{eq:gens}) correspond to the following action of an element $U \in G = \mathrm{SU}(N)$ on the wavefunction:
\begin{equation}
    (U \psi)(X) = f(U^{-1} X U) | (U M U^{-1}) (U^{-1} X U) \rangle,
\end{equation}
that is, the group acts by matrix conjugation. The wavefunction is required to be invariant under the group action, i.e. $U \psi = \psi$ for any $U \in G$.

Gauge invariance allows us to evaluate the wavefunction using a representative for each orbit of the gauge group. Let $\wtd{X}$ be the representative in the gauge orbit of $X$. Gauge invariance of the wavefunction implies that there must exist functions $\wtd{f}$ and $\wtd{M}$ such that
\begin{equation}
    f(X) = \wtd{f}(\wtd{X}), \quad |M(X)\rangle = |U \wtd{M}(\wtd{X}) U^{-1}\rangle \; \text{ where } \; X = U \wtd{X} U^{-1} \,. \label{eq:gauge_invariant_wavefunc}
\end{equation}
The functions $\wtd{f}$ and $\wtd{M}$ take gauge representatives as inputs, or may be thought as gauge invariant functions. The wavefunction we use will be in the form (\ref{eq:gauge_invariant_wavefunc}). The functions $\wtd{f}$ and $\wtd{M}$ will be parametrized by neural networks, as we describe in the following section \ref{sec:archi}.

We proceed to describe the gauge fixing we use to select the representative for each orbit, as well as the measure factor associated with this choice. The $\mathrm{SU}(N)$ gauge representative $\wtd{X}$ will be such that
\begin{enumerate}
\item $X^i = U \wtd{X}^i U^{-1}$ for $i = 1, 2, 3$ and some unitary matrix $U$.

\item $\wtd{X}^1$ is diagonal and $\wtd{X}^1_{1 1} \leq \wtd{X}^1_{2 2} \leq \ldots \leq \wtd{X}^1_{N N}$. \label{item:2}

\item $\wtd{X}^2_{i (i + 1)}$ is purely imaginary with the imaginary part positive for $i = 1, 2, \ldots, N - 1$. \label{item:3}
\end{enumerate}
The third condition is needed to fix the $\mathrm{U}(1)^{N - 1}$ residual gauge freedom after diagonalizing $X^1$. The representative $\wtd{X}$ is well-defined except on a subspace of measure zero where the matrices are degenerate. Then $\wtd{X}$ can be represented as a vector in $\mathbb{R}^{2 (N^2 - 1)}$ with a positivity constraint on some components. The change of variables from $X$ to $\wtd{X}$ leads to a measure factor given by the volume of the gauge orbit:
\be
d^{3 (N^2 - 1)} X = \Delta(\wtd{X})\, d^{2 (N^2 - 1)} \wtd{X} \,, 
\ee
with
\begin{equation}\label{eq:measure}
        \Delta(\wtd{X}) \propto \prod_{i \neq j = 1}^N \left|\wtd{X}^1_{i i} - \wtd{X}^1_{j j}\right| \prod_{i = 1}^{N - 1} \left|\wtd{X}^2_{i (i + 1)}\right|.
\end{equation}
Keeping track of this measure (apart from an overall prefactor) will be important for proper sampling in the Monte Carlo algorithm. The derivation of (\ref{eq:measure}) is shown in Appendix \ref{app:gauge}.

\section{Architecture design for matrix quantum mechanics}
\label{sec:archi}



In this work we propose a variational Monte Carlo method with importance sampling to approximate the ground state of matrix quantum mechanics theories, leading to an upper bound on the ground state energy. The importance sampling is implemented with generative flows. The basic workflow is sketched as follows:

\begin{enumerate}
\item Start with a wavefunction $\psi_\theta$ with variational parameters $\theta$. In our case $\theta$ will characterize neural networks. 

\item Write the expectation value of the Hamiltonian to be minimized as 
\begin{equation}
    E_\theta = \langle \psi_\theta | H | \psi_\theta \rangle = \int d X\, |\psi_\theta(X)|^2 H_X[\psi_\theta] = \E_{X \sim |\psi_\theta|^2}[H_X[\psi_\theta]]\,. \label{eq:variational_energy}
\end{equation}
In the mini-BMN case $X$ denotes three traceless Hermitian matrices (indices omitted) and $H_X[\psi_\theta]$ is the energy density at $X$. Notationally $\E_{X \sim p(X)}$ is the expectation value, with the random variable $X$ drawn from the probability distribution $p(X)$. 

\item Generate random samples according to the wavefunction probabilities $X \sim p_\theta(X) = |\psi_\theta(X)|^2$, and evaluate their energy densities $H_X[\psi_\theta]$. The variational energy (\ref{eq:variational_energy}) can then be estimated as the average of energy densities of the samples.

\item Update the parameters $\theta$ (via stochastic gradient descent) to minimize $E_\theta$:
\begin{equation}
    \theta_{t + 1} = \theta_t - \alpha \nabla_{\theta_t} E_{\theta_t},
\end{equation}
where $t = 1, 2, \ldots$ denotes the steps of training and the parameter $\alpha > 0$ sets the learning rate. The gradient of energy is estimated from Monte Carlo samples:
\begin{equation}
    \nabla_\theta E_\theta = \E_{X \sim p_\theta}[\nabla_{\theta} H_X[\psi_\theta]] + \E_{X \sim p_\theta}[\nabla_\theta \left(\ln p_\theta(X)\right) \left(H_X[\psi_\theta] - E_\theta\right)].
\end{equation}
The method is applicable even if the probabilities are available only up to an unknown normalization factor.

\item Repeat steps 3 and 4 until $E_\theta$ converges. Observables of physical interest are evaluated with respect to the optimal parameters after training. 

\end{enumerate}

In the following we discuss details of parametrizing and sampling from gauge invariant wavefunctions with fermions. Technicalities concerning the evaluation of $H_X[\psi_\theta]$ are spelled out in Appendix \ref{app:obs}. More details concerning the training are given in Appendix \ref{app:train}. Benchmarks are presented at the end of this section.

\subsection{Parametrizing and sampling the gauge invariant wavefunction}

We first describe how gauge invariance is incorporated into the variational Monte Carlo algorithm. As just discussed, an important step is to sample according to $X \sim |\psi(X)|^2$. From (\ref{eq:gauge_invariant_wavefunc}), for a gauge invariant wavefunction $|\psi(X)|^2 = |\wtd{f}(\wtd{X})|^2$. However, in sampling $\wtd{X}$ we must keep track of the measure factor $\Delta(\wtd{X})$ in (\ref{eq:measure}). This is done as follows:
\begin{enumerate}
\item Sample $\wtd{X}$ according to $p(\wtd{X}) = \Delta(\wtd{X}) |\wtd{f}(\wtd{X})|^2$.
\item Generate Haar random elements $U \in \mathrm{SU}(N)$.
\item Output samples $X = U \wtd{X} U^{-1}$.
\end{enumerate}
The correctness of this procedure is shown in Appendix \ref{app:gauge}. 

Conversely at the evaluation stage, $\psi(X)$ can be computed in the following steps for gauge invariant wavefunctions (\ref{eq:gauge_invariant_wavefunc}):
\begin{enumerate}
    \item Gauge fix $X  = U \wtd{X} U^{-1}$ as discussed in the last section.
    \item Compute $\wtd{M}(\wtd{X})$ and $\wtd{f}(\wtd{X})$. Details of the structure of $\wtd{M}$ and $\wtd{f}$ will be discussed below.
    \item Return $\psi(X) = \wtd{f}(\wtd{X}) |U \wtd{M}(\wtd{X}) U^{-1} \rangle$ according to (\ref{eq:gauge_invariant_wavefunc}).
\end{enumerate}

We now describe the implementation of $\wtd{M}$ and $\wtd{f}$ as neural networks. The basic building block, a multilayer fully-connected (also called dense) neural network, is an elemental architecture capable of parametrizing complicated functions efficiently \cite{Goodfellow-et-al-2016}. The neural network defines a function $F: x \mapsto y$ mapping an input vector $x$ to an output vector $y$ via a sequence of affine and nonlinear transformations:
\begin{equation}
    F = A^{m}_\theta \circ \tanh \circ A^{m - 1}_\theta \circ \tanh \circ \dots \circ \tanh \circ A^{1}_\theta \,. \label{eq:dense}
\end{equation}
Here $A^1_\theta(x) = M^1_\theta x + b^1_\theta$ is an affine transformation, where the weights $M^1_\theta$ and the biases $b^1_\theta$ are trainable parameters. The hyperbolic tangent nonlinearity then acts elementwise on $A^1_\theta(x)$.\footnote{We experimented with different activation functions; the final result is not sensitive to this choice.} Similar mappings are applied $m$ times, allowing $M^i_\theta$ and $b^i_\theta$ to be different for different layers $i$, to produce the output vector $y$. The mapping $F:x \mapsto y$ is  nonlinear and capable of approximating any square integrable function if the number of layers and the dimensions of the affine transformations are sufficiently large \cite{NIPS2017_7203}. 

The function $\wtd{M}(\wtd{X})$ is implemented as such a multilayer fully-connected neural network, mapping from vectorized $\wtd{X}$ to  $\wtd{M}$ in (\ref{eq:fermionic_states}), i.e., $\mathbb{R}^{2 (N^2 - 1)} \to \mathbb{R}^{D R\, 2 (N^2 - 1)}$.
The implementation of $\wtd{f}(\wtd{X})$ is more interesting, as both evaluating $\wtd{f}(\wtd{X})$ and sampling from the distribution $p(\wtd{X}) = \Delta(\wtd{X}) |\wtd{f}(\wtd{X})|^2$ are necessary for the Monte Carlo algorithm. Generative flows are powerful tools to efficiently parameterize and sample from complicated probability distributions. The function $\wtd{f}(\wtd{X}) = \sqrt{p(\wtd{X}) / \Delta(\wtd{X})}$, so we can focus on sampling and evaluating $p(\wtd{X})$, which will be implemented by generative flows.

Two generative flow architectures are implemented for comparison: a normalizing flow and a masked autoregressive flow. The normalizing flow starts with a product of simple univariate probability distributions $p(x) = p_1(x_1) \ldots p_M(x_M)$, where the $p_i$ can be different. Values of $x$ sampled from this distribution are passed through an invertible multilayer dense network as in (\ref{eq:dense}). The probability distribution of the output $y$ is then
\begin{equation}
    q(y) = p(x) \left|\det \frac{D y}{D x}\right|^{-1} = p(F^{-1}(y)) |\det D F|^{-1}. \label{eq:flow}
\end{equation}

The masked autoregressive flow generates samples progressively. It requires an ordering of the components of the input, say $x_1, x_2, \ldots, x_M$. Each component is drawn from a parametrized distribution $p_i(x_i; F_i(x_1, \ldots, x_{i - 1}))$, where the parameter depends only on previous components. Thus $x_1$ is sampled independently and for other components, the dependence $F_i$ is given by (\ref{eq:dense}). The overall probability is the product
\begin{equation}
    q(x) = \prod_{i = 1}^M p_i(x_i; F_i(x_1, \ldots, x_{i - 1})).
\end{equation}

When $p_i(x_i)$ are chosen as normal distributions, both flows are able to represent any multivariate normal distribution exactly. Features of the wavefunction (such as polynomial or exponential tails) can be probed by experimenting with different base distributions $p_i(x_i)$. Choices of the base distributions and performances of the two flows are assessed in the following benchmark subsection and also in Appendix \ref{app:train}. We will use both types of flow in the numerical results of section \ref{sec:geom}.

\subsection{Benchmarking the architecture} \label{sec:benchmark}

In \cite{Anous:2017mwr} the Schr\"{o}dinger equation for the $N=2$ mini-BMN model was solved numerically. Comparison with the results in that paper will allow us to benchmark our architecture, before moving to larger values of $N$. In \cite{Anous:2017mwr} the Schr\"{o}dinger equation is solved in sectors with a fixed fermion number
\begin{equation}
    R = \sum_{A \alpha} \lambda^{\alpha \dagger}_A \lambda_{A \alpha}, \qquad [R, H] = 0,
\end{equation}
and total SO(3) angular momentum $j = 0, 1 / 2$. We do not constrain $j$, but do fix the number of fermions in the variational wavefunction.

\begin{figure}
    \centering
    \includegraphics[width=0.75\textwidth]{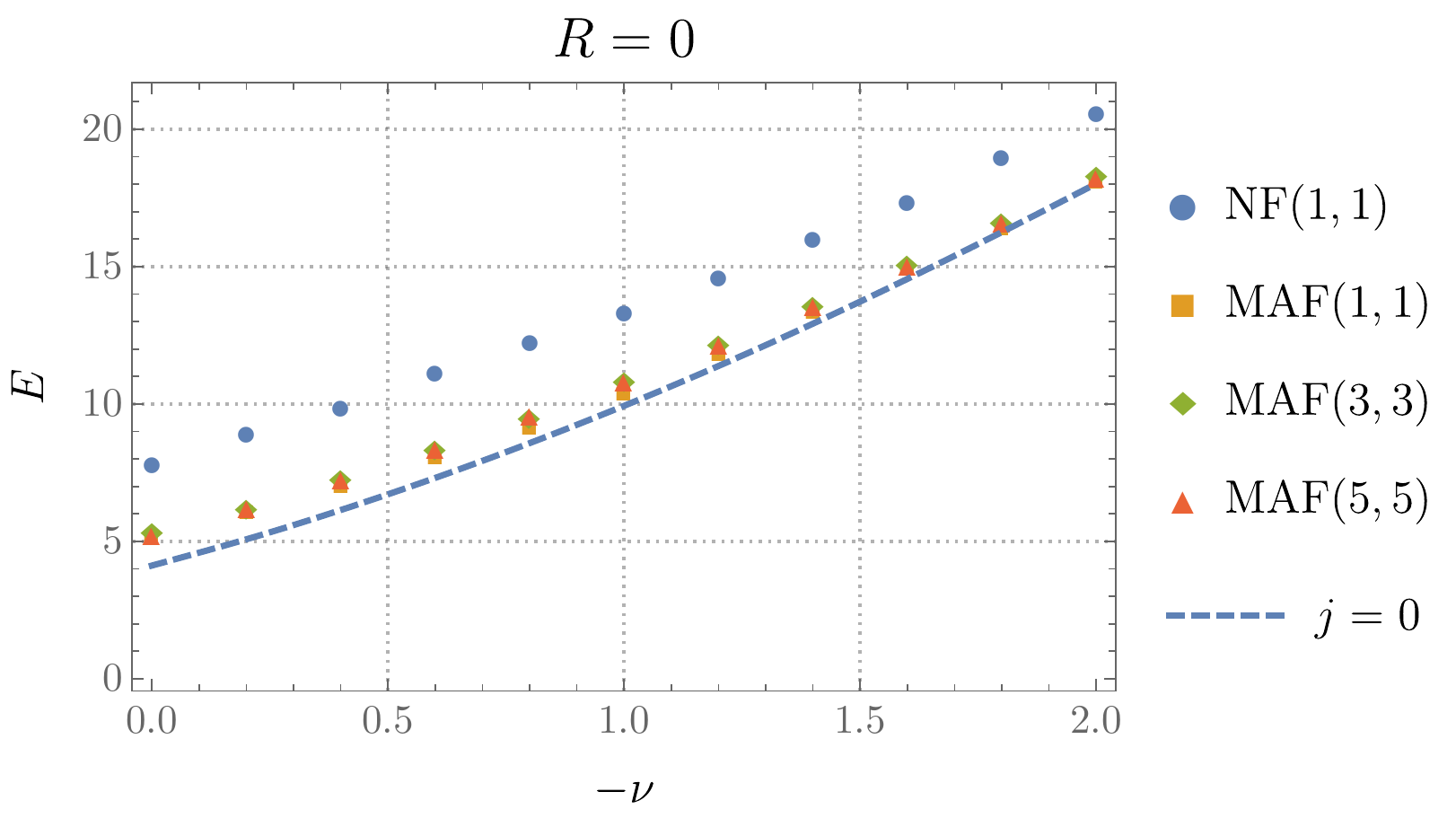}
    \includegraphics[width=0.75\textwidth]{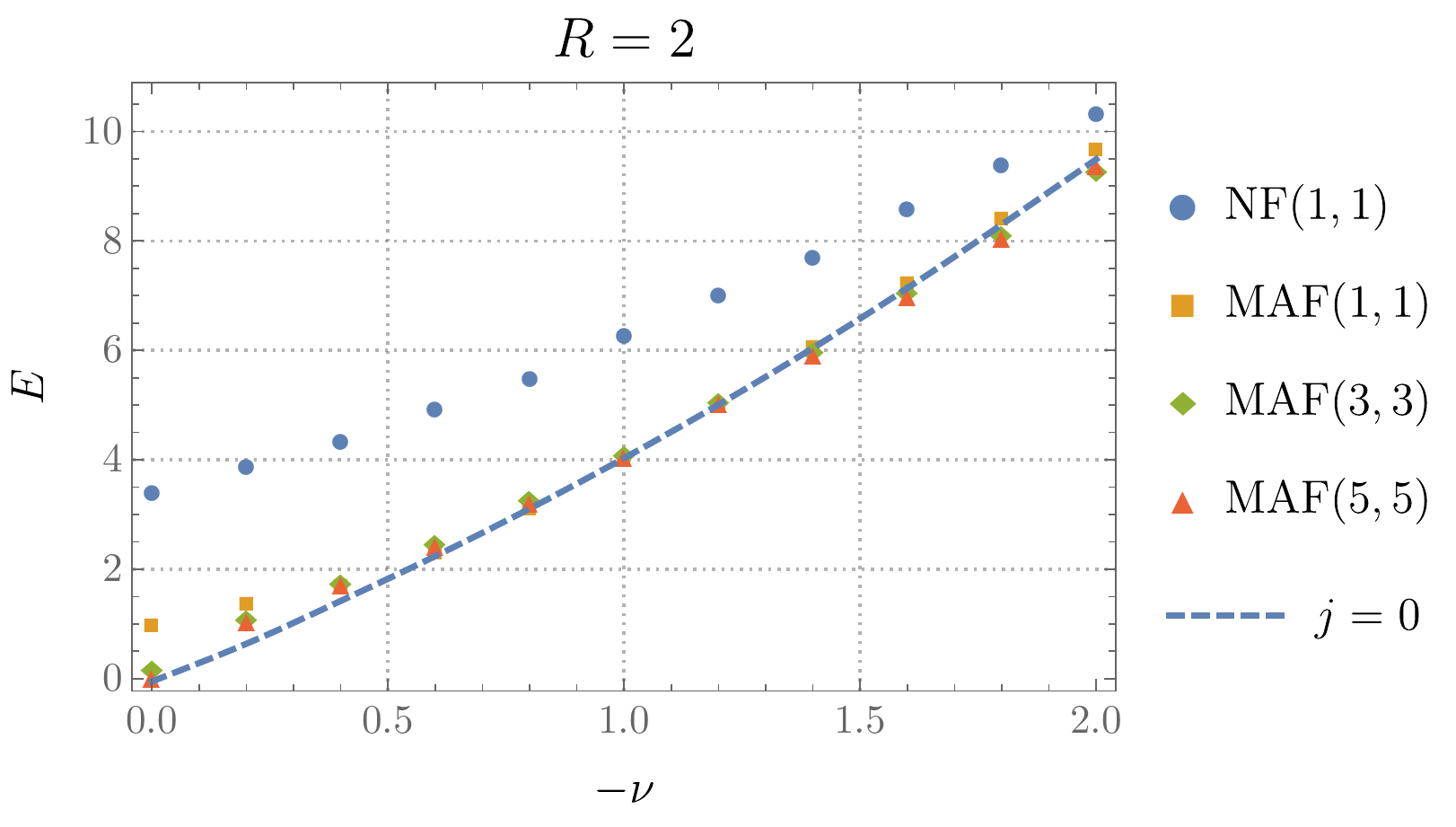}
    \caption{{\bf Benchmarking the architecture:} Variational ground state energies for the mini-BMN model with $N = 2$ and fermion numbers $R = 0$ and $R = 2$ (shown as dots) compared to the exact ground state energy in the $j = 0$ sector, obtained in \cite{Anous:2017mwr} (shown as the dashed curve). Uncertainties are at or below the scale of the markers; in particular the variational energies slightly below the dashed line are within numerical error of the line. NF stands for normalizing flows and MAF for masked autoregressive flows. As described in the main text, the numbers in the brackets are firstly the number of layers in the neural networks, and secondly the number of generalized normal distributions in each base mixed distribution.}
    \label{fig:benchmark}
\end{figure}

The variational energies obtained from our machine learning architecture with $R = 0$ and $R = 2$ are shown as a function of $\nu$ in Fig.\,\ref{fig:benchmark}. We take negative $\nu$ to compare with the results given in \cite{Anous:2017mwr}, which uses an opposite sign convention.\footnote{There is a particle-hole symmetry of the Hamiltonian (\ref{eq:hamitonian_full}) via $\nu \to - \nu$, $\lambda \to \lambda^\dagger$, $\lambda^\dagger \to \lambda$ and $X \to - X$.} The masked autoregressive flow yields better (lower) variational energies. These energies are seen to be close to the $j = 0$ results obtained in \cite{Anous:2017mwr}. The variational results seem to be asymptotically accurate as $|\nu| \to \infty$, while remaining a reasonably good approximation at small $\nu$. Small $\nu$ is an intrinsically more difficult regime, as the potential develops flat directions (visualized in \cite{Anous:2017mwr}) and hence the wavefunction is more complicated, possibly with long tails. In the 
`supersymmetric' $R=2$ sector, where quantum mechanical effects at small $\nu$ are expected to be strongest,
further significant improvement at the smallest values of $\nu$ is seen with deeper autoregressive networks and more flexible base distributions, as we describe shortly. Analogous improvements in these regimes will also be seen at larger $N$ in Sec.~\ref{sec:susy} and Appendix \ref{app:train}.


In Fig.\,\ref{fig:benchmark} the base distributions $p_i(x_i)$, introduced in the previous subsection, are chosen to be a mixture of $s$ generalized normal distributions:
\begin{equation}
    p_i(x_i) = \sum_{r = 1}^s k^i_r \frac{\beta^i_r}{2 \alpha^i_r \Gamma(1 / \beta^i_r)} e^{-(|x_i - \mu^i_r| / \alpha^i_r)^{\beta^i_r}}, \quad \sum_{r = 1}^s k^i_r = 1 \,. \label{eq:mixed_dist1}
\end{equation}
Here the $k^i_r$ are positive weights for each generalized normal distribution in the mixture. In (\ref{eq:mixed_dist1}) the $k^i_r$, $\alpha^i_r$, $\beta^i_r$ and $\mu^i_r$ are learnable (i.e. variational) parameters. For autoregressive flows these parameters further depend on $x_j$, with $1 \leq j < i$, according to (\ref{eq:dense}).

Due to the gauge fixing conditions \ref{item:2} and \ref{item:3} in section \ref{sec:fix}, some components $x_i$ are constrained to be positive. In the normalization flow this is implemented by an additional map $x_i \mapsto \exp(x_i)$. For the autoregressive flows we have a more refined control over the base distributions; in this case, for components $x_i$ that must be positive, we draw from Gamma distributions instead:
\begin{equation}
    p_i(x_i > 0) = \sum_{r = 1}^s k^i_r \frac{(\beta^i_r)^{\alpha^i_r}}{ \Gamma(\alpha^i_r)} (x_i)^{\alpha^i_r - 1} e^{-\beta^i_r x_i}, \quad \sum_{r = 1}^s k^i_r = 1. \label{eq:mixed_dist2}
\end{equation}
Where again the $k^i_r$, $\alpha^i_r$ and $\beta^i_r$ depend on $x_j$, with $1 \leq j < i$, according to (\ref{eq:dense}).

In Fig.\,\ref{fig:benchmark} we have shown mixtures with $s = 1, 3, 5$ distributions. The number of layers in (\ref{eq:dense}) has been increased with $s$ to search for potential improvements in the space of variational wavefunctions. As noted, the only improvement within the autoregressive flows in going beyond one layer and one generalized normal distribution is seen at the smallest values of $\nu$ with $R=2$. On the other hand, the gap between the variational energies of the two types of flows in Fig.\,\ref{fig:benchmark} suggests that the wavefunction is complicated in this regime, so that the more sophisticated MAF architecture shows an advantage. The recursive nature of the MAF flows means that they are already `deep' with only a single layer. The complexity of the small $\nu$ wavefunction should be contrasted with the fuzzy sphere phase at large positive $\nu$ discussed in the following section \ref{sec:geom} and shown in e.g. Figs.\,\ref{fig:r} and \ref{fig:e} below. The wavefunction in this semiclassical regime is almost Gaussian, and indeed the NF(1, 1) and MAF(1, 1) flows give similar energies when initialized near fuzzy sphere configurations. The NF architecture in fact gives slightly lower energies in this regime, so we have used normalizing flows in Figs.\,\ref{fig:r} and \ref{fig:e} for the fuzzy sphere.

The numerics above and below are performed with $D=4$ in (\ref{eq:fermionic_states}), so that the fermionic wavefunction $|M(X)\rangle$ is a sum of four free fermion states for each value of the bosonic coordinates $X$. In 
Appendix \ref{app:train} we see that increasing $D$ above one lowers the variational energy at small $\nu$, indicating that the fermionic states are not Hartree-Fock in this regime.

\section{The emergence of geometry}
\label{sec:geom}

\subsection{Numerical results, bosonic sector}

The architecture described above gives a variational wavefunction for low energy states of the mini-BMN model. With the wavefunction in hand, we can evaluate observables. We will start with the purely bosonic sector of the model (i.e. $R=0$). Then we will add fermions. An important difference between the bosonic and supersymmetric cases will be that the semiclassical fuzzy sphere state is metastable in the bosonic theory but stable in the supersymmetric theory.

\begin{figure}[h]
    \centering
    \includegraphics[width=0.8\textwidth]{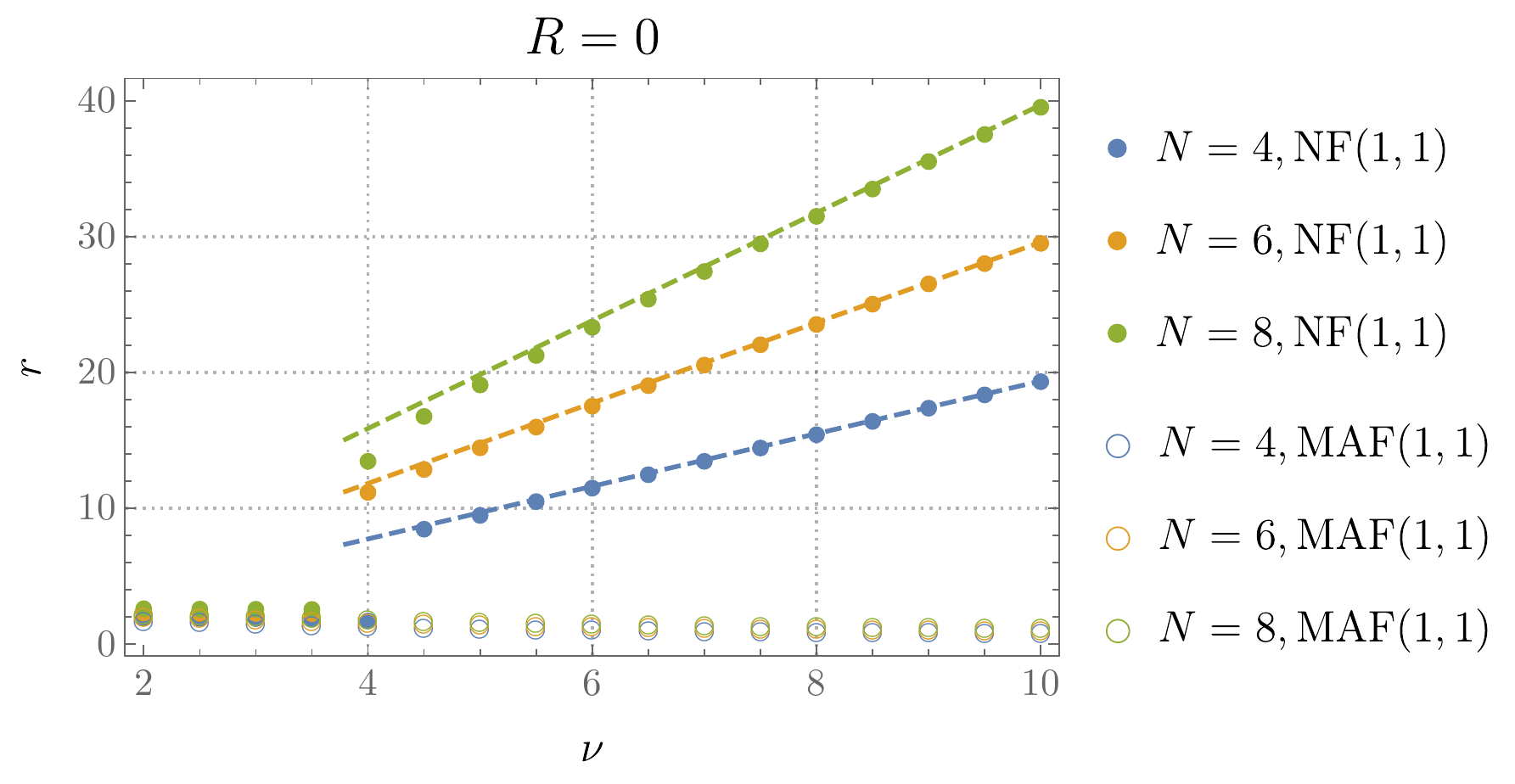}
    \caption{Expectation value of the radius in the zero fermion sector of the mini-BMN model, for different $N$ and $\nu$. The dashed lines are the semiclassical values (\ref{eq:rclassical}). Solid dots are initialized near the fuzzy sphere configuration, and the open markers are initialized near zero. We have used normalizing and autoregressive flows, respectively, as these produce more accurate variational wavefunctions in the two different regimes.}
    \label{fig:r}
\end{figure}

\begin{figure}[h!]
    \centering
    \includegraphics[width=0.8\textwidth]{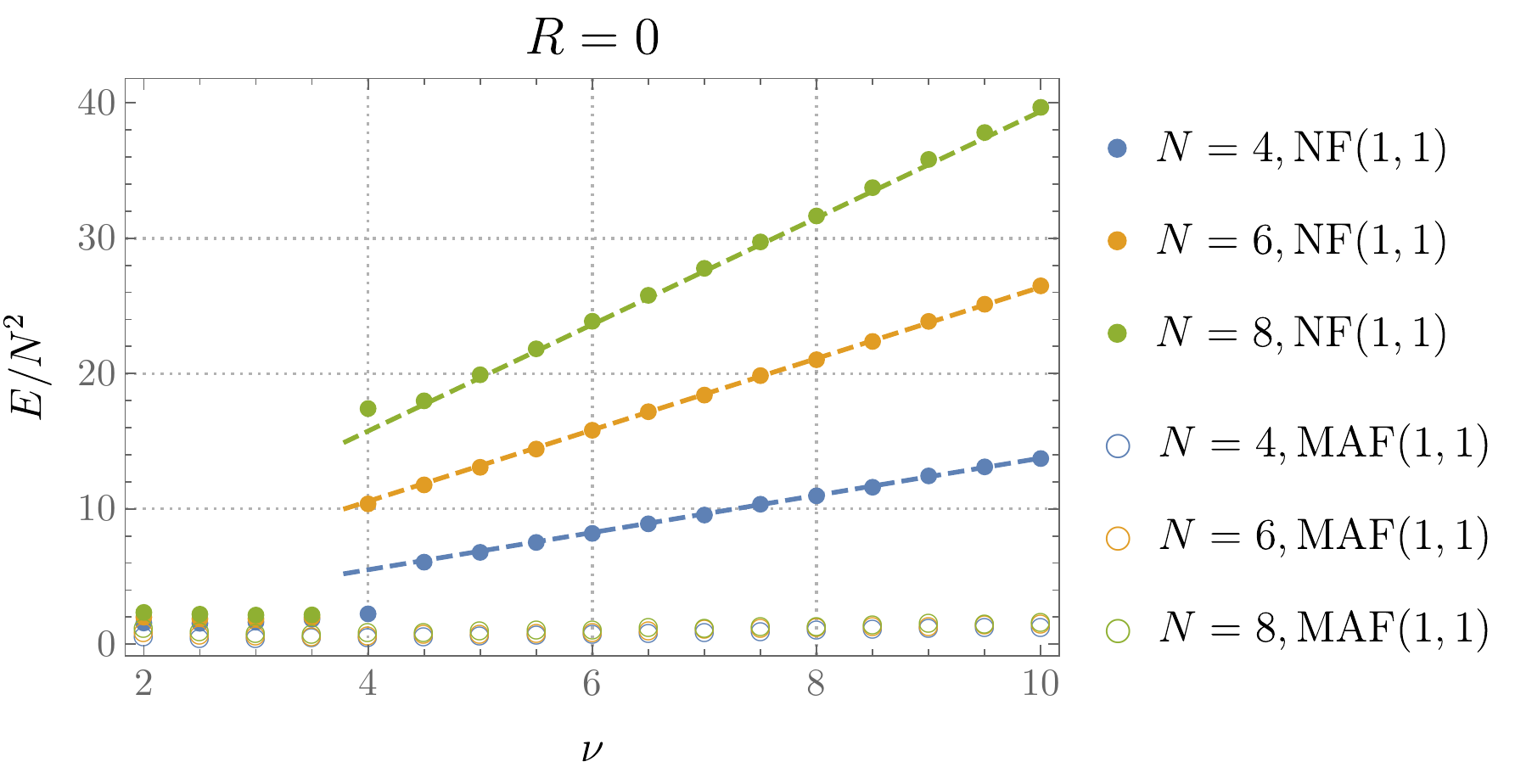}
    \caption{Variational energies in the zero fermion sector of the mini-BMN model, for different $N$ and $\nu$. The dashed lines are semiclassical values: $E =  -\frac{3}{2} \nu (N^2-1) + \left. \Delta E \right|_\text{bos}$, with $\left. \Delta E \right|_\text{bos}$ given in (\ref{eq:bos}). As in  Fig.~\ref{fig:r}, solid dots are initialized near the fuzzy sphere configuration, and the open markers are initialized near zero.}
    \label{fig:e}
\end{figure}

Figure \ref{fig:r} shows the expectation value of the radius 
\begin{equation}
    r = \sqrt{\frac{1}{N} \tr (X_1^2 + X_2^2 + X_3^2)} \,,
\end{equation}
for runs initialized close to a fuzzy sphere configuration (solid) and close to zero (open). For large $\nu$ a fuzzy sphere state with large radius is found, in addition to a `collapsed' state without significant spatial extent. Below $\nu_\text{c} \approx 4$, the fuzzy sphere state ceases to exist.
The nature of the transition at $\nu_\text{c}$ can be understood from the variational energy of the states, plotted in Figure \ref{fig:e}. The bosonic semiclassical fuzzy sphere state is seen to be metastable at large $\nu$, as the collapsed state has lower energy. For $\nu < \nu_\text{c}$ the fuzzy sphere is no longer even metastable. We will gain a semiclassical understanding of this transition in section \ref{sec:semiclassmain} shortly.

\begin{figure}
    \centering
    \includegraphics[width=0.8\textwidth]{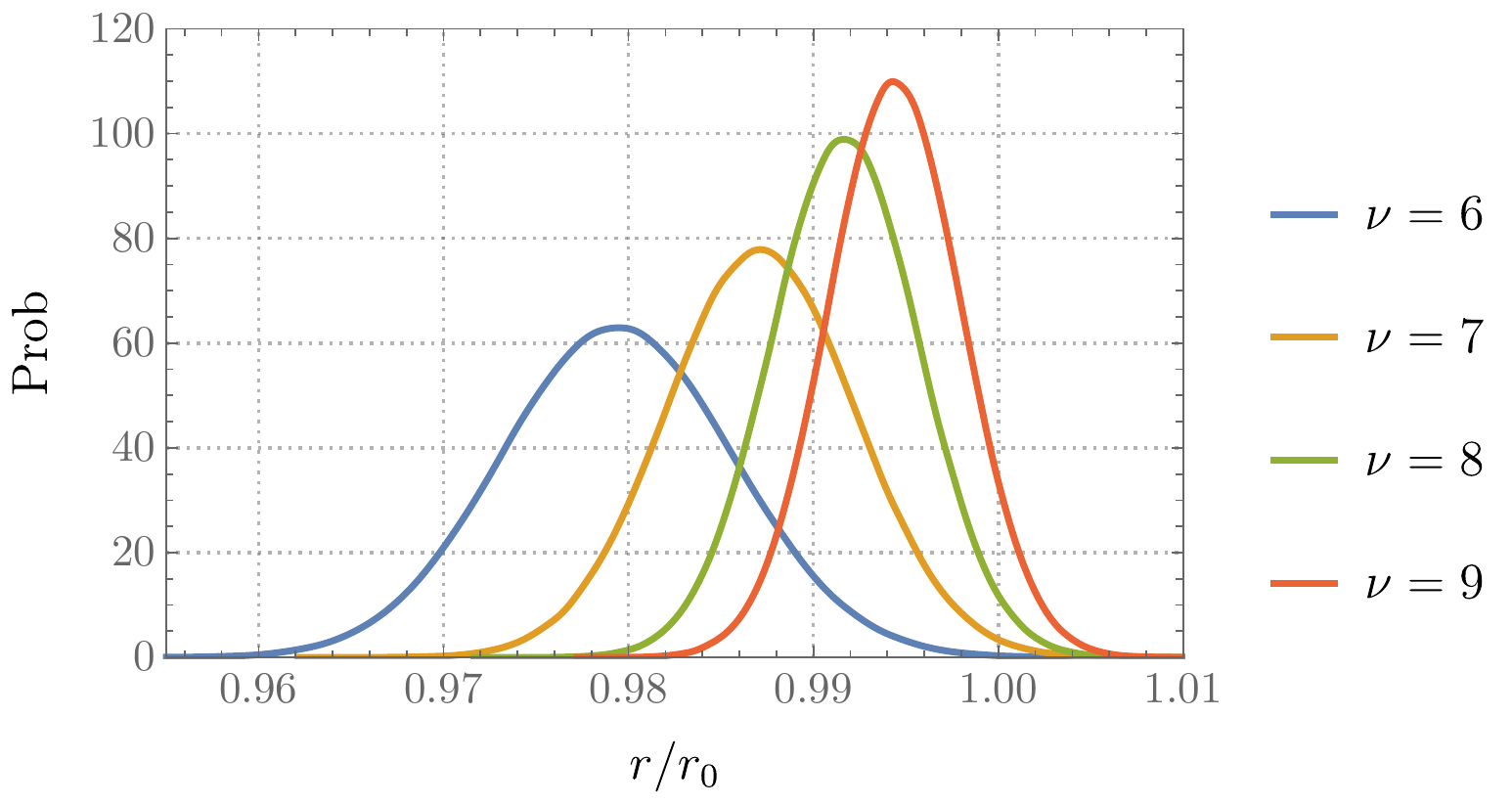}
    \caption{Probability distribution, from the variational wavefunction, for the radius in the fuzzy sphere phase for $N = 8$ and different $\nu$. The horizontal axis is rescaled by the semiclassical value of the radius $r_0$, given in (\ref{eq:rclassical}) below. The width of the distribution in units of the classical radius becomes smaller as $\nu$ is increased.}
    \label{fig:radius_dist}
\end{figure}

Figures \ref{fig:r} and \ref{fig:e} show that the radius and energy of the fuzzy sphere state are accurately described by semiclassical formulae (derived in the following section) for all $\nu > \nu_\text{c}$. In particular this means that $E/N^3$ and $r/N$ are rapidly converging towards their large $N$ values. Figure \ref{fig:radius_dist} further shows that the probability distribution for the radius $r$ becomes strongly peaked about its semiclassical expectation value at large $\nu$.

Analogous behavior to that shown in Figures \ref{fig:r} and \ref{fig:e} has previously been seen in classical Monte Carlo simulations of a thermal analogue of our quantum transition \cite{Azuma:2004zq,CastroVillarreal:2004vh,DelgadilloBlando:2007vx}. These papers study the thermal partition function of models similar to (\ref{eq:HamB}) in the classical limit, i.e. without the $\Pi^2$ kinetic energy term. The fuzzy geometry emerges in a first order phase transition as a low temperature phase in these models.  We will see that in our quantum mechanical context the geometric phase is associated with the presence of a specific boundary-law entanglement.

\subsection{Semiclassical analysis of the fuzzy sphere}
\label{sec:semiclassmain}

The results above describe the emergence of a (metastable) geometric fuzzy sphere state at $\nu > \nu_\text{c}$. In this section we recall that in the $\nu \to \infty$ limit the fluctuations of the geometry are classical fields. For finite $\nu > \nu_\text{c}$ the background geometry is well-defined at large $N$, but fluctuations will be described by an interacting (noncommutative) quantum field theory.

In the large $\nu$ limit, the wavefunction can be described semiclassically \cite{Jatkar:2001uh, Dasgupta:2002hx}. We will now briefly review this limit, with details given in the Appendix \ref{sec:semiclass}. These results provide a further useful check on the numerics, and will guide our discussion of entanglement in the following section \ref{sec:entangle}.

The minima of the classical potential occur at:
\be\label{eq:classical}
[X^i,X^j] = i \nu \epsilon^{ijk} X^k \,.
\ee
These are supersymmetric solutions of the classical theory, annihilated by the supercharges (\ref{eq:supercharge}) in the classical limit, and therefore have vanishing energy. The solutions of equations (\ref{eq:classical}) are
\be\label{eq:XJ}
X^i = \nu J^i \,,
\ee
where the $J^i$ are representations of the $\mathfrak{su}(2)$ algebra, $[J^i, J^j] = i \epsilon^{i j k} J^k$. We will be interested here in maximal, $N$-dimensional irreducible representations. (Reducible representations can also be studied, corresponding to multiple polarized D branes.)

The $\mathfrak{su}(2)$ Casimir operator suggests a notion of `radius' given by
\begin{align}\label{eq:rclassical}
    r^2 = \frac{1}{N} \sum_{i = 1}^3 \tr (X^i)^2 = \frac{\nu^2 (N^2 - 1)}{4}.
\end{align}
Indeed, the algebra generated by the $X^i$ matrices tends towards the algebra of functions on a sphere as $N \to \infty$ \cite{Hoppe:1988gk,DEWIT1988545}. At finite $N$, a basis for this space of matrices is provided by the matrix spherical harmonics $\hat{Y}_{j m}$. These obey
\begin{equation}
    \sum_{i = 1}^3 [J^i, [J^i, \hat{Y}_{j m}]] = j (j + 1) \hat{Y}_{j m}, \qquad [J^3, \hat{Y}_{j m}] = m \hat{Y}_{j m} \,.
\end{equation}
We construct the $\hat{Y}_{j m}$ explicitly in Appendix \ref{sec:semiclass}. The $j$ index is restricted to $0 \leq j \leq j_\text{max} = N-1$. The space of matrices therefore defines a regularized or `fuzzy' sphere \cite{Madore:1991bw}.

Matrix spherical harmonics are useful for parametrizing fluctuations about the classical state (\ref{eq:XJ}). Writing
\be\label{eq:pert}
X^i = \nu J^i + \sum_{jm} y^i_{jm} \hat Y_{jm} \,,
\ee
the classical equations of motion can be perturbed about the fuzzy sphere background to give linear equations for the  parameters $y^i_{jm}$. The solutions of these equations define the classical normal modes. We find the normal modes in
Appendix \ref{sec:semiclass}, proceeding as in \cite{Jatkar:2001uh, Dasgupta:2002hx}.
The normal mode frequencies are found to be $\nu \omega$ with
\begin{flalign}
    \omega^2 = 0  & \qquad \text{multiplicity } N^2-1 \,,\nonumber \\
    \omega^2 = j^2 & \qquad \text{multiplicity } 2(j-1)+1 \,, \label{eq:mid}\\
    \omega^2 = (j + 1)^2 & \qquad \text{multiplicity } 2(j+1)+1 \,. \nonumber
\end{flalign}
Recall that $1 \leq j \leq j_\text{max} = N-1$. The three different sets of frequencies in (\ref{eq:mid}) correspond to the group theoretic $\mathfrak{su}(2)$ decomposition $j \otimes 1 = (j-1) \oplus j \oplus (j+1)$. Here $j$ is the `orbital' angular momentum and the $1$ is due to the vector nature of the $X^i$. We will give a field theoretic interpretation of these modes shortly. The modes give the following semiclassical contribution to the energy of the fuzzy sphere state
\be
\left. \Delta E \right|_\text{bos} = \frac{|\nu|}{2} \sum |\omega| = \frac{4 N^3 + 5 N - 9}{6} |\nu| \,. \label{eq:bos}
\ee
This energy is shown in Figure \ref{fig:e}. The scaling as $N^3$ arises because there are $N^2$ oscillators, with maximal frequency of order $N$. This semiclassical contribution will be cancelled out in the supersymmetric sector studied in section \ref{sec:susy} below.

The normal modes (\ref{eq:mid}) can be understood by mapping the matrix quantum mechanics Hamiltonian onto a  noncommutative gauge theory. The analogous mapping for the classical model has been discussed in \cite{Iso:2001mg}.
We carry out this map in Appendix \ref{sec:semiclass}. The original Hamiltonian (\ref{eq:HamB}) becomes the following noncommutative $\mathrm{U}(1)$ gauge theory on a unit spatial $S^2$ (setting the sphere radius to one in the field theory description will connect easily to the quantized modes in (\ref{eq:mid})):
\begin{equation} \label{eq:hamil_main}
    H = \nu \int d \Omega\, \left(\frac{1}{2} (\pi^i)^2 + \frac{1}{4} (f^{i j})^2\right) + \text{const} \,.
\end{equation}
The noncommutative star product $\star$ is defined in the Appendix and
\begin{align}\label{eq:fmain}
    f^{i j} \equiv i \left(L^i a^j - L^j a^i\right) + \epsilon^{i j k} a^k + i \sqrt{\frac{4 \pi}{N \nu^3}} [a^i, a^j]_\star \,,
\end{align}
where the derivatives generate rotations on the sphere
$L^i = -i \epsilon_{i j k} x^j \partial_k$ and $[f, g]_\star \equiv f \star g - g \star f$.
In (\ref{eq:hamil_main}) and (\ref{eq:fmain}) the vector potential $a^i$ can be decomposed into two components tangential to the sphere, that become the two dimensional gauge field, and a component transverse to the sphere, that becomes a scalar field. This decomposition is described in Appendix \ref{sec:semiclass}. The normal modes (\ref{eq:mid}) are coupled fluctuations of the gauge field and the transverse scalar field. The zero modes in (\ref{eq:mid}) are pure gauge modes, given in (\ref{eq:gaugemain}) below. In (\ref{eq:fmain})
the effective coupling controlling quantum field theoretic interactions is seen to be $1/ (N\nu)^{3/2}$. The extra $1/N$ arises because the commutator $[a^i, a^j]_\star$ vanishes as $N \to \infty$, see Appendix \ref{sec:semiclass}. Corrections to the Gaussian fuzzy sphere state are therefore controlled by a different coupling than
that of the `t Hooft expansion (recall $\lambda = N/\nu^3$).

The $\mathrm{SU}(N)$ gauge symmetry generators (\ref{eq:gens}) are realized in an interesting way in the non-commutative field theory description. We see in Appendix \ref{sec:semiclass} that upon mapping to non-commutative fields, the gauge transformations become
\begin{equation}\label{eq:gaugemain}
    \delta a^i = - i L^i y - \sqrt{\frac{4 \pi}{N \nu^3}} (\bm{n} \times \nabla y \cdot \nabla) a^i \,.
\end{equation}
Here $\bm{n}$ is the normal vector and $y(\theta, \phi)$ a local field on the sphere.
The first term in (\ref{eq:gaugemain}) is the usual $\mathrm{U}(1)
$ transformation. The second term describes a coordinate transformation with infinitesimal displacement $\bm{n} \times \nabla y$. Indeed, it is known that non-commutative gauge theories mix internal and spacetime symmetries, which in this case are area-preserving diffeomorphisms of the sphere
\cite{Paniak:2002fi,Lizzi:2001nd}. The emergent $\mathrm{U}(1)$ non-commutative gauge theory thereby realizes the large $N$ limit of the microscopic $\mathrm{SU}(N)$ gauge symmetry, as area-preserving diffeomorphisms \cite{Hoppe:1988gk,DEWIT1988545}.

\begin{figure}[ht]
    \vskip 0.1in
    \centering
    \includegraphics[width=0.8\textwidth]{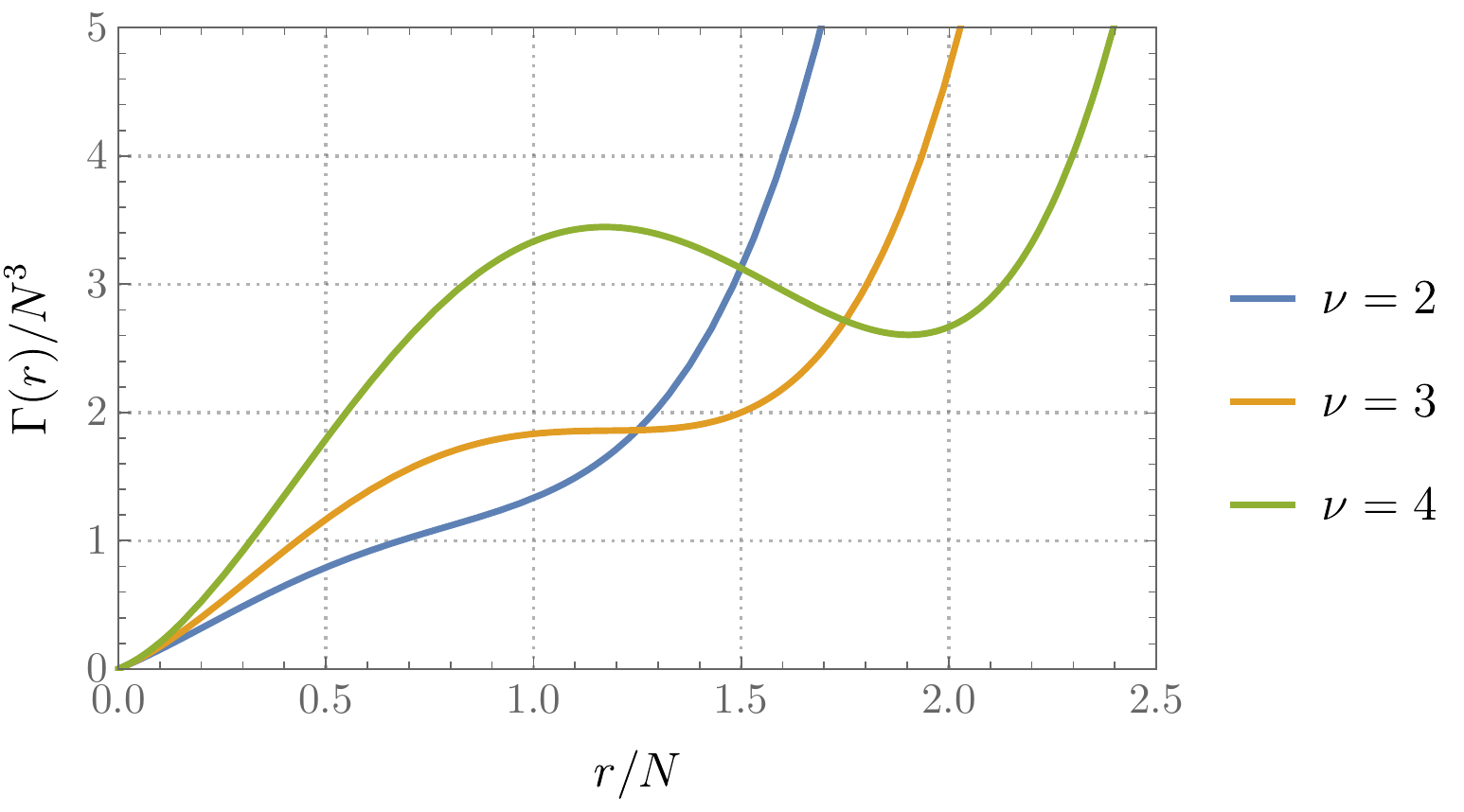}
    \caption{One-loop effective potential $\Gamma(r)$ for the radius of the bosonic ($R=0$) fuzzy sphere as $N \to \infty$. The fuzzy sphere is only metastable when $\nu > \nu^\text{1-loop}_{\text{c},N=\infty} \approx 3.03$, see Appendix \ref{sec:semiclass}.\label{fig:effective}}
    \vskip -0.1in
\end{figure}

The fluctuation modes about the fuzzy sphere background allow a one-loop quantum effective potential for the radius to be computed in Appendix \ref{sec:semiclass}. The potential at $N \to \infty$ is shown in Fig. \ref{fig:effective}. At large $\nu$ the effective potential shows a metastable minimum at $r \sim N \nu / 2$. For $\nu < \nu^\text{1-loop}_{\text{c},N=\infty}$ this minimum ceases to exist. The large $N$, one-loop analysis therefore qualitatively reproduces the behavior seen in Figs. \ref{fig:r} and \ref{fig:e}. The quantitative disagreement is mainly due to finite $N$ corrections. The transition is only sharp as $N \to \infty$.

\subsection{Numerical results, supersymmetric sector}
\label{sec:susy}

\begin{figure}[h!]
    \vskip 0.2in
    \centering
    \includegraphics[width=0.8\textwidth]{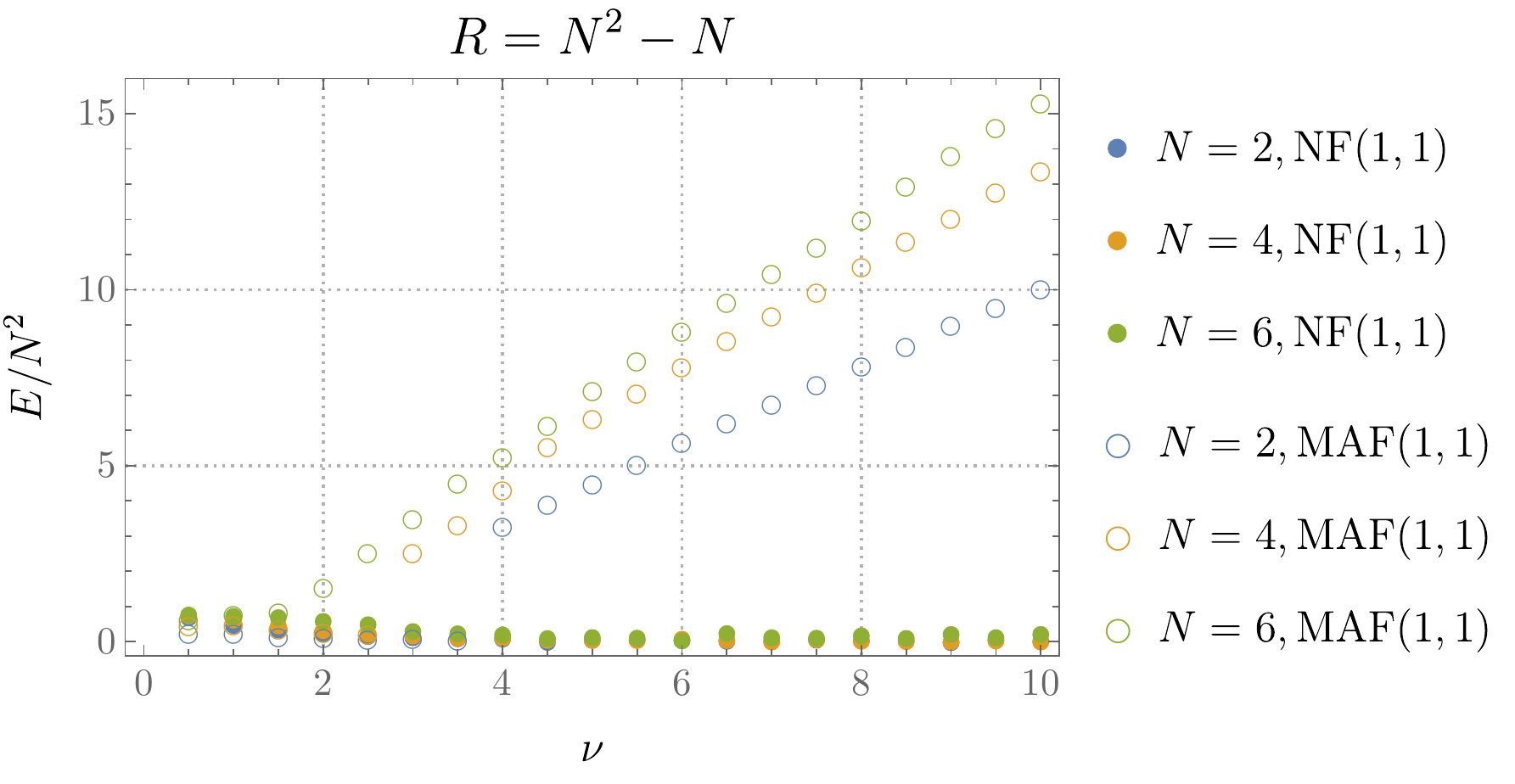}
    \caption{Variational energies in the SUSY sector of the mini-BMN model, for different $N$ and $\nu$. Solid dots are initialized near the fuzzy sphere configuration, and the open markers are initialized near zero.   We are using normalizing and autoregressive flows, respectively, as these produce more accurate variational wavefunctions in the two different regimes.}
    \label{fig:susy_e}
    \vskip -0.1in
\end{figure}

\begin{figure}[h!]
    \vskip 0.2in
    \centering
    \includegraphics[width=0.8\textwidth]{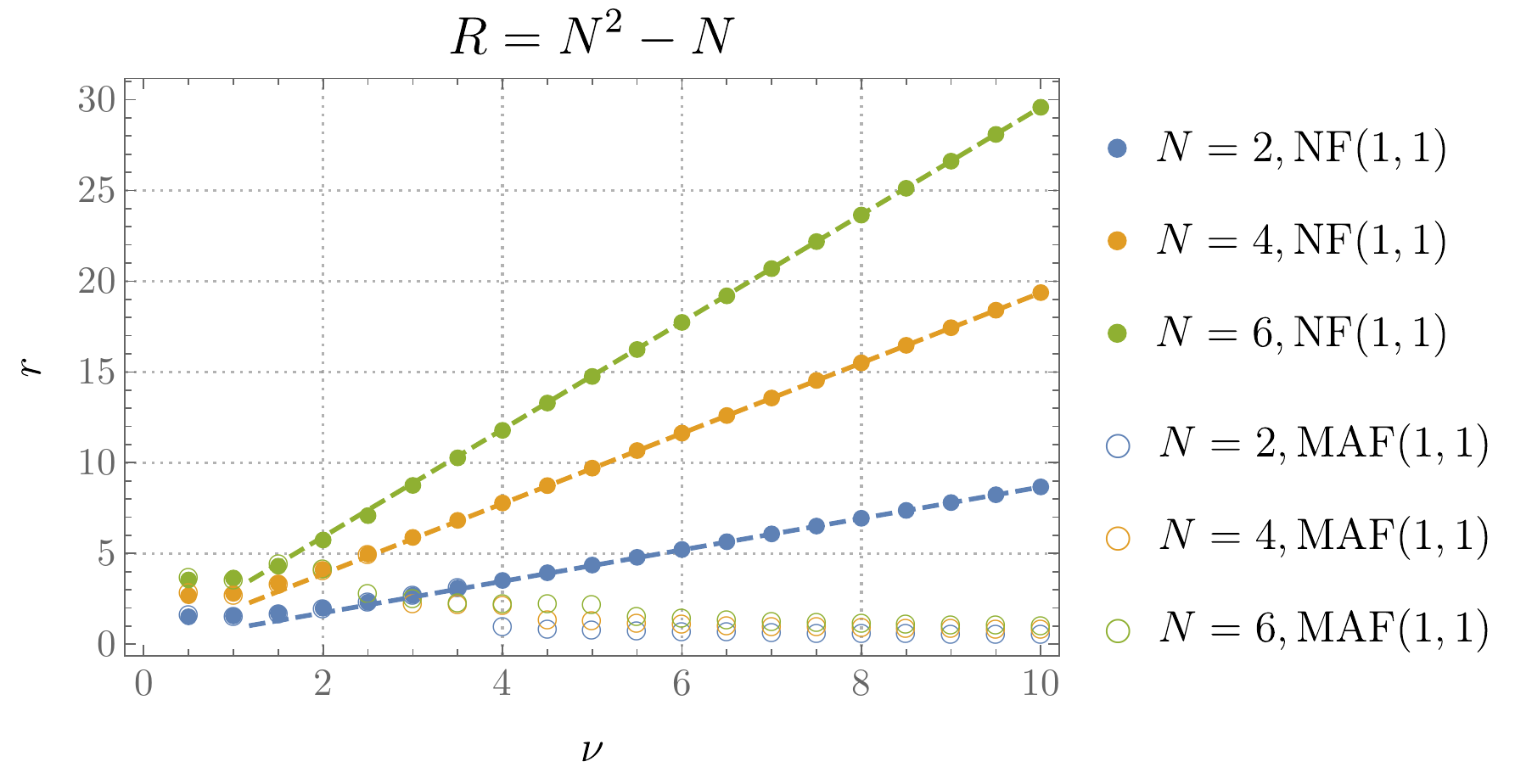}
    \caption{Expectation value of radius in the SUSY sector of the mini-BMN model, for different $N$ and $\nu$. Solid dots are initialized near the fuzzy sphere configuration, and the open markers are initialized near zero. The dashed lines are the semiclassical values (\ref{eq:rclassical}).}
    \label{fig:susy_r}
\end{figure}

We now consider states with fermion number $R = N^2-N$. The fuzzy sphere background is now supersymmetric at large positive $\nu$ \cite{Asplund:2015yda}. The contribution of the fermions to the ground state energy is seen in Appendix \ref{sec:semiclass} to cancel the bosonic contribution (\ref{eq:bos}) at one loop:
\be
-\frac{3}{2} \nu (N^2-1) + \left. \Delta E \right|_\text{fer} + \left. \Delta E \right|_\text{bos} = 0 \,.
\ee
In Figure \ref{fig:susy_e} the variational upper bound on the energy of the fuzzy sphere state remains close to zero for all values of $\nu$.
Figure \ref{fig:susy_r} shows the radius as a function of $\nu$. Probing the smallest values of $\nu$ requires a more powerful wavefunction ansatz than those of Figs. \ref{fig:susy_e} and \ref{fig:susy_r}. We will consider that regime shortly.

In contrast to the states with zero fermion number in Figure \ref{fig:e}, here the fuzzy sphere is seen to be the stable ground state at large $\nu$. However, the fuzzy sphere appears to merge with the collapsed state below a value of $\nu$ that decreases with $N$. This is physically plausible: while the classical fuzzy sphere radius 
$r^2 \sim \nu^2 N^2$ decreases at small $\nu$, quantum fluctuations of the collapsed state are expected to grow in space as $\nu \to 0$. This is because the flat directions in the classical potential of the $\nu = 0$ theory, given by commuting matrices, are not lifted in the presence of supersymmetry \cite{deWit:1997ab}. Eventually, the fuzzy sphere should be subsumed into these quantum fluctuations. This smoother large $N$ evolution towards small $\nu$ (relative to the bosonic sector) is mirrored in the thermal behavior of classical supersymmetric models \cite{Anagnostopoulos:2005cy, Ydri:2012bq}. 

\begin{figure}[h!]
    \vskip 0.2in
    \centering
    \includegraphics[width=0.8\textwidth]{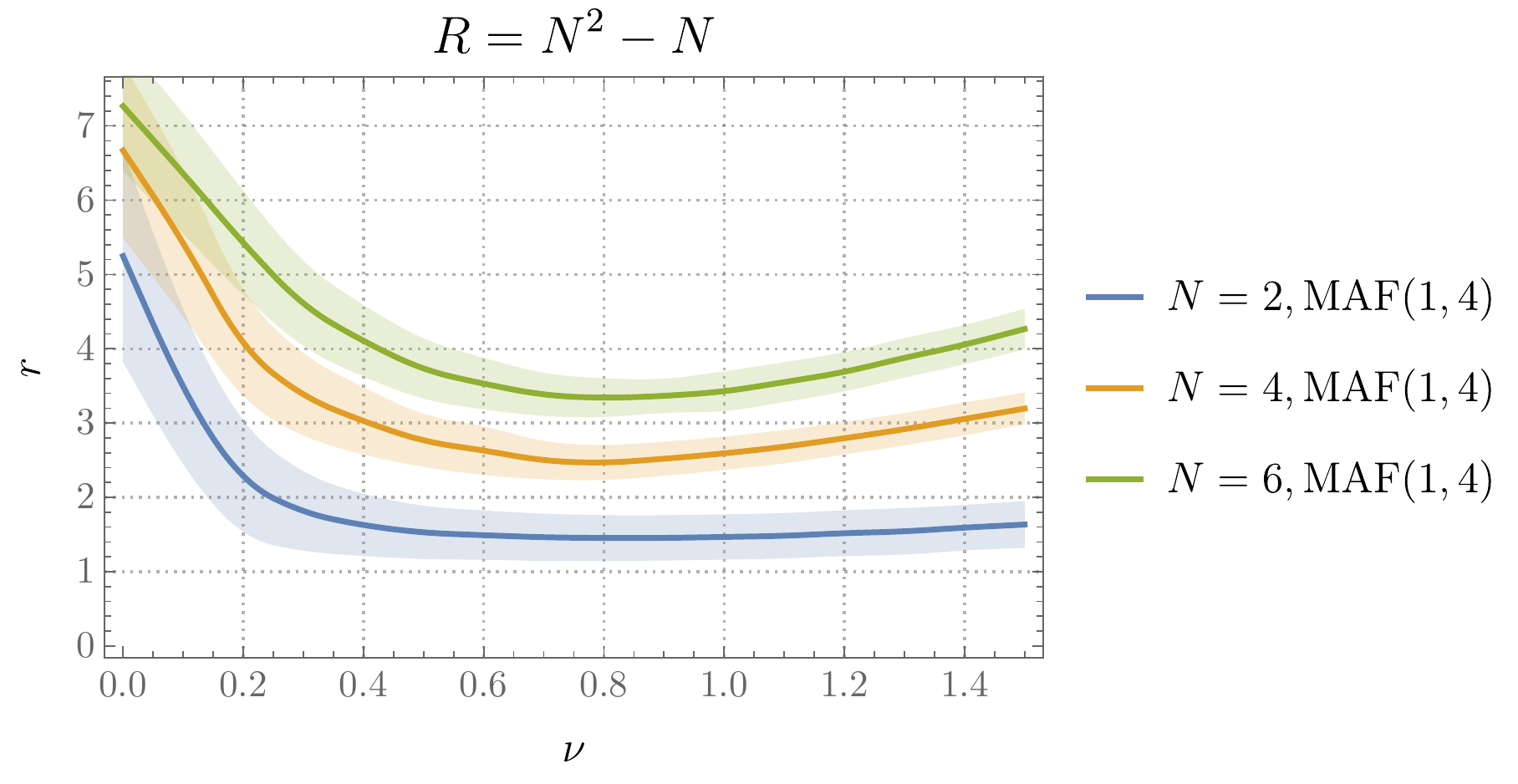}
    \caption{Distribution of radius for different $N$ and small $\nu$. Bands show the standard deviation of the quantum mechanical distribution of $r = \sqrt{\frac{1}{N} \sum \tr X_i^2}$, not to be confused with numerical uncertainty of the average. Recall that the numbers in the brackets are firstly the number of layers in the neural networks, and secondly the number of generalized normal distributions in each base mixed distribution.}
    \label{fig:radius_small}
\end{figure}

Indeed, exploring the small $\nu$ region with more precision we observe a physically expected feature.
In Fig.\,\ref{fig:radius_small} we see that as $\nu$ decreases towards zero, the radius not only ceases to follow the semiclassical decreasing behavior, but turns around and starts to increase. The variance in the distribution of the radius is also seen to increase towards small $\nu$, revealing the quantum mechanical nature of this regime. These behaviors (non-monotonicity of radius and increasing variance) are expected --- and proven for $N=2$ --- because the flat directions of the classical potential at $\nu = 0$ mean that the extent of the wavefunction is set by purely quantum mechanical effects in this limit.

The small $\nu$ regime here is furthermore an opportunity to test the versatility of our variational ansatz away from semiclassical regimes.
In Appendix \ref{app:train} we see that for small $\nu$ MAFs achieve much lower energies than NFs. Increasing the number of distributions in the mixture 
and the number $D$ of free fermions states in (\ref{eq:fermionic_states}) further lowers the energy. These facts mirror the behavior we found in our $N=2$ benchmarking in Sec.\,\ref{sec:benchmark} at small $\nu$, increasing our confidence in the ability of the network to capture this regime for large $N$ also. The error in a variational ansatz is, as always, not controlled and therefore further exploration of this regime is warranted before very strong conclusions can be drawn. We plan to revisit this regime in future work, to search for the possible presence of emergent `throat' geometries as we discuss in Sec.\,\ref{sec:disc} below.

\section{Entanglement on the fuzzy sphere}
\label{sec:entangle}

In this section we will see that the large $\nu$ fuzzy sphere state discussed above contains boundary-law entanglement. To compute the entanglement, one must first define a factorization of the Hilbert space. For our emergent space at finite $N$ and $\nu$ the geometry is both fuzzy and fluctuating, and hence lacks a canonical spatial partition. The fuzziness of the sphere is captured by a toy model of a free field on a sphere with an angular momentum cutoff. Recall
from the previous section \ref{sec:geom} that the noncommutative nature of the fuzzy sphere amounts to  an angular momentum cutoff $\jm = N - 1$. We will start, then, by defining a partition of the space of functions with such a cutoff.

\subsection{Free field with an angular momentum cutoff}

Consider a free massive complex scalar field $\fd(\theta, \phi)$ on a unit two-sphere with the following Hamiltonian:
\begin{equation}
    H = \int_{S^2} d\Omega \, [|\pi|^2 + |\nabla \fd|^2 + \mu^2 |\fd|^2] \,. \label{eq:free_hamil}
\end{equation}
Here $\pi$ is the field conjugate to $\fd$. We impose a cutoff $j \leq \jm$ on the angular momentum, rending the quantum mechanical problem well-defined. 
The fields can therefore be decomposed into a sum of spherical harmonic modes: 
\be
    \fd(\theta, \phi) = \modesum a_{j m} Y_{j m}(\theta, \phi) \, .
\ee
The `wavefunctional' of the quantum field $\fd(\theta, \phi)$ is then a mapping from coefficients $a_{j m}$ to complex amplitudes. The ground state wavefunctional of the Hamiltonian (\ref{eq:free_hamil}) is 
\begin{equation}\label{eq:psi1}
    \psi(a_{j m}) \propto e^{- \sum_{j m} \sqrt{j(j + 1) + \mu^2} |a_{j m}|^2} \, .
\end{equation}

To calculate entanglement for quantum states a factorization of the Hilbert space $\cH = \cH_1 \otimes \cH_2$ is prescribed. To motivate the construction of such a factorization in the fuzzy sphere case, we now review a general framework of defining entanglement in (factorizable) quantum field theories. In quantum mechanics, a quantum state is a function from the configuration space $Q$ to complex numbers, and the Hilbert space of all quantum states is commonly the square integrable functions $\cH = L^2(Q)$. In quantum field theories, the space $Q$ is furthermore a linear space of functions on some geometric manifold $M$, and thus an orthogonal decomposition $Q = Q_1 \oplus Q_2$ induces a factorization of $\cH = L^2(Q_1) \otimes L^2(Q_2)$, which can be exploited to define entanglement. 

To define entanglement it then suffices to find an orthogonal decomposition of the space of fields on the fuzzy sphere. Without an angular momentum cutoff, i.e. with $\jm \to \infty$, there is a natural choice for any region $A$ on the sphere, which sets $Q_1$ to be all functions supported on $A$, and $Q_2$ all functions supported on $\bar{A}$, the complement of $A$. Any function $f$ on $M$ can be uniquely written as a sum of $f_1 \in Q_1$ and $f_2 \in Q_2$, where $f_1 = f \chi_A$ and $f_2 = f (1 - \chi_A)$. Here $\chi_A$ is the function on the sphere that is 1 on $A$ and 0 otherwise. Note that the map of multiplication by $\chi_A$, $f \mapsto f \chi_A$, acts as the projection $Q_1 \oplus Q_2 \to Q_1$. Conversely, given any orthogonal projection operator $P : Q \to Q$, we can decompose $Q = \mathrm{im} P \oplus \ker P$. 

When the cutoff $\jm$ is finite, multiplication by $\chi_A$ will generally take the function out of the subspace of functions with $j \leq \jm$. However, we can still do our best to approximate the projector $P_A^\infty$ of multiplication by $\chi_A$, as defined in the previous paragraph, with a projector $P^\jm_A$ that lives in the subspace with $j \leq \jm$. Formally let $Q^\jm$ be the space of functions on the sphere spanned by $Y_{j m}(\theta, \phi)$ with $j \leq \jm$. Define the orthogonal projector $P_A^\jm : Q^\jm \to Q^\jm$ to minimize the distance $\|P_A^\jm - P_A^\infty\|$. The projector $P^\jm_A$ annihilates all functions in the orthogonal complement of $Q^\jm$, when viewed as an operator acting on $Q^\infty$. It is convenient to choose $\|\cdot\|$ to be the Frobenius norm, and in Appendix \ref{app:ent} an explicit formula for $P^\jm_A$ is obtained.

The projector $P^\jm_A$ then defines a factorization of the Hilbert space $L^2(Q^\jm) = L^2(\mathrm{im} \, P^\jm_A) \otimes L^2(\ker P^\jm_A)$ for any region $A$, and entanglement can be evaluated in the usual way. In particular, the second R\'enyi entropy of a pure state $|\psi\rangle$ on a region $A$ is
\begin{align}
    S_2(\rho_A) &= - \ln \int d x_A d x_{\bar{A}} d x'_A d x'_{\bar{A}}\, \psi(x_A + x_{\bar{A}}) \psi^*(x'_A + x_{\bar{A}}) \psi(x'_A + x'_{\bar{A}}) \psi^*(x_A + x'_{\bar{A}}) \nonumber \\
    &= - \ln \int d x d x'\, \psi(x) \psi^*(P x' + (I - P) x) \psi(x') \psi^*(P x + (I - P) x'), \label{eq:entropy_free}
\end{align}
where $x_A = P x$ and $x_{\bar{A}} = (I - P) x$ are integrated over $\im P$ and $\ker P$, for $P = P_A^\jm$, and $x_A$ and $x_{\bar{A}}$ can be more compactly combined into a field $x$ with $j \leq \jm$. Note that the various $x$'s in (\ref{eq:entropy_free}) denote functions on the sphere.

The projector $P^\jm_A$ is found to have two important geometric features:
\begin{enumerate}
\item The trace of the projector, which counts the number of modes in a region, is proportional to the size of the region. Specifically, at large $\jm$, $\tr P^\jm_A \propto j_\text{max}^2 \left|A\right|$ as is seen numerically in Fig.\,\ref{fig:free_fields1} and understood analytically in Appendix \ref{app:ent}.

\begin{figure}[h]
    \vskip 0.2in
    \centering
    \includegraphics[width=0.7\textwidth]{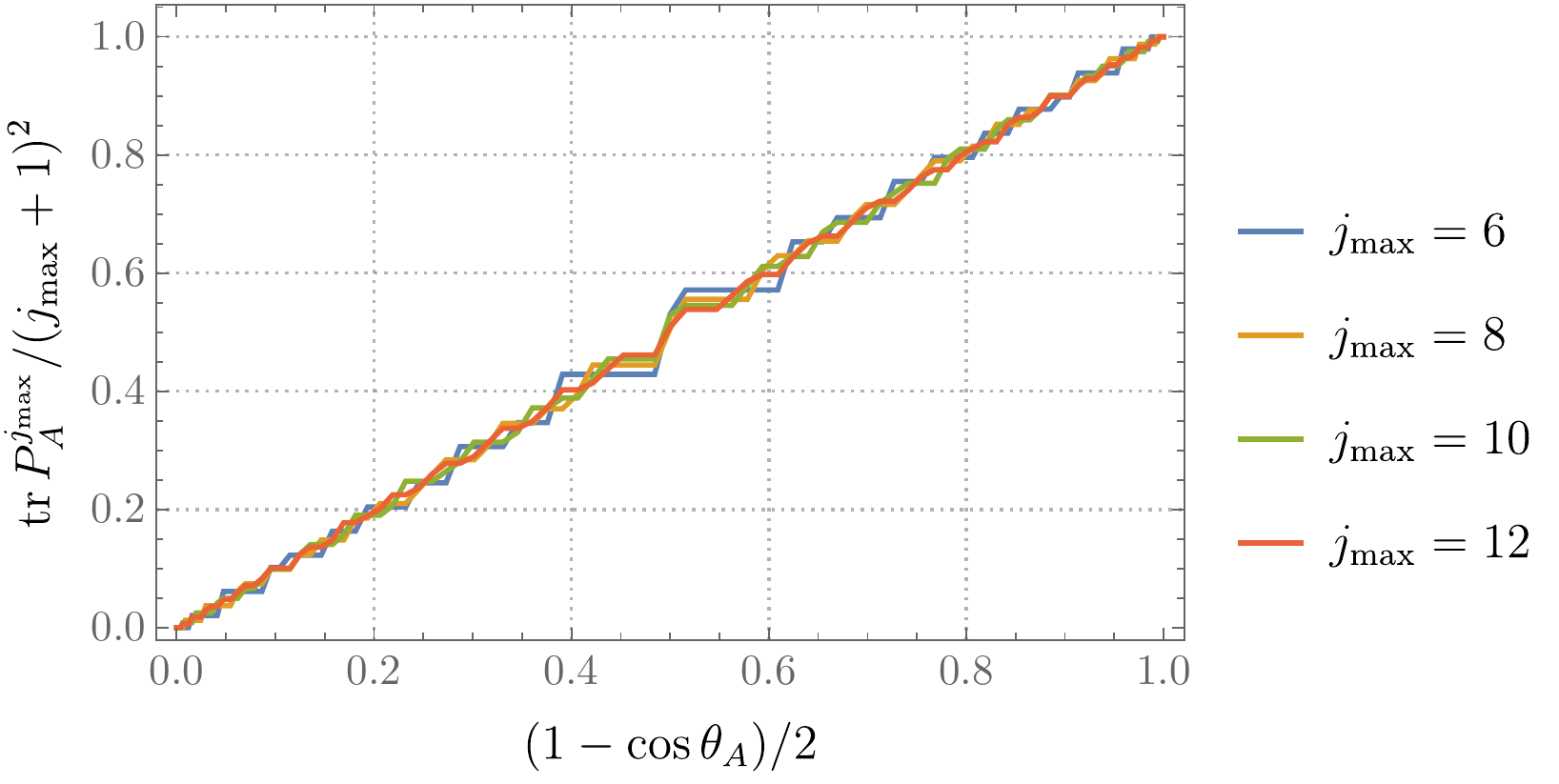}
    \caption{Trace of the projector versus fractional area of the region (a spherical cap with polar angle $\theta_A$), with different angular momentum cutoffs $\jm$. A linear proportionality is observed at large $\jm$. The discreteness in the plot arises because the finite $j_\text{max}$ space of functions cannot resolve all angles.}
    \label{fig:free_fields1}
    \vskip -0.1in
\end{figure}

\item The second R\'enyi entropy defined by the projector follows a boundary law. At large $\jm$, with the mass fixed to $\mu=1$, the entropy
$S_2 \approx 0.03 \, \jm \left|\partial A\right|$ as is seen numerically in Fig.\,\ref{fig:free_fields2} and understood analytically in Appendix \ref{app:ent}.

\begin{figure}[h!]
    \vskip 0.2in
    \centering
    \includegraphics[width=0.7\textwidth]{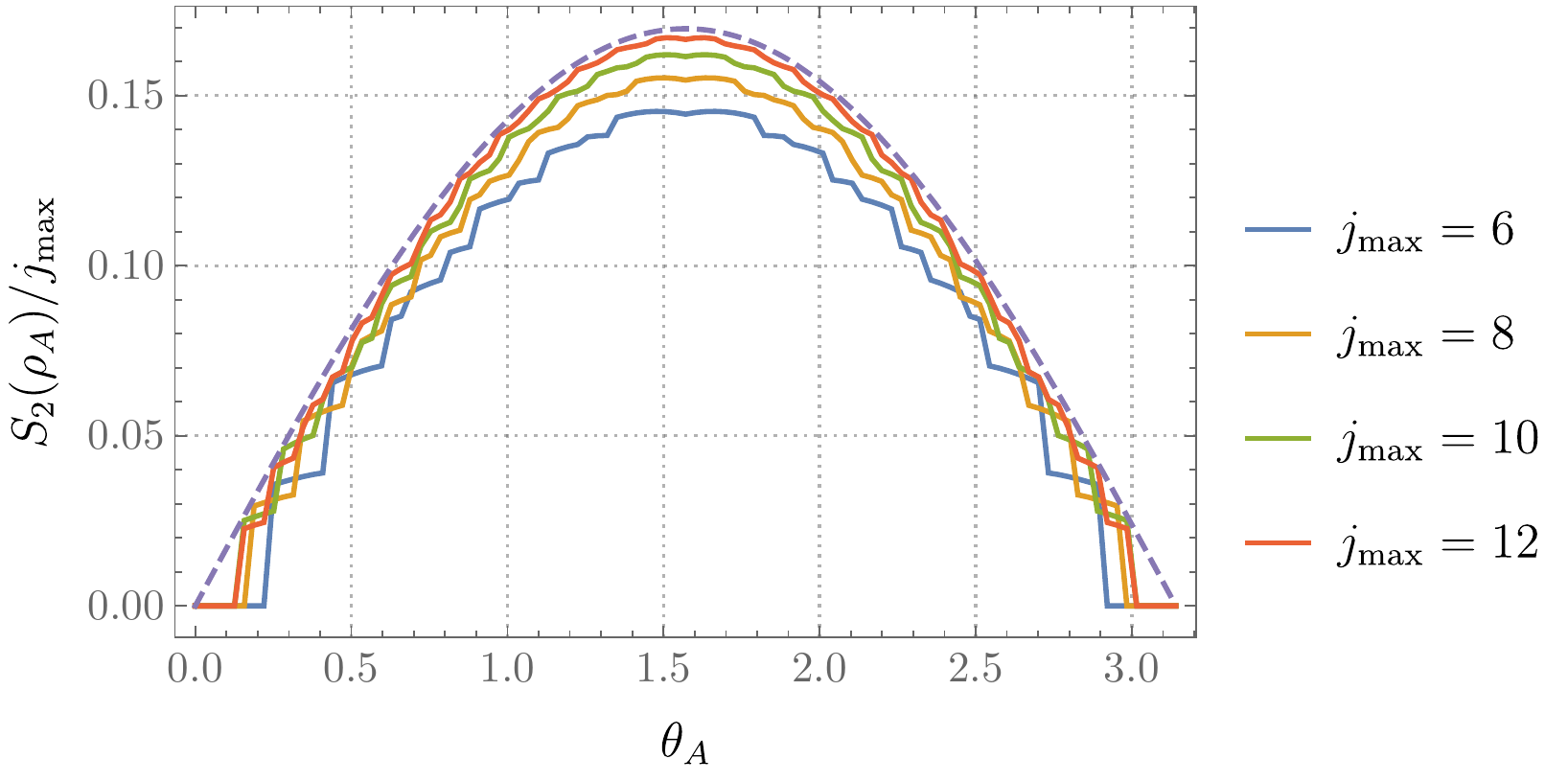}
    \caption{The second R\'enyi entropy for a complex scalar free field (with mass $\mu = 1$) versus the polar angle $\theta_A$ of a spherical cap. The entropy with different cutoffs $\jm$ is shown. At large $\jm$ the curve approaches the boundary law $0.03 \times 2 \pi \sin \theta_A$, shown as a dashed line. Discreteness in the plot is again due to the finite $j_\text{max}$ space of functions.}
    \label{fig:free_fields2}
    \vskip -0.1in
\end{figure}

\end{enumerate}

This boundary entanglement law in Fig. \ref{fig:free_fields2} is of course precisely the expected entanglement in the ground state of a local quantum field \cite{Bombelli:1986rw,Srednicki:1993im}. As the cutoff $\jm$ is removed, the entanglement grows unboundedly.

The partition we have just defined can now be adapted to the fluctuations about the large $\nu$ fuzzy sphere state in the matrix quantum mechanics model. We do this in the following subsection. Intuitively, we would like to replace the $j(j+1) + \mu^2$ spectrum of the free field in the wavefunction (\ref{eq:psi1}) with the matrix mechanics modes (\ref{eq:mid}). Recall that the matrix modes are cut off at angular momentum $j_\text{max} = N - 1$.

\subsection{Fuzzy sphere in the mini-BMN model}


Now we address two additional subtleties that arise when adapting the free field ideas above to the mini-BMN fuzzy sphere. Firstly, the mini-BMN theory is an $\mathrm{SU}(N)$ gauge theory. It is known that entanglement in gauge theories may depend upon the choice of gauge-invariant algebras associated to spatial regions \cite{Casini:2013rba}. Different prescriptions correspond to different boundary or gauge conditions \cite{Lin:2018bud}. However for a fuzzy geometry, the boundaries of regions and gauge edge modes are not sharply defined. To introduce the fewest additional degrees of freedom, we choose to factorize the physical Hilbert space, instead of an extended one \cite{Donnelly:2011hn, Donnelly:2014gva}, to evaluate entanglement in the mini-BMN model. This is similar to the `balanced center' procedure in \cite{Casini:2013rba}, where edge modes are absent.\footnote{It should, nonetheless, be possible to identify meaningful $\mathrm{SU}(N)$ `edge modes' that would reproduce the edge mode contribution of the emergent Maxwell field. This is an especially interesting question in the light of the fact that the microscopic $\mathrm{SU}(N)$ gauge symmetry also acts as an area-preserving diffeomorphism on the emergent fields in (\ref{eq:gaugemain}). This is left for future work.}

Secondly, the emergent fields include fluctuations of the geometry itself. The factorization that we have discussed in the previous subsection is tailored to a region on the sphere, and does not need to approximate a spatial region in other geometries. The partition is even less meaningful in non-geometric regions of the Hilbert space. The variational wavefunction we have constructed can be used to compute entanglement for any given factorization of the Hilbert space, but it is unclear that preferred factorizations exist away from geometric limits. In this work we will focus on the entanglement in the $\nu \to \infty$ limit where the fields are infinitesimal, and hence do not backreact on the spherical geometry. In this limit the factorization is precisely --- up to issues of gauge invariance --- that of the free-field case discussed in the previous subsection.

The matrices corresponding to the infinitesimal fields on the fuzzy sphere are, cf. (\ref{eq:pert}),
\begin{equation}
    A^i = X^i - \nu J^i,
\end{equation}
which should be thought of as living in the tangent space at $X^i = \nu J^i$. At large $\nu$ the wavefunction is strongly supported on the classical configuration and hence in this limit the infinitesimal description is accurate. Gauge transformations then act as
\begin{equation}
    A^i \to A^i + i \epsilon [Y, \nu J^i] + \ldots, \label{eq:inf_gauge}
\end{equation}
where $\epsilon$ is infinitesimal and $Y$ is an arbitrary Hermitian matrix. The $\epsilon [Y, A^i]$ term is omitted in (\ref{eq:inf_gauge}) as it is of higher order. Gauge invariance of the state is manifested as
\begin{equation}
    \psi(\nu J^i + A^i) = \psi(\nu J^i + A^i + i \epsilon [Y, \nu J^i]). \label{eq:inf_gauge_inv}
\end{equation}

Physical states are wavefunctions on gauge orbits $[A^i]$, the set of infinitesimal matrices differing from $A^i$ by a gauge transformation (\ref{eq:inf_gauge}). Similarly to the discussion of free fields above, a partition of the space of gauge orbits is specified by a projector $P$. We will now explain how this projector is constructed. Given a projector $P'$ acting on infinitesimal matrices $ A^i$, a projector acting on gauge orbits can be defined as
\begin{equation}
    P([A^i]) = [P'(A^i)]. \label{eq:def_proj}
\end{equation}
However, for $P$ to be well-defined, $P'$ must preserve gauge directions:
\begin{equation}
    P'(A^i + i \epsilon [Y, \nu J^i]) = P'(A^i) + i \epsilon [Y', \nu J^i], \label{eq:req_proj}
\end{equation}
for any $A^i$, $Y$ and some $Y'$ dependent on $Y$. Let $V$ be the subspace of gauge directions:
\begin{equation}
    V = \{i [Y, J^i] : Y \text{ is Hermitian}\},
\end{equation}
then (\ref{eq:req_proj}) is equivalent to the requirement that $P'(V) \subset V$.
The strategy for finding the projector $P$ is to solve for the projector $P'$ that minimizes $\|P' - \chi_A\|$ subject to the constraint that (\ref{eq:req_proj}) is satisfied. Then $P$ is defined via $P'$ as in (\ref{eq:def_proj}).

The problem of minimizing $\|P' - \chi_A\|$ for orthogonal projectors $P'$ such that $P'(V) \subset V$ is exactly solvable as follows. The condition that $P'(V) \subset V$ is equivalent to imposing that $P' = P_V \oplus P_{V_\perp}$, where $P_V$ is some projector in the subspace $V$ and $P_{V_\perp}$ in its orthogonal complement $V_\perp$. And $\|P' - \chi_A\|$ is minimized if and only if $\|P_V - \left. \chi_A\right|_V\|$ and $\|P_{V_\perp} - \left. \chi_A\right|_{V_\perp}\|$ are both minimized. Via the correspondence between matrix spherical harmonics $\hat{Y}_{j m}$ and spherical harmonic functions $Y_{j m}(\theta, \phi)$ in Appendix \ref{sec:semiclass}, both of these minimizations become the same problem as in the free field case, with a detailed solution in Appendix \ref{app:ent}.

The second R\'enyi entropy, in terms of gauge orbits, is evaluated similarly to (\ref{eq:entropy_free}):
\begin{align} \label{eq:entropy_matrix}
    S_2(\rho_A) &= - \ln \int d [A] d [A'] \, \Delta([A]) \Delta([A']) \nonumber \\
    &\times \psi_\inv([A]) \psi_\inv^*(P [A'] + (I - P) [A]) \psi_\inv([A']) \psi_\inv^*(P [A] + (I - P) [A']),
\end{align}
where $\Delta$ are measure factors for gauge orbits and $\psi_\inv([A]) = \psi(\nu J + A)$. Recall that $\psi$ is gauge invariant according to (\ref{eq:inf_gauge_inv}). The formula (\ref{eq:entropy_matrix}) as displayed does not involve any gauge choice. However, there are some gauges where evaluating (\ref{eq:entropy_matrix}) is particularly convenient. The gauge we choose for this purpose, which is different from that in section \ref{sec:fix}, is that $A \in V_\perp$, i.e., the fields are perpendicular to gauge directions. In this gauge measure factors are trivial and the projector is simply $P_{V_\perp}$ that minimizes $\|P_{V_\perp} - \left. \chi_A\right|_{V_\perp}\|$:
\begin{align} \label{eq:entropy_fixed}
    S_2(\rho_A) &= - \ln \int_{V_\perp} d A d A' \, \nonumber \\
    & \times \psi_\perp(A) \psi_\perp^*(P_{V_\perp} A' + (I - P_{V_\perp}) A) \psi_\perp(A') \psi_\perp^*(P_{V_\perp} A + (I - P_{V_\perp}) A'),
\end{align}
where $\psi_\perp(A)$ is defined as $\psi(\nu J + A)$ for $A \in V_\perp$.\footnote{We can find a gauge transformation $U \in \mathrm{SU}(N)$ mapping any matrices $X^i$ into this perpendicular gauge as follows. We are looking for $\wtd{X}^i = U X^i U^{-1}$, such that $\wtd{X}^i - \nu J^i \in V_\perp$. This means that $\sum_i \tr \left( [Y, J^i]^\dagger (\wtd{X}^i - \nu J^i)\right) = 0$ for any Hermitian matrix $Y$. Equivalently, $\sum_i \tr \left( J^i [Y, \wtd{X}^i] \right) = 0$ for any $Y$. This is achieved by numerically finding the $U$ that maximizes the overlap $\sum_i \tr \left(J^i U X^i U^{-1}\right)$.}

\begin{figure}[h!]
    \vskip 0.2in
    \centering
    \includegraphics[width=0.8\textwidth]{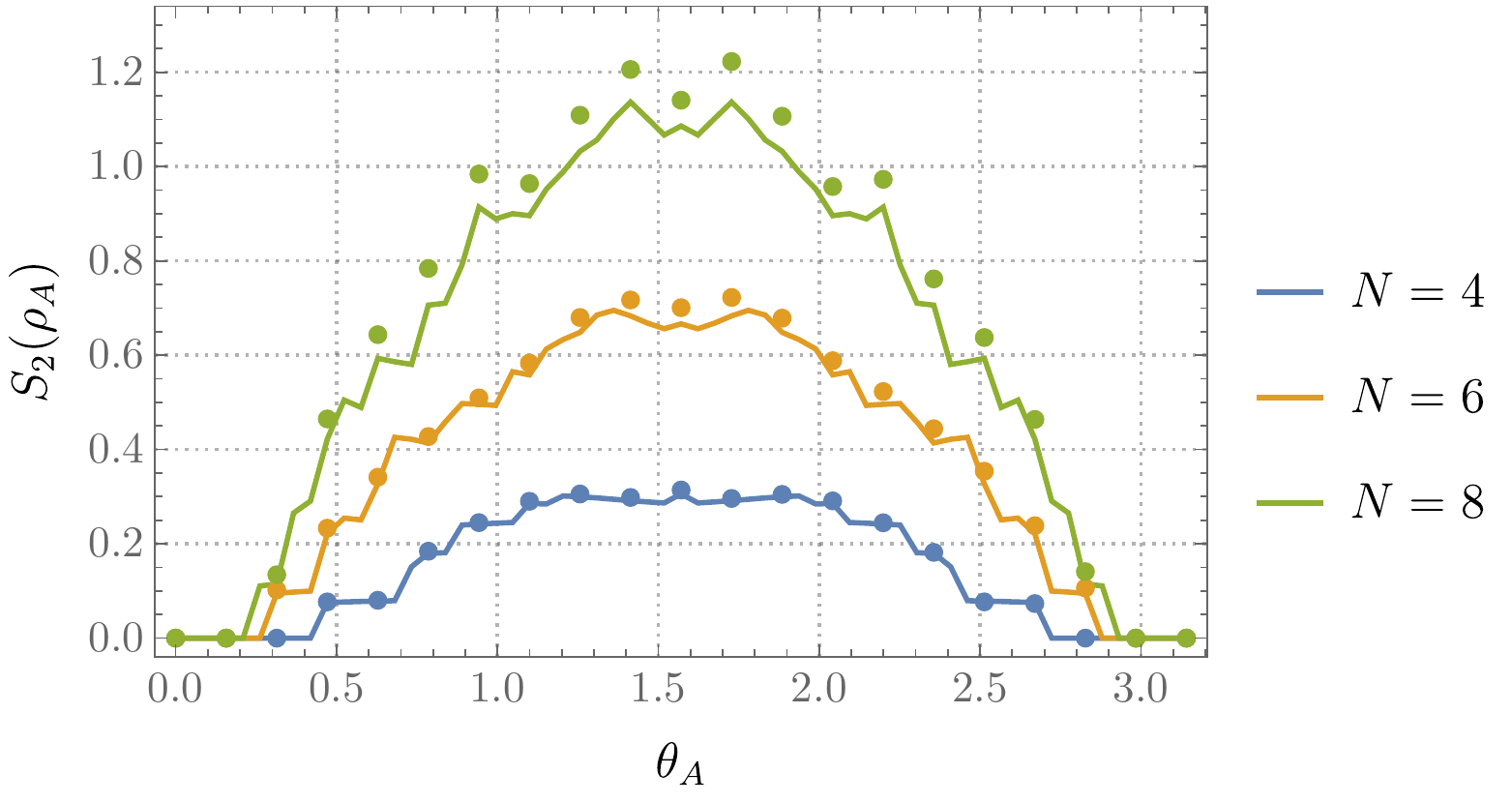}
    \caption{The second R\'enyi entropy for a spherical cap on the matrix theory fuzzy sphere versus the polar angle $\theta_A$ of the cap. Solid curves are exact values at $\nu = \infty$ and dots are numerical values from variational wavefunctions at $\nu = 10$ for different $N$. The wavefunctions are NF(1, 1) in the zero fermion sector as shown in Figs.\,\ref{fig:r} and \ref{fig:e}.}
    \label{fig:fuzzy_ent}
    \vskip -0.1in
\end{figure}

The bosonic fuzzy sphere wavefunction can be written in the $\nu \to \infty$ limit as follows. As in (\ref{eq:pert}), the perturbations can be decomposed as
$A^i = \sum_{a} \delta x_a \sum_{jm} y^{i}_{jma} \hat Y_{jm} \,,$
where the $y^{i}_{jma}$ diagonalize the potential energy at quadratic order in $A$ so that $V = \frac{\nu^2}{2} \sum_a \omega_a^2 (\delta x_a)^2 + \cdots$ (see Appendix \ref{sec:semiclass}). The wavefunction is then, analogously to (\ref{eq:psi1}),
\begin{equation}
    \psi_\perp(A) \propto e^{ - \frac{|\nu|}{2} \sum_a |\omega_a| (\delta x_a)^2 }. \label{eq:exact_wavefunc}
\end{equation}
The frequencies are given by (\ref{eq:mid}), excluding the pure gauge zero modes. Using this wavefunction, the R\'enyi entropy (\ref{eq:entropy_fixed}) can be computed exactly and is shown as a solid line in Fig.\,\ref{fig:fuzzy_ent}. As $N \to \infty$ these curves approach a boundary law
\be\label{eq:Nscaling}
S_2(\rho_A) \approx 0.03 \, N  \left|\partial A\right| \,.
\ee
Here $\left|\partial A\right| = 2 \pi \sin \theta_A$ is again the circumference of the spherical cap $A$ (in units where the sphere has radius one, consistent with the field theoretic description in (\ref{eq:hamil_main})).
The result (\ref{eq:Nscaling}) is the same as that of the toy model in Fig.\,\ref{fig:free_fields2}, with $j_\text{max}$ now set by the microscopic matrix dynamics to be $N - 1$.\footnote{A (simpler) instance of entanglement revealing the inherent graininess of a spacetime built from matrices is two dimensional string theory \cite{Das:1995jw,Hartnoll:2015fca}.} This regulated boundary-law entanglement underpins the emergent locality on the fuzzy sphere at large $N$ and $\nu$. Recall from the discussion around (\ref{eq:hamil_main}) that there are only two emergent fields on the sphere: a Maxwell field and a scalar field. The perpendicular gauge choice we have made translates into the Coulomb gauge for the emergent Maxwell field, cf. the discussion around (\ref{eq:gaugemain}) above. The factor of $N$ in (\ref{eq:Nscaling}) is due to the microscopic cutoff at a scale $L_\text{fuzz} \sim L_\text{sph}/N$.

Previous works on the entanglement of a free field on a fuzzy sphere involved similar wavefunctions but a different factorization of the Hilbert space, which was inspired instead by coherent states \cite{Dou:2006ni, Karczmarek:2013jca, Okuno:2015kuc, Chen:2017kfj}. Those results did not always produce boundary-law entanglement. Here we see that the UV/IR mixing in noncommutative field theories does not preclude a partition of the large $N$ and large $\nu$ Hilbert space with a boundary-law entanglement.

We can also evaluate the entropy (\ref{eq:entropy_fixed}) using the large $\nu$ variational wavefunctions, without assuming the asymptotic form (\ref{eq:exact_wavefunc}). The results are shown as dots in Fig.\,\ref{fig:fuzzy_ent}. However, we stress that only the $\nu \to \infty$ limit has a clear physical meaning, where fluctuations are infinitesimal. The variational results are close to the exact values in Fig.\,\ref{fig:fuzzy_ent}, showing that the neural network ansatz captures the entanglement structure of these matrix wavefunctions. 

The results in this section are for the bosonic fuzzy sphere.
The projection we have introduced in order to partition the space of matrices can be extended in a similar, but more involved, way to factorize the fermionic Hilbert space.

\section{Discussion}
\label{sec:disc}

We have seen that neural network variational wavefunctions capture in detail the physics of a semiclassical spherical geometry that emerges in the mini-BMN model (\ref{eq:fermionic_potential}) at large $\nu$. Away from the semiclassical limit, the spherical geometry either abruptly or gradually collapses towards a new state. In Fig. \ref{fig:radius_small} we saw that in the `supersymmetric' sector this new state was characterized by an increase in both the expectation value and quantum mechanical variance of the radius as $\nu \to 0$.
To understand the physics of this process, and to start thinking about the nature of the collapsed state as $\nu \to 0$, it is helpful to consider the string theoretic embedding of the model.

The mini-BMN model can be realized in string theory as the description of $N$ D-particles in an AdS$_4$ spacetime. Let us review some aspects of this realization \cite{Asplund:2015yda}. The parameter
\be\label{eq:string1}
\frac{1}{\nu^3} \sim g_\text{s} \left(\frac{L_\text{AdS}}{L_\text{s}}\right)^3 \,. 
\ee
Here $L_\text{AdS}$ is the AdS radius, $L_\text{s}$ is the string length and $g_\text{s}$ is the string coupling. The proportionality in (\ref{eq:string1}) depends on the volume, in units of the string length, of internal cycles wrapped by the branes in the compactification down to AdS$_4$. In particular, the mass of a single D-particle goes like $1/g_\text{s}$ times the wrapped internal volume. The strength of the gravitational backreaction of $N$ coincident D-particles is then controlled by $G_N \cdot N/g_\text{s}$. Here $G_N \sim g_\text{s}^2$ is the four dimensional Newton constant, where we have suppressed a factor of the volume of the compactification manifold. Therefore, if we keep the AdS radius fixed in string units, gravitational backreation becomes important when $g_\text{s} N \sim N/\nu^3 \gtrsim 1$. Up to factors of the volume of compactification cycles, this is equivalent to the statement that the dimensionless 't Hooft coupling $\lambda=N/\nu^3$, introduced below (\ref{eq:HamB}), becomes large.

For $N/\nu^3 \lesssim 1$, then, the D-particles can be treated as light probes on the background AdS spacetime. The fuzzy sphere configuration describes a polarization of the D-particles into spherical `dual giant gravitons'. From the string theory perspective, this polarization is driven by the 4-form flux $\Omega\sim 1/L_\text{AdS}$ supporting the background AdS$_4$ spacetime. Together with the discussion in the previous paragraph on the strength of the gravitational interaction, we can write the heuristic relation $N/\nu^3 \sim \text{gravity}/\text{flux}$. At large $\nu$ the flux wins out and semiclassical fuzzy spheres can exist, but at small $\nu$ gravitational forces cause the spheres to collapse. The entanglement and emergent locality that we have described in this paper is that of the polarized spheres, whose excitations are described by the usual gauge fields and transverse scalar fields of string theoretic D-branes.

For $N/\nu^3 \gg 1$ it is possible that the strongly interacting, collapsed D-particles will develop a geometric `throat', in the spirit of the canonical holographic correspondence \cite{Maldacena:1997re}.
It is not well-understood when such a throat would be captured by the mini-BMN matrix quantum mechanics. The variational wavefunctions that we have developed here provide a new window into this problem. In particular, we hope to investigate the small $\nu$ collapsed state in more detail in the future, with the objective of revealing any entanglement associated to emergent local dynamics in the throat spacetime. If the emergent dynamics includes gravity, there are two potentially interesting complications. Firstly, the entanglement of bulk fields may be entwined with entanglement due to the `stringy' degrees of freedom that seem to be manifested in the Bekenstein-Hawking entropy of black holes as well as in the Ryu-Takayanagi formula \cite{Susskind:1994sm,Fiola:1994ir,Bianchi:2012ev,Faulkner:2013ana}. Secondly, and perhaps relatedly, it may become crucial to understand the `edge mode' contribution to the entanglement, that we have avoided in our discussion here \cite{Donnelly:2016auv, Harlow:2016vwg}.

More generally, the methods we have developed will be applicable to a wide range of quantum problems of interest in the holographic correspondence.
The benefit of the variational neural network approach is direct access to properties of the zero temperature quantum mechanical state. Optimizing the numerical methods and variational ansatz further, and with more computational power, it should not be difficult to work with larger values of $N$. In addition to understanding the emergence of spacetime from first principles, it should also be possible to study, for example, the microstates and dynamics of quantum black holes.

\section*{Acknowledgements}

It is a pleasure to thank Frederik Denef and Xiaoliang Qi for helpful discussions, Aitor Lewkowycz, Raghu Mahajan and Edward Mazenc for comments on the draft, and
Tarek Anous for sharing his code with us. We also thank Zhaoheng Guo and Yang Song for collaboration on a related project. SAH is partially funded by DOE award de-sc0018134. XH is supported by a Stanford Graduate Fellowship. Computational work was performed on the Sherlock cluster at Stanford University, with the TensorFlow code for the project available online.

\newpage

\providecommand{\href}[2]{#2}\begingroup\raggedright\endgroup

\newpage
\appendix

\section{Geometry of the gauge} \label{app:gauge}

\subsection*{Gauge invariant sampling}

In the procedure of sampling bosonic matrices $X$ according to the wavefunction probability distribution $|\psi(X)|^2 = |f(X)|^2$, it is asserted in the main text that $X \sim |f(X)|^2$ if we let $X = U \wtd{X} U^{-1}$ where $U$ is a Haar random element in $\mathrm{SU}(N)$ and the representative of the gauge orbit $\wtd{X} \sim \Delta(\wtd{X})|\wtd{f}(\wtd{X})|^2$. A proof of this assertion, along with a more precise definition of the gauge orbit measure $\Delta$, is presented here. 

To simplify notation, denote $\wtd{X} \sim \wtd{p}(\wtd{X})$. If the random variable $X = U \wtd{X} U^{-1}$, it follows the probability distribution
\begin{equation}
    p(X = X_0) = \int d U d \wtd{X}\, \wtd{p}(\wtd{X}) \delta(U \wtd{X} U^{-1} = X_0),
\end{equation}
where the integral over $\mathrm{SU}(N)$ is with respect to the normalized Haar measure, and $\delta$ is the Dirac delta distribution. For almost any $X_0$, there is a unique gauge representative $\wtd{X}_0$, with a discrete set of $U_i \in \mathrm{SU}(N)$ ($i = 1, 2, \ldots, N$), such that $U_i \wtd{X}_0 U_i^{-1} = X_0$. These unitaries differ by an overall phase (powers of $\exp (i 2 \pi / N)$). Hence
\begin{equation}
    p(X = X_0) = \wtd{p}(\wtd{X}_0) \sum_{i = 1}^N |J^{-1}(\wtd{X}_0, U_i)|,
\end{equation}
where $J$ is the Jacobian determinant of the map $(\wtd{X}, U) \mapsto U \wtd{X} U^{-1}$. As will be seen in the next subsection, $J(\wtd{X}, U) = J(\wtd{X})$ does not depend on the unitary $U$. So if we assign
\begin{equation}
    \Delta(\wtd{X}) = N^{-1} |J(\wtd{X})|, \label{eq:orbit_measure}
\end{equation}
and note $\wtd{p}(\wtd{X}) = \Delta(\wtd{X}) |\wtd{f}(\wtd{X})|^2$,
\begin{equation}
    p(X = X_0) = N^{-1} |J(\wtd{X}_0)| |\wtd{f}(\wtd{X}_0)|^2 \sum_{i = 1}^N |J^{-1}(\wtd{X}_0)| = |\wtd{f}(\wtd{X}_0)|^2 = |f(X_0)|^2,
\end{equation}
for a gauge invariant wavefunction (\ref{eq:gauge_invariant_wavefunc}). This is the desired result. 

\subsection*{Derivation of the gauge orbit measure}

From (\ref{eq:orbit_measure}), the gauge orbit measure $\Delta$ is given by the Jacobian determinant $J$ of the map $X: (\wtd{X}, U) \mapsto U \wtd{X} U^{-1}$. Recall that for a general mapping $F$ between smooth manifolds of equal dimension $S \to T$, the Jacobian determinant can be written in terms of the pullback of the volume form
\begin{equation}
    F^*(\omega_T) = J \omega_S, \label{eq:jacobian}
\end{equation}
where $\omega_S$ and $\omega_T$ are volume forms on $S$ and $T$. That is, $J$ is the ratio of the volume element after and before the mapping. If $x_i$ and $y_i$ are two orthonormal coordinate systems at $x \in S$ and $y = F(x) \in T$, in terms of the wedge product,
\begin{equation}
    \omega_S = \bigwedge_i d x_i, \quad \omega_T = \bigwedge_i d y_i, \quad F^*(d y_i) = \sum_j \frac{\partial y_i}{\partial x_j } d x_j \,.
\end{equation}
Therefore equation (\ref{eq:jacobian}) can be expressed more explicitly as
\begin{equation}
    \bigwedge_i \sum_j \frac{\partial y_i}{\partial x_j } d x_j = J \bigwedge_i d x_i \quad \Leftrightarrow \quad J = \det \frac{\partial y_i}{\partial x_j}.
\end{equation}

We would like to show firstly that $J(\wtd{X}, U)$ does not depend on $U$. Note that the map $X: (\wtd{X}, U) \mapsto U \wtd{X} U^{-1}$ is equivariant with respect to the following actions of $G = \mathrm{SU}(N)$: for any $U' \in G$, in the base space $U' \cdot (\wtd{X}, U) = (\wtd{X}, U' U)$, and in the target space $U' \cdot X = U' X U'^{-1}$. And the two actions preserve the volume forms, because the Haar measure is left invariant and the metric $\tr d X^\dagger d X$ is invariant under matrix conjugation. Hence the Jacobian $J(\wtd{X}, U) = J(\wtd{X})$ is independent of $U$. 

We will obtain the Jacobian by explicitly computing the pullback of the volume form at $X$.
As the Jacobian does not depend on $U$, it is convenient to evaluate it at $U = I$. To further simplify the computation, we shall complexify the cotangent spaces, which does not change the Jacobian determinant. The $\mathfrak{su}(N)$ real Lie algebra is complexified to $\mathfrak{sl}(N)$, and the following basis $\{D_i, E_{i j}\}$ of $\mathfrak{sl}(N)$ is employed. The basis is orthonormal with respect to the matrix inner product $\tr X^\dagger Y$:
\begin{enumerate}
\item For $1 \leq i \leq N - 1$, $D_i$ is a diagonal matrix with $(D_i)_{j j} = 1 / \sqrt{i (i + 1)}$ for $1 \leq j \leq i$, $(D_i)_{j j} = -(j - 1) / \sqrt{i (i + 1)}$ for $j = i + 1$ and $(D_i)_{j j} = 0$ for $j > i + 1$. 

\item For $1 \leq i, j \leq N$ and $i \neq j$, $E_{i j}$ is the matrix that has only one nonzero entry $(E_{i j})_{i j} = 1$. 
\end{enumerate}
A general element in the complexified cotangent space of $\wtd{X}$ is (with the gauge choice defined in the main text)
\begin{align}
    d \wtd{X}^1 &= \sum_{i = 1}^{N - 1} D_i d \wtd{c}_i^1, \quad d \wtd{X}^3 = \sum_{i = 1}^{N - 1} D_i d \wtd{c}_i^3 + \sum_{1 \leq i \neq j \leq N} E_{i j} d \wtd{e}_{i j}^3,\nonumber \\
    d \wtd{X}^2 &= \sum_{i = 1}^{N - 1} D_i d \wtd{c}_i^2 + \sum_{i = 1}^{N - 1} \frac{1}{\sqrt{2}}\left(E_{i (i + 1)} - E_{(i + 1) i}\right) d \wtd{e}^2_{i (i + 1)} + \sum_{1 \leq i \neq j \leq N}^{|i - j| \neq 1} E_{i j} d \wtd{e}_{i j}^2, \label{eqs:tildeX}
\end{align}
where the superscript $i = 1, 2, 3$ denotes three bosonic matrices. The equations (\ref{eqs:tildeX}) thus define a basis $\{d \wtd{c}^1_i, d \wtd{c}^2_i, d \wtd{e}^2_{i (i + 1)}, d \wtd{e}^2_{i j}, d \wtd{c}^3_i, d \wtd{e}^3_{i j}\}$ of the complexified cotangent space of $\wtd{X}$.

The complexified cotangent space of $\mathrm{SU}(N)$ at $U = I$ is isomorphic to the Lie algebra $\mathfrak{sl}(N)$, so that (introducing basis forms $dc_i, de_{ij}$):
\begin{equation}
    -i d U = \sum_{i = 1}^{N - 1} D_i d c_i + \sum_{1 \leq i \neq j \leq N} E_{i j} d e_{i j}. \label{eq:U}
\end{equation}
The differential of the map $X: (\wtd{X}, U) \mapsto U \wtd{X} U^{-1}$ at $U = I$ is
\begin{equation}
    d X = [d U, \wtd{X}] + d \wtd{X}, \label{eq:diff}
\end{equation}
and the cotangent space of $X$ is complexified to three copies of $\mathfrak{sl}(N)$, so that (introducing basis forms $dc^k_i, de^k_{ij}$):
\begin{equation}
    d X^k = \sum_{i = 1}^{N - 1} D_i d c_i^k + \sum_{1 \leq i \neq j \leq N} E_{i j} d e_{i j}^k.
\end{equation}
Substituting (\ref{eqs:tildeX}) and (\ref{eq:U}) into (\ref{eq:diff}), recalling that $\wtd{X}^1$ is diagonal, and equating the expressions for $d X^1$ we have 
\begin{equation}
    d c^1_i = d \wtd{c}^1_i, \quad d e^1_{i j} = i \left(\wtd{X}^1_{j j} - \wtd{X}^1_{i i}\right) d e_{i j}. \label{eq:diff1}
\end{equation}
Equating the expressions for $d X^2$ gives, with terms that drop out of the final result omitted:
\begin{align}
    d c^2_i &= d \wtd{c}^2_i + (\text{terms with } d e), 
    \quad d e^2_{i j} = d \wtd{e}^2_{i j} + (\text{terms with } d c, d e), 
    \nonumber \\
    d e^2_{i (i + 1)} &= + i \wtd{X}^2_{i (i + 1)} \sqrt{\frac{i + 1}{i}} d c_i + \frac{1}{\sqrt{2}} d \wtd{e}^2_{i (i + 1)} + (\text{terms with } d c_{i - 1}, d e), \nonumber \\ 
    d e^2_{(i + 1) i} &= - i \wtd{X}^2_{(i + 1) i} \sqrt{\frac{i + 1}{i}} d c_i - \frac{1}{\sqrt{2}} d \wtd{e}^2_{i (i + 1)} + (\text{terms with } d c_{i - 1}, d e), \label{eq:diff2}
\end{align}
where the expression for $d e^2_{i j}$ holds for $|i - j| \neq 1$ and the prefactor $i$ in the expressions for $d e^2_{i (i + 1)}$ and $d e^2_{(i + 1) i}$ is the imaginary unit. Subscripts are omitted if that term with any subscript is unimportant, e.g., $d e$ means linear combinations of $d e_{i j}$ for $1 \leq i \neq j \leq N$. Similarly 
\begin{equation}
    d c^3_i = d \wtd{c}^3_i + (\text{terms with } d e), \quad d e^3_{i j} = d \wtd{e}^3_{i j} + (\text{terms with } d c, d e). \label{eq:diff3}
\end{equation}

The Jacobian determinant $J$ is evaluated as, schematically, 
\begin{equation}
    d c^1_i \wedge d e^1_{i j} \wedge d c^2_i \wedge d e^2_{i j} \wedge d c^3_i \wedge d e^3_{i j} = J\, d \wtd{c}^1_i \wedge d \wtd{c}^2_i \wedge d \wtd{e}^2_{i (i + 1)} \wedge d \wtd{e}^2_{i j} \wedge d \wtd{c}^3_i \wedge d \wtd{e}^3_{i j} \wedge d c_i \wedge d e_{i j}, \label{eq:diff_jacobian}
\end{equation}
where $d e^1_{i j}$ denotes $\bigwedge_{i j} d e^1_{i j}$ for $1 \leq i \neq j \leq N$, for example. Substitution of (\ref{eq:diff1}), (\ref{eq:diff2}) and (\ref{eq:diff3}) into the left-hand side of (\ref{eq:diff_jacobian}) yields a sum of wedge products of differentials. The wedge product is nonzero only if each factor on the right-hand side of (\ref{eq:diff_jacobian}) appears exactly once. Now observe that $d e_{i j}$ already appears in $d e^1_{i j}$ in (\ref{eq:diff1}), hence all $d e_{i j}$ terms in other factors can be safely ignored. 

With the $d e_{i j}$ ignored, $d e^2_{i (i + 1)} \wedge d e^2_{(i + 1) i}$ is proportional to $d c_i \wedge d \wtd{e}^2_{i (i + 1)}$ for $i = 1$, because for any differential $d a$, $d a \wedge d a = 0$. Then remaining factors of $d c_1$ and $d \wtd{e}^2_{1 2}$ can be ignored. Next, for $i = 2$, $d e^2_{i (i + 1)} \wedge d e^2_{(i + 1) i}$ must be proportional to $d c_i \wedge d \wtd{e}^2_{i (i + 1)}$ as well, up to terms that can be ignored. In the end we have (note that $\wtd{X}^2_{i (i + 1)} = - \wtd{X}^2_{(i + 1) i}$ is purely imaginary)
\begin{equation} \label{eq:factors}
    \bigwedge_{i = 1}^{N - 1} d e^2_{i (i + 1)} \wedge d e^2_{(i + 1) i} = \sqrt{2^{N - 1} N} \bigwedge_{i = 1}^{N - 1} \Im{\wtd{X}^2_{i (i + 1)}} d c_i \wedge d \wtd{e}^2_{i (i + 1)} + (\text{terms with } d e).
\end{equation}

Now terms with $d c_i$ can be ignored as well as they appear in (\ref{eq:factors}). With the $d c_i$ and $d e_{i j}$ ignored, $d c^1_i$, $d c^2_i$, $d e^2_{i j}$ for $|i - j| \neq 1$, $d c^3_i$ and $d e^3_{i j}$ on the left-hand side of (\ref{eq:diff_jacobian}) can be replaced by $d \wtd{c}^1_i$, $d \wtd{c}^2_i$, $d \wtd{e}^2_{i j}$, $d \wtd{c}^3_i$ and $d \wtd{e}^3_{i j}$, respectively, in the light of (\ref{eq:diff1}), (\ref{eq:diff2}) and (\ref{eq:diff3}). The Jacobian is then a product of the factors in (\ref{eq:diff1}) and (\ref{eq:factors}). Thus overall the gauge orbit measure is
\begin{equation}
    \Delta \propto |J| \propto \prod_{i \neq j = 1}^N \left|\wtd{X}^1_{i i} - \wtd{X}^1_{j j}\right| \prod_{i = 1}^{N - 1} \left|\wtd{X}^2_{i (i + 1)}\right|.
\end{equation}

\section{Evaluation of observables} \label{app:obs}

The physical observables that we are interested in fall into roughly three categories: \emph{(i)} bosonic potentials; \emph{(ii)} fermionic bilinears; \emph{(iii)} casimirs of Lie group actions. Efficient numerical recipes for evaluating these observables via Monte Carlo simulation are discussed in this Appendix. Monte Carlo requires that the integrals are written as the average over samples $\E_{X \sim |f|^2}[\cdot]$.

Bosonic potentials are real functions of bosonic matrix coordinates $V(X)$, and they are straightforward to evaluate:
\begin{equation}
    \langle \psi | \hat{V}_1 | \psi \rangle \equiv \int d X\, |f(X)|^2 V(X) = \E_{X \sim |f|^2}[V(X)].
\end{equation}
Fermionic bilinears and casimirs are more elaborate to compute. The final results are (\ref{eq:mc_fermionic_bilinear}) and (\ref{eq:mc_casimir}) with detailed derivations presented below.

\subsection*{Fermionic bilinears}

Expectation values of fermionic bilinears $B(\lambda^\dagger, \lambda, X)$ are
\begin{equation}
    \langle \psi | \hat{V}_2 | \psi \rangle \equiv \int d X\, |f(X)|^2 \langle M(X) | B(\lambda^\dagger, \lambda, X) | M(X) \rangle.
\end{equation}
The problem is thus essentially to evaluate fermionic bilinears in the fermionic state $|M(X)\rangle$, which can furthermore be reduced to calcuating
\begin{equation}
    \langle M^r | B(\lambda^\dagger, \lambda, X) |M^s\rangle, \label{exp:matrix_elements}
\end{equation}
where $|M^r\rangle$ is the free fermion state
\begin{equation}
    |M^r \rangle \equiv \prod_{a = 1}^R \Big(\sum_{\alpha = 1}^2 \sum_{A = 1}^{N^2 - 1} M_{A \alpha}^{r a} \lambda_A^{\alpha \dagger}\Big) |0\rangle. \label{eq:free_state}
\end{equation}

The question is more generally formulated as follows: let $M$ be a complex matrix of size $R \times P$ and denote its corresponding free fermion state as
\begin{equation}
    |M \rangle = \prod_{a = 1}^R \Big(\sum_{p = 1}^P M_{a p} \lambda_p^\dagger\Big) |0 \rangle, \label{eq:general_free_state}
\end{equation}
then what are the matrix elements
$
    \langle M' | B(\lambda^\dagger, \lambda, X) | M \rangle
$?
The starting point is the Slater determinant:
\begin{equation}
    \langle M' | M \rangle = \det ( M M'^\dagger), \label{eq:slater_det}
\end{equation}
and note that 
\begin{equation}
    \langle M' | \lambda_p^\dagger \lambda_q | M \rangle = \delta_{p q} \langle M' | M \rangle  - \langle M' | \lambda_q \lambda_p^\dagger | M \rangle, \label{eq:commutation}
\end{equation}
where the first term on the right-hand side can be evaluated from (\ref{eq:slater_det}). The second term in (\ref{eq:commutation}) can be read as the overlap between free fermion states $\lambda_q^{ \dagger} | M'\rangle$ and $\lambda_p^{\dagger} | M\rangle$ and thus (\ref{eq:slater_det}) is again applicable:
\begin{align}
    s^2 \langle M' | \lambda_q \lambda_p^\dagger | M \rangle &= \det \left(
    \begin{array}{cc}
        s^2 \delta_{p q}  & s M'^\dagger_{p :} \\
         s M_{: q} & M M'^\dagger
    \end{array} \right) \nonumber \\
    &= \det \left(\begin{array}{cc}
        1  & s M'^\dagger_{p :} \\
        s M_{: q} & M M'^\dagger
    \end{array} \right) + (s^2 \delta_{p q} - 1) \det (M M'^\dagger) \nonumber \\
    &= \det (M M'^\dagger - s^2 M_{: q} M'^\dagger_{p :}) + (s^2 \delta_{p q} - 1) \det (M M'^\dagger). \label{eq:ttt}
\end{align}
A dummy variable $s$ is introduced for later convenience. Using (\ref{eq:ttt}) in (\ref{eq:commutation})
\begin{equation}
    s^2 \langle M' | \lambda_p^\dagger \lambda_q | M \rangle =  \det (M M'^\dagger) - \det (M M'^\dagger - s^2 M_{: q} M'^\dagger_{p :}).
\end{equation}
Differentiate both sides with respect to $s^2$ to obtain a more compact expression:
\begin{equation}
    \langle M' | \lambda_p^\dagger \lambda_q | M \rangle = \tr \left[\adj (M M'^\dagger) M_{: q} M'^\dagger_{p :} \right],
\end{equation}
where $\adj A = (\det A) A^{-1}$ is the adjucate of $A$. For an arbitrary bilinear $W$,
\begin{equation}
     \sum_{p q} \langle M' | \lambda_p^\dagger W_{q p} \lambda_q | M \rangle = \det (M M'^\dagger) \tr \left[(M M'^\dagger)^{-1} M W M'^\dagger \right]. \label{eq:free_bilinear}
\end{equation}

Back to the original problem of calculating (\ref{exp:matrix_elements}). Equation (\ref{eq:free_bilinear}) is applicable if we regard the index $p$ in (\ref{eq:general_free_state}) as running over both the indices $\alpha$ and $A$ in (\ref{eq:free_state}). Define the overlap matrix
\begin{equation}
    (O^{r s})^{a b} \equiv \sum_{\alpha = 1}^2 \sum_{A = 1}^{N^2 - 1} (M^{r a}_{A \alpha})^* M^{s b}_{A \alpha},
\end{equation}
then 
\begin{equation}
    \langle M^r | B(\lambda^\dagger, \lambda, X) |M^s\rangle = \sum_{a b = 1}^R (\adj O^{r s})_{b a} B(M^{r a\dagger}, M^{s b}, X),
\end{equation}
where the fermionic operators in the bilinear are replaced by complex matrices so that the expression is a complex number. Finally summing over $r$ and $s$,
\begin{equation}
    \langle \psi | \hat{V}_2 | \psi \rangle = \E_{X \sim |f|^2} \left[\sum_{r s = 1}^D \sum_{a b = 1}^R (\adj O^{r s}(X))_{b a} B(M^{r a\dagger}(X), M^{s b}(X), X)\right]. \label{eq:mc_fermionic_bilinear}
\end{equation}

\subsection*{Casimirs}

The observables discussed above do not involve derivatives. Derivatives show up in kinetic terms, for example, and can be understood in a geometric way. For an action of a Lie group $G$ on the wavefunction $\psi$, a casimir term can be defined as
\begin{equation}
    \langle \psi | \hat{V}_3 | \psi \rangle \equiv \sum_A \int d X\, \langle d_A \psi(X) | d_A \psi(X) \rangle, \label{eq:casimir}
\end{equation}
where the summation is over an orthonormal basis of the Lie algebra and 
\begin{equation}
    |d _A \psi(X) \rangle \equiv \left.\frac{d}{d s} (e^{i s T_A} \psi)(X) \right|_{s = 0}. \label{eq:d_state}
\end{equation}

As an example, consider the group of translations of bosonic coordinates $X \to X + \delta X$ that acts on the wavefunction as 
\begin{equation}
    (e^{i s T_A} \psi)(X) = \psi(X - s T_A), \quad \left.\frac{d}{d s} (e^{i s T_A} \psi)(X) \right|_{s = 0} = - \sum_{i j} T_{A i j}  \frac{\partial \psi}{\partial X_{i j}},
\end{equation}
and thus in this case
\begin{align}
    \langle \psi | \hat{V}_3 | \psi \rangle &= \sum_{A i j i' j'} \int d X\, T^*_{A i'j'} T_{A i j} \Big\langle \frac{\partial \psi}{\partial X_{i' j'}} \Big| \frac{\partial \psi}{\partial X_{i j}} \Big\rangle = \sum_{i j} \int d X\, \Big\langle \frac{\partial \psi}{\partial X_{i j}} \Big| \frac{\partial \psi}{\partial X_{i j}} \Big\rangle,
\end{align}
which is the usual kinetic term. If $G = \mathrm{SU}(N)$ with the adjoint action on matrices, the observable (\ref{eq:casimir}) is the casimir of the gauge group, and if $G = \mathrm{SO}(3)$ in the mini-BMN model, the observable measures the angular momentum quantum number of the state. 

The summation and the integral in (\ref{eq:casimir}) are estimated from Monte Carlo samples as:
\begin{equation}
    \langle \psi | \hat{V}_3 | \psi \rangle = \E_{|T_A|^2 = \dim G, X \sim |f|^2} \Big[|f(X)|^{-2} \langle d_A \psi(X) | d_A \psi (X) \rangle\Big], \label{eq:mc_casimir}
\end{equation}
where $f = |\psi|$, $|T_A|^2 = \dim G$ means that the expectation value averages over all Lie algebra elements $T_A$ with norm $\sqrt{\dim G}$.

\section{Semiclassical analysis of the fuzzy sphere}
\label{sec:semiclass}

\subsection*{Correspondence between matrices and fields on the emergent sphere}

A mapping from any $N$-by-$N$ complex matrix $A$ to a function $f_A(\theta, \phi)$ is constructed as follows. The construction is motivated by the following principles: \emph{(i)} the map $A \mapsto f_A(\theta, \phi)$ should be linear; \emph{(ii)} the map should preserve the inner products:
\begin{equation}
    \frac{1}{N} \tr (A^\dagger A') = \frac{1}{4 \pi} \int d \Omega\, f_A^*(\theta, \phi) f_{A'}(\theta, \phi). \label{eq:inner_product}
\end{equation}
Here $\int d \Omega$ is the integral over a $4 \pi$ solid angle; 
\emph{(iii)} the map should preserve the $\mathfrak{su}(2)$ action:
\begin{equation}
    f_{[J^i, A]} (\theta, \phi) = (L^i f_A)(\theta, \phi). \label{eq:su2_action}
\end{equation}
As in the main text, the $J^i$ are generators of the $N$ dimensional irreducible representation of $\mathfrak{su}(2)$ and
the $L^i$ are generators for rotations of functions on a sphere:
\begin{equation} \label{eq:def_Li}
    L^i = -i \epsilon_{i j k} x^j \frac{\partial}{\partial x^k},
\end{equation}
and $(x^1, x^2, x^3) = (\sin \theta \cos \phi, \sin \theta \sin \phi, \cos \theta)$.

Requirements \emph{(i)} and \emph{(ii)} can be accomplished by mapping an orthonormal basis of matrices to an orthonormal basis of functions on the sphere. In the light of \emph{(iii)}, we choose spherical harmonics $Y_{j m}(\theta, \phi)$ ($j \geq 0$, $|m| \leq j$) as the basis of functions:
\begin{equation}
    \sum_{i = 1}^3 L^i L^i Y_{j m} = j (j + 1) Y_{j m}, \quad L^3 Y_{j m} = m Y_{j m},
\end{equation}
and they are orthonormal with respect to the inner product in (\ref{eq:inner_product}):
\begin{equation}
\frac{1}{4 \pi} \int d \Omega\, Y^*_{j m}(\theta, \phi) Y_{j' m'}(\theta, \phi) = \delta_{j j'} \delta_{m m'}.
\end{equation}

To construct matrix counterparts of spherical harmonics $\hat{Y}_{j m}$, we note that
\begin{equation}
    Y_{j (m + 1)} = \frac{L^+ Y_{j m}}{\sqrt{(j - m)(j + 1 + m)}},
\end{equation}
where $L^\pm = L^1 \pm i L^2$, so \emph{(iii)} requires (denote $J^\pm = J^1 \pm i J^2$)
\begin{equation}
    \hat{Y}_{j (m + 1)} = \frac{[J^+, \hat{Y}_{j m}]}{\sqrt{(j - m)(j + 1 + m)}}, \label{eq:recursion_matrix_sh}
\end{equation}
which fixes all the matrices $\hat{Y}_{j m}$ given $\hat{Y}_{j (-j)}$. The $\mathfrak{su}(2)$ representation further requires that
$L^- Y_{j (-j)} = 0$ and $L^+ Y_{j j} = 0$,
which translates to the matrix side as $[J^-, Y_{j (-j)}] = 0$ and $[J^+, Y_{j j}] = 0$. 
Thus for some normalizing factor $C$,
\begin{equation}
    \hat{Y}_{j (-j)} = C (J^-)^j. \label{eq:init_matrix_sh}
\end{equation}
The matrix $J^-$ is nilpotent with order $N$: $(J^-)^N = 0$. Therefore the matrices in (\ref{eq:init_matrix_sh}) are restricted to $j \leq N - 1$. For $j \leq N - 1$, the numerical factor $C$ is chosen such that
\begin{equation}
    \frac{1}{N} \tr \hat{Y}^\dagger_{j (-j)} \hat{Y}_{j (-j)} = 1. \label{eq:normalization}
\end{equation}
The sign of $C$ is not fixed by the three requirements, and we pick $C > 0$ in correspondence with spherical harmonics $Y_{j (-j)} \propto (x^1 - i x^2)^j$. 

It is straightforward to verify that 
\begin{equation}
    \sum_{i = 1}^3 [J^i, [J^i, \hat{Y}_{j m}]] = j (j + 1) \hat{Y}_{j m}, \quad [J^3, \hat{Y}_{j m}] = m \hat{Y}_{j m},
\end{equation}
given the $\mathfrak{su}(2)$ algebra and eqs.\,(\ref{eq:recursion_matrix_sh}) and (\ref{eq:init_matrix_sh}). Hence the matrices $\hat{Y}_{j m}$ form an eigenbasis of adjoint actions of $J^3$ and the casimir $(J^i)^2$, and are therefore orthogonal. They are normalized as well because of (\ref{eq:normalization}). The map $A \mapsto f_A(\theta, \phi)$ is then defined on the basis as $\hat{Y}_{j m} \mapsto Y_{j m}(\theta, \phi)$, fulfilling the requirements \emph{(i)} to \emph{(iii)}. 

Under the correspondence $\hat{Y}_{j m} \mapsto Y_{j m}(\theta, \phi)$, $N$-by-$N$ matrices describe fields on a sphere with angular momentum cutoff $\jm = N - 1$. Furthermore (\ref{eq:inner_product}) connects matrix observables and averages of fields on the emergent sphere. For instance, the classical fuzzy sphere solution sets $X^i = \nu J^i$, and we would like to interpret $f_{X^i}(\theta, \phi)$ as coordinates $x^i$ of the point on the sphere at angle $(\theta, \phi)$. Thus according to (\ref{eq:inner_product}), the radius of the emergent sphere (for irreducible representation $J^i$) is
\begin{align}
    r^2 &= \frac{1}{4 \pi} \sum_{i = 1}^3 \int d \Omega\, f_{X^i}(\theta, \phi)^2 = \frac{1}{N} \sum_{i = 1}^3 \tr (X^i)^2 \nonumber\\
    &= \frac{\nu^2}{N} \sum_{i = 1}^3 \tr (J^i)^2 = \frac{\nu^2 (N^2 - 1)}{4}.
\end{align}

\subsection*{Noncommutative gauge theory on the fuzzy sphere}

In the last subsection we have discussed the correspondence between matrix degrees of freedom and fields on the fuzzy sphere. Given that correspondence the matrix Hamiltonian (\ref{eq:hamitonian_full}) can be cast into a quantum field theory on the sphere. The caveat is that the fields on the sphere are not commutative, due to the noncommutative nature of matrix multiplication. 

To be more precise, we define the `star product' of the fields as induced from their corresponding matrix multiplications: 
\begin{equation} \label{eq:star_product}
    (f \star g)(\theta, \phi) \equiv \frac{1}{N} \sum_{j m} \tr \left( \hat{Y}^\dagger_{j m} \hat{f} \hat{g}\right) Y_{j m}(\theta, \phi),
\end{equation}
where $\hat{f}$ and $\hat{g}$ are the matrix counterparts of functions $f(\theta, \phi)$ and $g(\theta, \phi)$ via the correspondence between matrix spherical harmonics and spherical harmonics on the sphere: $\hat{Y}_{j m} \leftrightarrow Y_{j m}(\theta, \phi)$. The prefactor is a result of the normalization (\ref{eq:normalization}). 

The star product is associative but noncommutative. In particular, the commutator of scalar functions may not vanish. For example, 
\begin{align} \label{eq:star_product_commutator}
    [Y_{j_1 m_1}, Y_{j_2 m_2}]_\star (\theta, \phi) &= \frac{1}{N} \sum_{j m} \tr \left( \hat{Y}^\dagger_{j m} [\hat{Y}_{j_1 m_1}, \hat{Y}_{j_2 m_2}]\right) Y_{j m}(\theta, \phi) \nonumber \\ &\equiv \sum_{j m} f^{j m}_{j_1 m_1 j_2 m_2} Y_{j m}(\theta, \phi),
\end{align}
where $[\cdot, \cdot]_\star$ is the commutator with the star product for multiplication. The structure constants $f$ in (\ref{eq:star_product_commutator}) are known to vanish as $1/N$ as $N \to \infty$ (see, e.g., the Appendix of \cite{Iso:2001mg}). The usual commutative product is recovered at $N = \infty$.

To repackage matrix degrees of freedom into emergent fields, expand the bosonic matrices around their classical values: 
\begin{equation} \label{eq:expand_fields}
    X^i = \nu J^i + A^i,
\end{equation}
where the $A^i$ are Hermitian matrices parametrizing fluctuations around the fuzzy sphere. Our re-writing of the Hamiltonian will be exact in $A$. The corresponding emergent fields $\wtd{a}^i(\theta, \phi)$ are as follows:
\begin{equation} \label{eq:matrix_field_exp}
     \wtd{a}^i(\theta, \phi) = \sum_{j m} a^i_{j m} Y_{j m}(\theta, \phi), \quad \text{if } A^i = \sum_{j m} a^i_{j m} \hat{Y}_{j m}.
\end{equation}
The conjugate momenta to the $A^i$ are
\begin{equation} \label{eq:matrix_conj_exp}
    \Pi^i_A = - \frac{i}{N} \sum_{j m} \hat{Y}^\dagger_{j m} \frac{\partial}{\partial a^i_{j m}} \,,
\end{equation}
obeying the canonical commutation relations $[A^i_{a b}, (\Pi^j_A)_{c d}] = i \delta^{i j} \delta_{a d} \delta_{b c}$. We will also want to introduce the momenta
\begin{equation} \label{eq:matrix_conj_exp2}
    \wtd{\pi}^i(\theta, \phi) = -\frac{i}{4 \pi} \sum_{j m} Y^*_{j m}(\theta, \phi) \frac{\partial}{\partial a^i_{j m}},
\end{equation}
which obey
\begin{equation}
     [\wtd{a}^i(\theta, \phi), \wtd{\pi}^k(\theta', \phi')] = \frac{i \delta^{i k}}{4\pi} \sum_{jm} Y_{j m}(\theta, \phi) Y^*_{j m}(\theta', \phi') .\label{eq:tttt}
\end{equation}
The $\wtd{\pi}^i$ therefore become the usual conjugate momenta when $\jm = \infty$, where the summation in (\ref{eq:tttt}) becomes
$4 \pi \delta(\cos \theta - \cos \theta') \delta(\phi - \phi')$. Hermiticity of the matrices $A^i$ and $\Pi^i_A$ is manifested as reality of the fields $\wtd{a}^i$ and $\wtd{\pi}^i$.

Substituting (\ref{eq:matrix_field_exp}), (\ref{eq:matrix_conj_exp}) and (\ref{eq:matrix_conj_exp2}) into the matrix Hamiltonian, the kinetic terms are 
\begin{equation} \label{eq:kinetic_fields}
    \frac{1}{2} \tr \left(\Pi^i \Pi^i\right) = \frac{1}{2} \tr \left(\Pi^i_A \Pi^i_A\right) = - \frac{1}{2 N} \sum_{i j m} \frac{\partial^2}{(\partial a^i_{j m})^2} = \frac{2 \pi}{N} \int d \Omega\, (\wtd{\pi}^i(\theta, \phi))^2.
\end{equation}
The bosonic potential in (\ref{eq:HamB}) can be written as a square:
\begin{equation} \label{eq:bosonic_potential_fields}
    V(X) = \frac{1}{4} \tr \left( i [X^i, X^j] + \nu \epsilon^{i j k} X^k\right)^2 \equiv \frac{\nu^2}{4} \tr \left(F^{i j}\right)^2,
\end{equation}
and substituting (\ref{eq:expand_fields}) into (\ref{eq:bosonic_potential_fields}):
\begin{equation} \label{eq:matrix_field_strength}
    F^{i j} = i \left([J^i, A^j] - [J^j, A^i]\right) + i \nu^{-1} [A^i, A^j] + \epsilon^{i j k} A^k.
\end{equation}
The corresponding field is (recall (\ref{eq:su2_action}) and (\ref{eq:star_product}))
\begin{equation}
    \wtd{f}^{i j}(\theta, \phi) = i \left(L^i \wtd{a}^j - L^j \wtd{a}^i\right) + \epsilon^{i j k} \wtd{a}^k + i \nu^{-1} [\wtd{a}^i, \wtd{a}^j]_\star,
\end{equation}
and the potential can now be written
\begin{equation} \label{eq:bosonic_pot_fields}
    V(X) = \frac{N \nu^2}{4} \int \frac{d \Omega}{4 \pi} \,(\wtd{f}^{i j}(\theta, \phi))^2.
\end{equation}
The fermionic potential in (\ref{eq:fermionic_potential}) is, in terms of $A^i$,
\be \label{eq:fermionic_matrix_pot}
    \nu \tr \left( \lambda^\dagger \sigma^k [J^k + \nu^{-1} A^k, \lambda] + \frac{3}{2} \lambda^\dagger \lambda\right) - \frac{3}{2} \nu (N^2 - 1) \,.
\ee
Let $\wtd{\psi}(\theta, \phi)$ be the fermionc field corresponding to $\lambda$, then (\ref{eq:fermionic_matrix_pot}) is recast into
\begin{equation} \label{eq:fermionic_pot_fields}
    \frac{N \nu}{4 \pi} \int d \Omega\, \left(- i \wtd{\psi}^\dagger \sigma^k D^k \wtd{\psi} + \frac{3}{2} \wtd{\psi}^\dagger \wtd{\psi}\right) + \text{const},
\end{equation}
where $D^k \wtd{\psi} \equiv i L^k \wtd{\psi} + i \nu^{-1} [\wtd{a}^k, \wtd{\psi}]_\star$. 

Collect all three parts (\ref{eq:kinetic_fields}), (\ref{eq:bosonic_pot_fields}) and (\ref{eq:fermionic_pot_fields}), and rescale the fields
\begin{equation} \label{eq:rescaling}
    \wtd{a}^i = \sqrt{\frac{4 \pi}{N \nu}} a^i, \quad \wtd{\pi}^i = \sqrt{\frac{N \nu}{4 \pi}} \pi^i, \quad \wtd{\psi} = \sqrt{\frac{4 \pi}{N}} \psi.
\end{equation}
The Hamiltonian for the emergent fields, which is equivalent to (\ref{eq:hamitonian_full}) for matrices, is then
\begin{equation} \label{eq:hamil_fields}
    H = \nu \int d \Omega\, \left(\frac{1}{2} (\pi^i)^2 + \frac{1}{4} (f^{i j})^2- i \psi^\dagger \sigma^k D^k \psi + \frac{3}{2} \psi^\dagger \psi\right) + \text{const},
\end{equation}
where
\begin{align} \label{eq:nc_hamil}
    f^{i j} &\equiv i \left(L^i a^j - L^j a^i\right) + \epsilon^{i j k} a^k + i \sqrt{\frac{4 \pi}{N \nu^3}}[a^i, a^j]_\star, \nonumber \\
    D^k \psi &\equiv i L^k \psi + i \sqrt{\frac{4 \pi}{N \nu^3}} [a^k, \psi]_\star.
\end{align}

The $\mathrm{SU}(N)$ gauge symmetry of the matrices leads to the noncommutative $\mathrm{U}(1)$ gauge symmetry of (\ref{eq:nc_hamil}). Under an infinitesimal $\mathrm{SU}(N)$ gauge transformation parametrized by a Hermitian matrix $Y$, $\delta X^i = i [Y, X^i]$, $\delta \lambda^\alpha = i [Y, \lambda^\alpha]$, and thus by (\ref{eq:expand_fields}), 
\begin{equation}
    \delta A^i = -i [\nu J^i, Y] + i [Y, A^i].
\end{equation}
Let $\wtd{y}(\theta, \phi)$ be the field corresponding to the matrix $Y$, then the gauge transformation of the noncommutative fields is ($\bm{n}$ is the radial vector and fields should be considered as defined on the unit sphere)
\begin{equation}
    \delta \wtd{a}^i = - i \nu L^i \wtd{y} - (\bm{n} \times \nabla \wtd{y} \cdot \nabla) \wtd{a}^i, \quad \delta \wtd{\psi}^\alpha = -(\bm{n} \times \nabla \wtd{y} \cdot \nabla) \wtd{\psi}^\alpha.
\end{equation}
Recall the rescaling (\ref{eq:rescaling}) and let $\wtd{y} = y \sqrt{4 \pi / N \nu^3} $,
\begin{equation}\label{eq:gg}
    \delta a^i = - i L^i y - \sqrt{\frac{4 \pi}{N \nu^3}} (\bm{n} \times \nabla y \cdot \nabla) a^i, \quad \delta \psi^\alpha = -\sqrt{\frac{4 \pi}{N \nu^3}} (\bm{n} \times \nabla y \cdot \nabla) \psi^\alpha.
\end{equation}
The first term in $\delta a^i$ is the usual $\mathrm{U}(1)
$ transformation. The second term, which can be obtained from the algebra in (\ref{eq:star_product_commutator}), describes a coordinate transformation with infinitesimal displacement $\bm{n} \times \nabla y$ \cite{DEWIT1988545}. Indeed, it is known that non-commutative gauge theories mix internal and spacetime symmetries, which in this case are area-preserving diffeomorphisms of the sphere \cite{Paniak:2002fi,Lizzi:2001nd}. The coordinate transformation in (\ref{eq:gg}) is area-preserving because $\nabla \cdot (\bm{n} \times \nabla y) = 0$.

In the commutative limit $\nu \to \infty$, the gauge field is decoupled from the fermions and the theory contains a U(1) gauge field on the sphere, with a real massive scalar and a massive Dirac fermion. To see more explicitly the field content of (\ref{eq:hamil_fields}) in this limit, note that $\bm{L} = -i \bm{n} \times \nabla$ and $f^{i j} = \epsilon^{i j k} \left((\bm{n} \times \nabla) \times \bm{a} + \bm{a}\right)^k$ when $\nu \to \infty$ ($\bm{a}$ is the three-dimensional vector notation for $a^i$). We then obtain
\begin{equation} \label{eq:curv}
    \frac{1}{4} (f^{i j})^2 = \frac{1}{2} \left|(\bm{n} \times \nabla) \times \bm{a} + \bm{a} \right|^2.
\end{equation}

The scalar field $\varphi$ is the radial component of the gauge field, and we denote the U(1) gauge field on the sphere as $\bm{b}$:
\begin{equation}
    \varphi = \bm{a} \cdot \bm{n}, \quad \bm{b} = \bm{a} \times \bm{n}.
\end{equation}
The U(1) curvature $f$ of the gauge field $\bm{b}$ defined on the sphere is 
\begin{equation}
    f = \bm{n} \cdot (\nabla \times \bm{b}) = 2 \bm{n} \cdot \bm{a} - \nabla \cdot \bm{a},
\end{equation}
and we have (after some vector calculus manipulations)
\begin{equation} \label{eq:vec}
    (\bm{n} \times \nabla) \times \bm{a} + \bm{a} = f \bm{n} + \nabla (\bm{n} \cdot \bm{a}) - \bm{n} (\bm{n} \cdot \bm{a}) = (f - \varphi) \bm{n} + \nabla \varphi. 
\end{equation}
Substituting (\ref{eq:vec}) into (\ref{eq:curv}), the commutative gauge theory can be rewritten as
\begin{align} \label{eq:hamil_fields_commutative}
    H = \nu \int d \Omega\,\left(\frac{1}{2} (\pi^a)^2 + \frac{1}{2} \pi^2 + \frac{1}{2} (f - \varphi)^2 + \frac{1}{2} (\nabla \varphi)^2
    - i \psi^\dagger (\bm{\sigma} \times \bm{n}) \cdot \nabla \psi + \frac{3}{2} \psi^\dagger \psi\right),
\end{align}
where $\pi^a$ and $\pi$ are the conjugate variables of $\bm{b}$ and $\varphi$, respectively, and $\bm{\sigma}$ is the vector of Pauli matrices. The fields in (\ref{eq:hamil_fields_commutative}) should be thought as living on the unit sphere.

\subsection*{Fluctuation spectrum around the classical fuzzy sphere}

The classical energy at the fuzzy sphere vanishes due to supersymmetry. In the following we analyze the spectrum of bosonic quadratic fluctuations near the fuzzy sphere configuration, and the spectrum of fermions, as the next order in a semiclassical expansion. The semiclassical correction to energy at this level is shown to be zero as well. 

The bosonic potential in (\ref{eq:HamB}) can be written as a square:
\begin{equation}
    V(X) = \frac{1}{2} \tr \left(\nu X^i + i \epsilon^{i}_{\, j k} X^j X^k\right)^2, \label{eq:bosonic_potential}
\end{equation}
and quadratic fluctuations around a classical solution are given by
\begin{align}
    \delta V(X) &= \frac{1}{2} \tr \left( \nu \delta X^i + i \epsilon^{i}_{\, j k} [X^j, \delta X^k]\right)^2 \nonumber \\
    & \equiv \sum_{a} \frac{1}{2} \nu^2 \omega_a^2 (\delta x_a)^2\,, \label{eq:quadratic}
\end{align}
where $\delta X^i = \sum_a \delta x_a Y^i_a$ and $Y^i_a$ are the normalized eigen-matrices:
\begin{equation}
    Y^i_a + i \epsilon^{i}_{\, j k} [J^j, Y^k_a] = \omega_a Y^i_a, \quad \sum_{i = 1}^3 \tr [(Y^i_a)^\dagger Y^i_b] = \delta_{ab}. \label{eq:eigs}
\end{equation}
Here we specialized to the background solution $X^j = \nu J^j$.

To solve the eigenvalue equation in (\ref{eq:eigs}), expand $Y^i$ (subscript $a$ omitted) into a sum of matrix spherical harmonics $Y^i = \sum_{j m} y^i_{j m} \hat{Y}_{j m}$, and note 
\begin{align}
    \sum_{i = 1}^3 [J^i, [J^i, \hat{Y}_{j m}]] &= j (j + 1) \hat{Y}_{j m}, \quad [J^+, \hat{Y}_{j m}] = \sqrt{(j - m)(j + m + 1)} \hat{Y}_{j (m + 1)},  \nonumber \\ [J^3, \hat{Y}_{j m}] &= m \hat{Y}_{j m}, \quad [J^-, \hat{Y}_{j m}] = \sqrt{(j + m)(j - m + 1)} \hat{Y}_{j (m - 1)}.
\end{align}
For convenience introduce the $\pm$ basis: $y^\pm = y^1 \pm i y^2$ and the indices must be raised with $g^{+ -} = g^{- +} = 2$ and $g^{3 3} = 1$ (other entries are zero). In this basis $\epsilon_{+ - 3} = i / 2$. Then (\ref{eq:eigs}) can be cast into equations for the coefficients $y^3_{j m}$ and $y^\pm_{j m}$:
\begin{equation}
    y^3_{j m} +  \frac{1}{2} \sqrt{(j + m + 1) (j - m)} y^+_{j (m + 1)} - \frac{1}{2} \sqrt{(j - m + 1) (j + m)} y^-_{j (m - 1)}= \omega y^3_{j m}, \label{eq:eigs_eq1}
\end{equation}
\begin{equation}
    (\omega \pm m) y^{\pm}_{j (m \pm 1)} = \pm \sqrt{(j \pm m + 1)(j \mp m)} y^3_{j m}. \label{eq:eigs_eq2}
\end{equation}

Equations (\ref{eq:eigs_eq1}) and (\ref{eq:eigs_eq2}) consist of three linear equations with three variables $y^3_{j m}$, $y^+_{j (m + 1)}$ and $y^-_{j (m - 1)}$. For there to be nonzero solutions, the determinant must be zero:
\begin{equation}
    \omega (\omega + j) (\omega - j - 1) = 0.
\end{equation}
Hence for $0 < j < N$, $|m| < j$, the eigenvalues are $\omega =0, -j, j + 1$. The edge cases $|m| = j, j + 1$ should be treated separately due to the additional constraint $y^\pm_{j m} = 0$ if $|m| > j$. The eigenvalue equation at $m = \pm j$ is instead $\omega (\omega - j - 1) = 0$,
and for $m = \pm (j + 1)$ it is $\omega - j - 1 = 0$.

The multiplicity of the eigenvalue $\omega = 0$ is $N^2 - 1$, which accounts for the $\mathrm{SU}(N)$ gauge degrees of freedom. The other eigenvalues are $\omega = - j$ for $1 \leq j \leq N - 1$ with multiplicity $2 j - 1$ and $\omega = j + 1$ for $1 \leq j \leq N - 1$ with multiplicity $2 j + 3$. The ground state energy of the bosonic oscillators (\ref{eq:quadratic}) is therefore
\begin{equation} \label{eq:bosonic_energy}
    \frac{|\nu|}{2} \sum_a |\omega_a| = \frac{|\nu|}{2} \sum_{j = 1}^{N - 1} [j (2 j - 1) + (j + 1) (2 j + 3)] = \frac{4 N^3 + 5 N - 9}{6} |\nu|.
\end{equation}

The spectrum of the fermionic bilinear is found similarly:
\begin{equation}
 (\sigma^k)^{\alpha}_{\,\beta} [J^k, \lambda^{\beta}] + \frac{3}{2} \lambda^\alpha = \omega \lambda^\alpha.
\end{equation}
Expand $\lambda^\alpha = \sum_{j m} y^{\alpha}_{j m} \hat{Y}_{j m}$ (note now $\alpha = \pm$ labels $\sigma^3 = \pm 1$ basis). The equations are
\begin{equation}
    \left(\omega - m - \frac{3}{2}\right) y^+_{j m} = \sqrt{(j + m + 1)(j - m)} y^-_{j(m + 1)}, \label{eq:fermion_eq1}
\end{equation}
\begin{equation}
    \left(\omega + m - \frac{1}{2}\right) y^{-}_{j(m + 1)} = \sqrt{(j + m + 1) (j - m)} y^+_{j m}.\label{eq:fermion_eq2}
\end{equation}

The eigenvalue equations (\ref{eq:fermion_eq1}) and (\ref{eq:fermion_eq2}) have nontrivial solutions when
\begin{equation}
    \left(\omega - j - \frac{3}{2}\right)\left(\omega + j - \frac{1}{2}\right) = 0,
\end{equation}
so that for $0 < j < N$ and $-j \leq m < j$ there are eigenvalues $\omega = j + 3 / 2$ and $\omega = - j + 1/ 2$. For $m = j$ or $m = - j - 1$ the eigenvalue equation is instead $\omega - j - 3 / 2 = 0$, as $y^-_{j (j + 1)} = y^+_{j (- j - 1)} = 0$ is imposed.

So the eigenvalues for $0 < j < N$ are $\omega = j + 3 / 2$ with multiplicity $2 j + 2$ and $\omega = - j + 1 / 2$ with multiplicity $2 j$. For $\nu > 0$ the $\omega = - j + 1 / 2$ modes are occupied with a total number of fermions:
\begin{equation} \label{eq:fermion_num1}
    \sum_{j = 1}^{N - 1} (2j) = N^2 - N.
\end{equation}
And the fermionic energy for $\nu > 0$ at this order is
\begin{equation}
    \nu \sum_{j = 1}^{N - 1} \left(- j + \frac{1}{2}\right) (2 j) - \frac{3}{2} \nu (N^2 - 1) = - \frac{4 N^3 + 5 N - 9}{6} \nu. \label{eq:positive_m}
\end{equation}

For $\nu < 0$ the $\omega = j + 3 / 2$ modes are occupied instead and the number of fermions is
\begin{equation} \label{eq:fermion_num2}
    \sum_{j = 1}^{N - 1} (2 j + 2) = N^2 + N - 2.
\end{equation}
We see that supersymmetry requires different number of occupied fermions in the case of $\nu > 0$ and $\nu < 0$. The fermionic energy for $\nu < 0$ is
\begin{equation}
    \nu \sum_{j = 1}^{N - 1}\left(j + \frac{3}{2}\right)(2 j + 2) - \frac{3}{2} \nu (N^2 - 1) = \frac{4 N^3 + 5 N - 9}{6} \nu. \label{eq:negative_m}
\end{equation}
In either case (\ref{eq:positive_m}) or (\ref{eq:negative_m}) the energy is $-(4 N^3 + 5 N - 9)|\nu| / 6$, which exactly cancels the bosonic contribution (\ref{eq:bosonic_energy}). 
Hence the semiclassical correction to the fuzzy sphere energy is zero at this order, for the specific number of fermions (\ref{eq:fermion_num1}) or (\ref{eq:fermion_num2}). 

\subsection*{One-loop effective potential and the estimate of $\nu_\text{c}$}

In the main text we observe a first-order phase transition near $\nu_\text{c} \approx 4$ when the bosonic fuzzy sphere phase becomes unstable. Here we give an estimate of $\nu_\text{c}$ from the bosonic one-loop effective potential for the radius, at $N = \infty$.

We start with the bosonic potential (\ref{eq:bosonic_potential}) with matrix sources $S_i$:
\begin{equation}
    V(X; S_i) = \frac{1}{2} \tr \left(\nu X^i + i \epsilon^{i}_{\, j k} X^j X^k\right)^2 + \tr S_i X^i, \label{eq:bosonic_potential_with_source}
\end{equation}
where the sources $S_i(\phi)$ are such that the local energy minimum is at $X^i = \phi J^i$. The parameter $\phi > 0$ is proportional to the radius:
\begin{equation}
    r = \frac{\phi}{2}\sqrt{N^2 - 1}.
\end{equation}
The classical contribution to the energy (\ref{eq:bosonic_potential_with_source}) at $X^i = \phi J^i$ is 
\begin{equation}
    E_0(S_i(\phi)) = \frac{N (N^2 - 1)}{8} (\nu - \phi)^2 \phi^2 + \tr S_i(\phi) \phi J^i. \label{eq:ground_state_energy}
\end{equation}

Quadratic fluctuations of (\ref{eq:bosonic_potential_with_source}) around the local minimum give:
\begin{align}
    \delta V(X) &= \frac{1}{2} \tr \left( \nu \delta X^i + i \phi \epsilon^{i}_{\, j k} [ J^j, \delta X^k]\right)^2 + i \epsilon^i_{\,j k} (\nu - \phi) \phi \tr \left(J^i \delta X^j \delta X^k\right). \label{eq:quadratic_source}
\end{align}
The norm of the spin matrices $J^i$ scales as $N$, and hence to leading order in $N$:
\begin{equation}
    \delta V(X) =  \frac{1}{2} \tr \left( i \phi \epsilon^{i}_{\, j k} [ J^j, \delta X^k]\right)^2 + \ldots.
\end{equation}
Diagonalizing this leading order piece as we did in the last subsection, the nonzero mode frequencies are now $\omega = - (j + 1) \phi$ for $0 < j < N$ with multiplicity $2 j - 1$ and $\omega = j \phi$ for $0 < j < N$ with multiplicity $2 j + 3$. So, the one-loop quantum correction to the ground state energy is
\begin{equation}
    \frac{1}{2} \sum_a |\omega_a| = \frac{1}{2} \sum_{j = 1}^{N - 1} \left[\left|- (j + 1) \phi\right| (2 j - 1) + \left|j \phi\right| (2 j + 3)\right] + \ldots = \frac{2}{3} \phi N^3 + \ldots.
\end{equation}
The one-loop effective potential $\Gamma(\phi) = E_0(S_i(\phi)) + \frac{1}{2} \sum_a |\omega_a| - \tr S_i(\phi) \phi J^i$ is then
\begin{equation}\label{eq:Gam}
    N^{-3} \Gamma(\phi; \nu) = \frac{1}{8} (\nu - \phi)^2 \phi^2 + \frac{2}{3} \phi + \ldots,
\end{equation}
where omitted terms are higher order in $N^{-1}$.
The critical value of $\nu$ is estimated as when the second order derivative of $\Gamma(\phi)$ at the fuzzy sphere solution vanishes:
\begin{equation}
    \Gamma'(\phi; \nu_\text{c}) = \Gamma''(\phi; \nu_\text{c}) = 0, \quad \Rightarrow \quad \nu_\text{c} \approx 3.03,\, \phi \approx 2.39.
\end{equation}

It is clear in (\ref{eq:Gam}) that, at large $N$, the leading quantum correction to the classical solution is suppressed by $\nu^{-3}$. This shows that the large $\nu$ limit rapidly becomes classical. The critical $\nu_\text{c}$ estimated above is at $N=\infty$, where the transistion is sharp.

\section{Training and tuning}
\label{app:train}

\begin{figure}[h!]
    \vskip 0.2in
    \centering
    \includegraphics[width=0.85\textwidth]{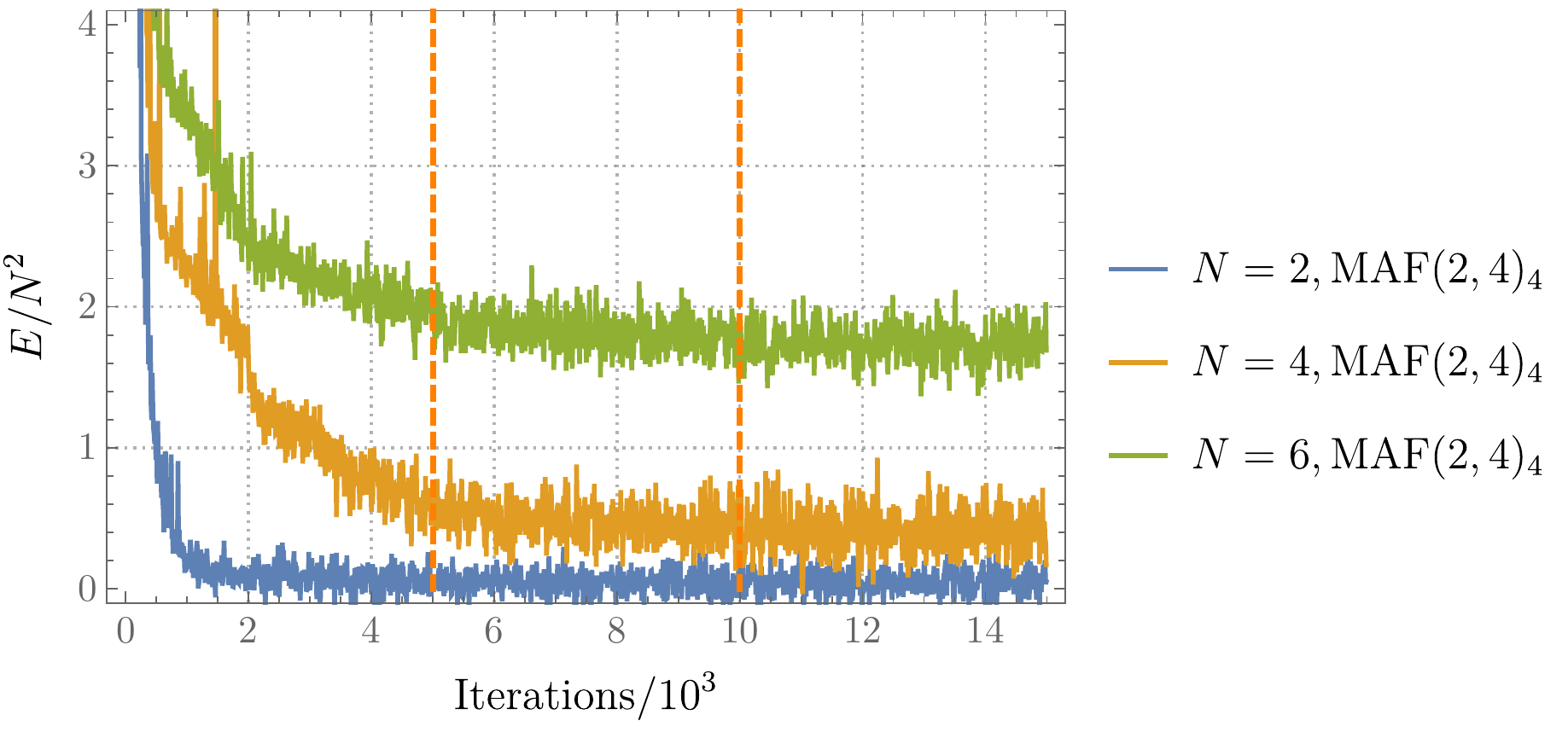}
    \caption{The variational energy as a function of training iterations for $N = 2, 4, 6$, with $\nu = 2$ and architecture MAF(2, 4) --- the subscript is $D = 4$ as in (\ref{eq:fermionic_states}). The dashed lines separate the three phases.}
    \label{fig:train}
    \vskip -0.1in
\end{figure}

Training of the model is divided into three epochs, each of which consists of 5000 iterations. The learning rate is set to be $10^{-3}$ for iterations from 1 to 5000, $2 \times 10^{-4}$ from 5001 to 10000 and $4 \times 10^{-5}$ from 10001 to 15000. In each iteration the energy is evaluated from a batch of $10^3$ random samples, and while the Monte Carlo energy fluctuates among iterations, its average value converges. Some typical training histories are shown in Fig.\,\ref{fig:train}.

\begin{figure}[ht!]
    \vskip 0.2in
    \centering
    \includegraphics[width=0.75\textwidth]{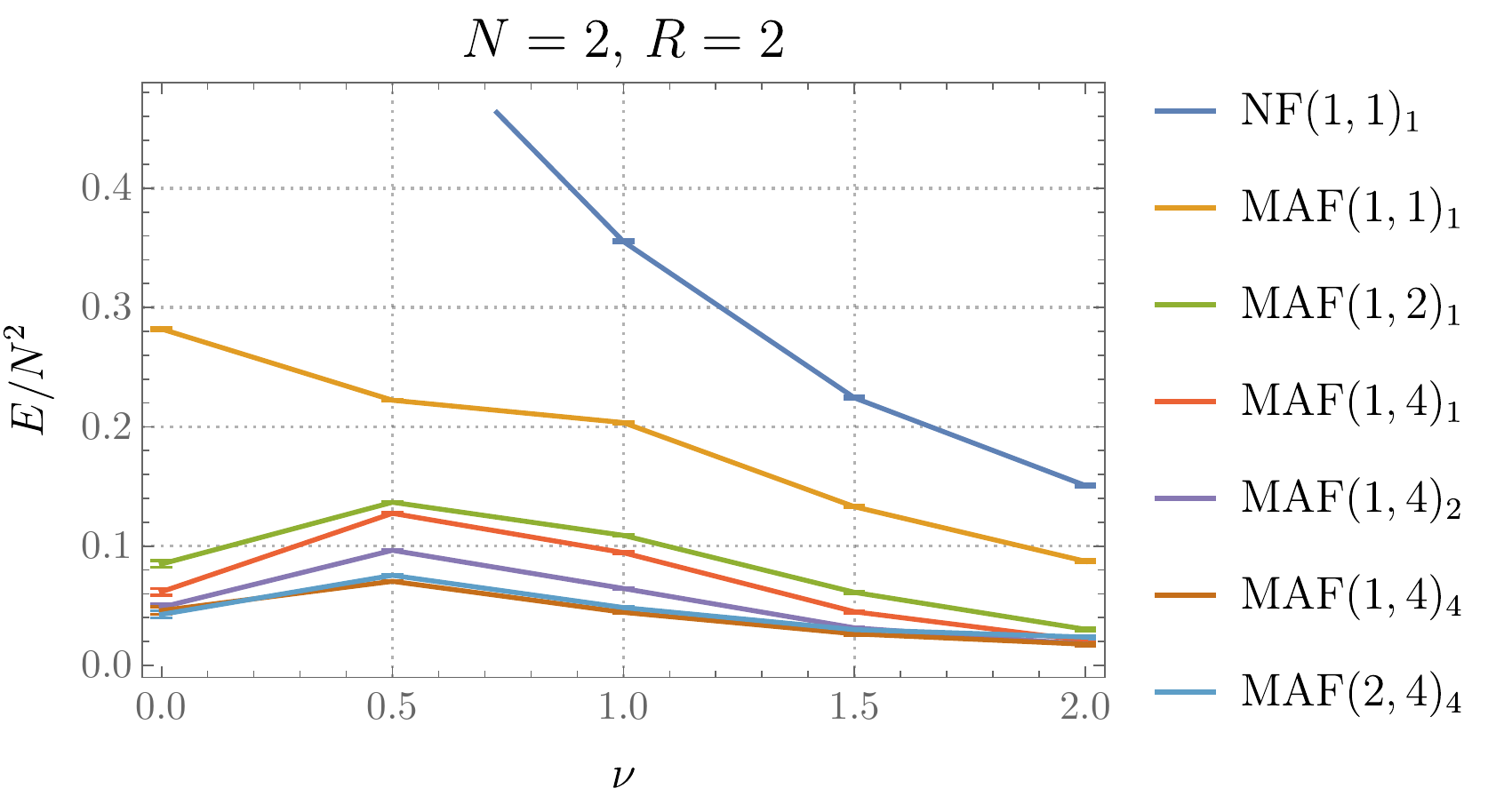}
    \includegraphics[width=0.75\textwidth]{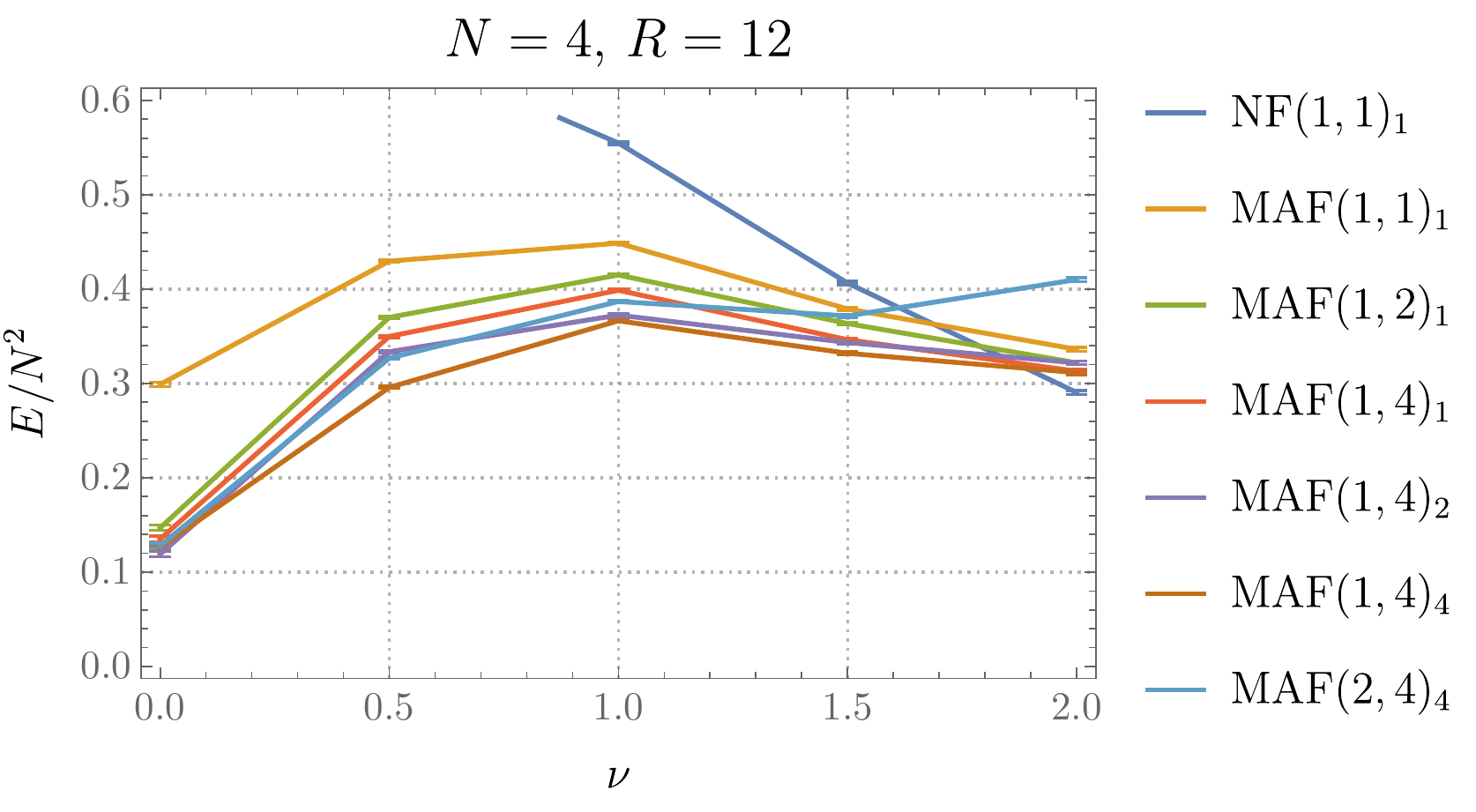}
    \includegraphics[width=0.75\textwidth]{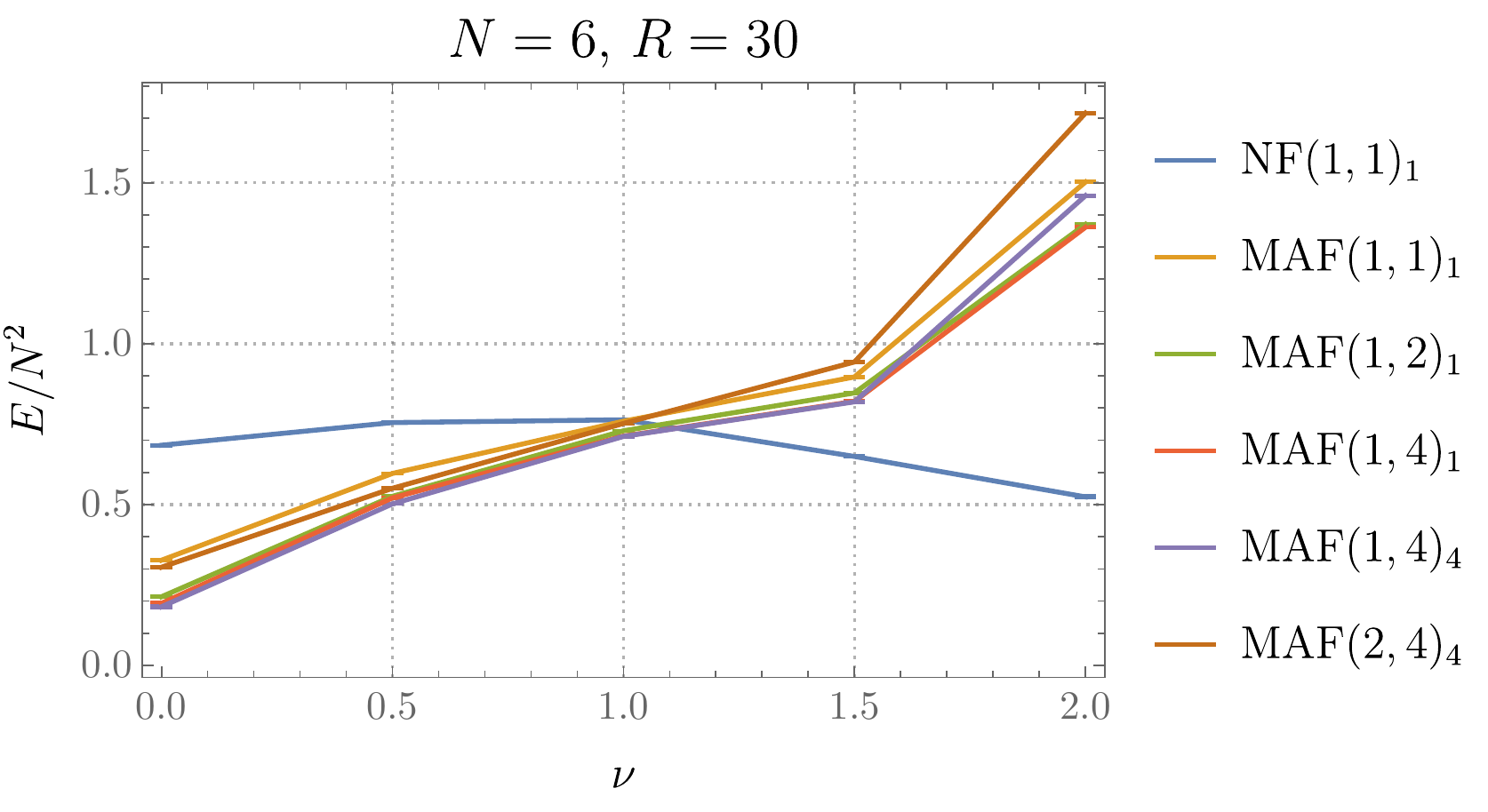}
    \caption{The variational energy for different $N$, $\nu$ and MAF architectures, in the supersymmetric sector. The wavefunctions are initialized near zero. Error bars (largely invisible) are Monte Carlo uncertainties of the final energy. }
    \label{fig:qmaf}
    \vskip -0.1in
\end{figure}

\begin{figure}[ht!]
    \vskip 0.2in
    \centering
    \includegraphics[width=0.75\textwidth]{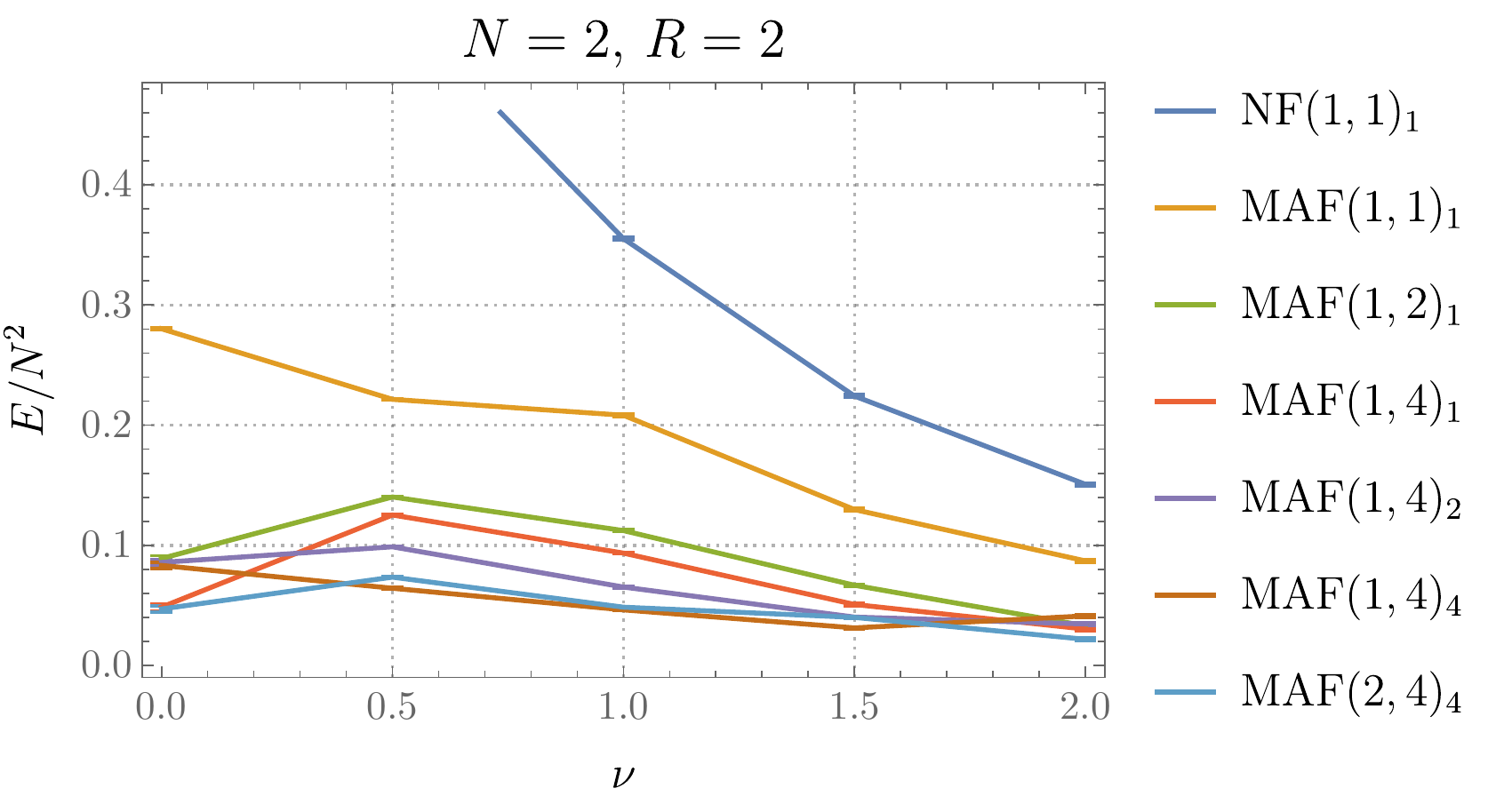}
    \includegraphics[width=0.75\textwidth]{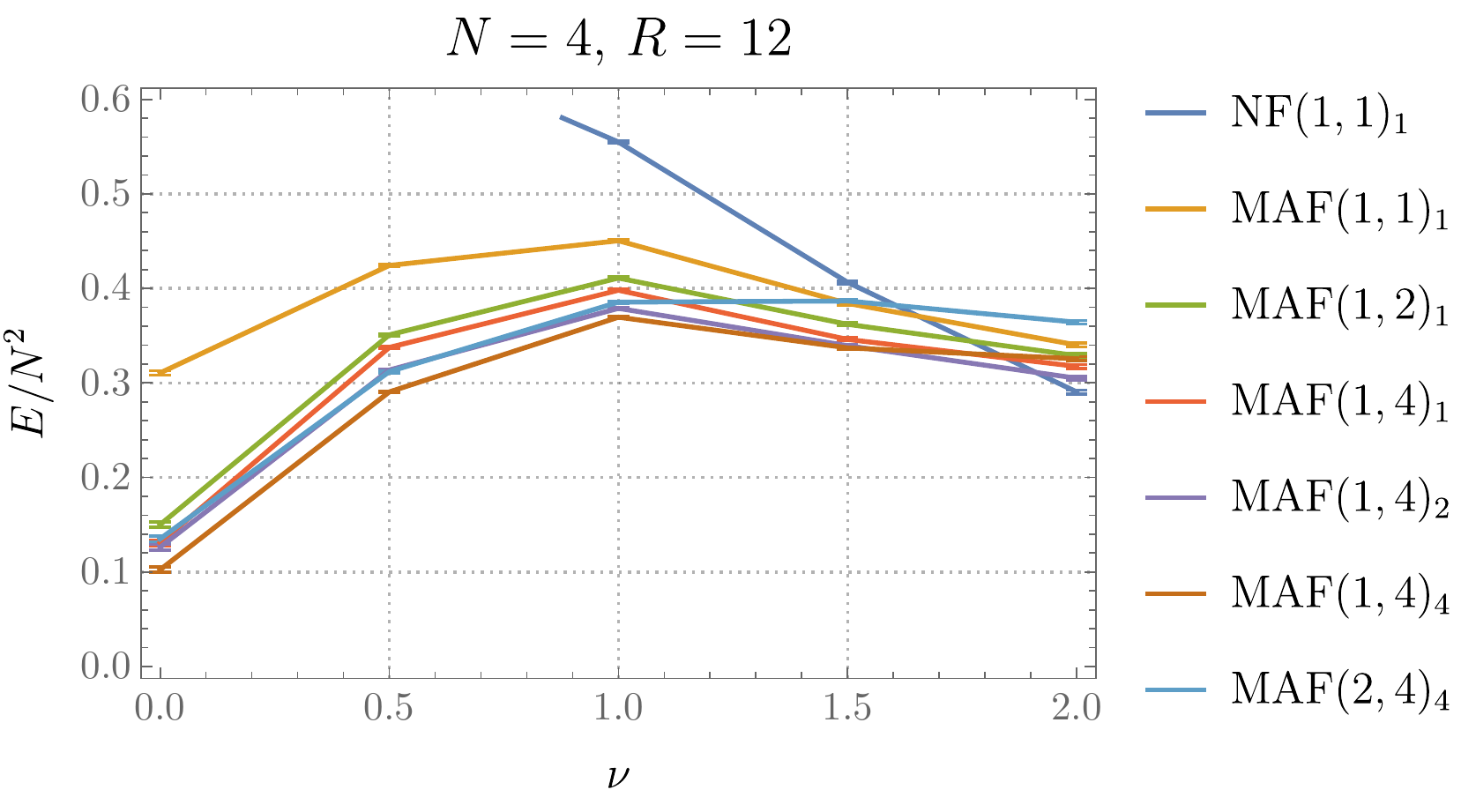}
    \includegraphics[width=0.75\textwidth]{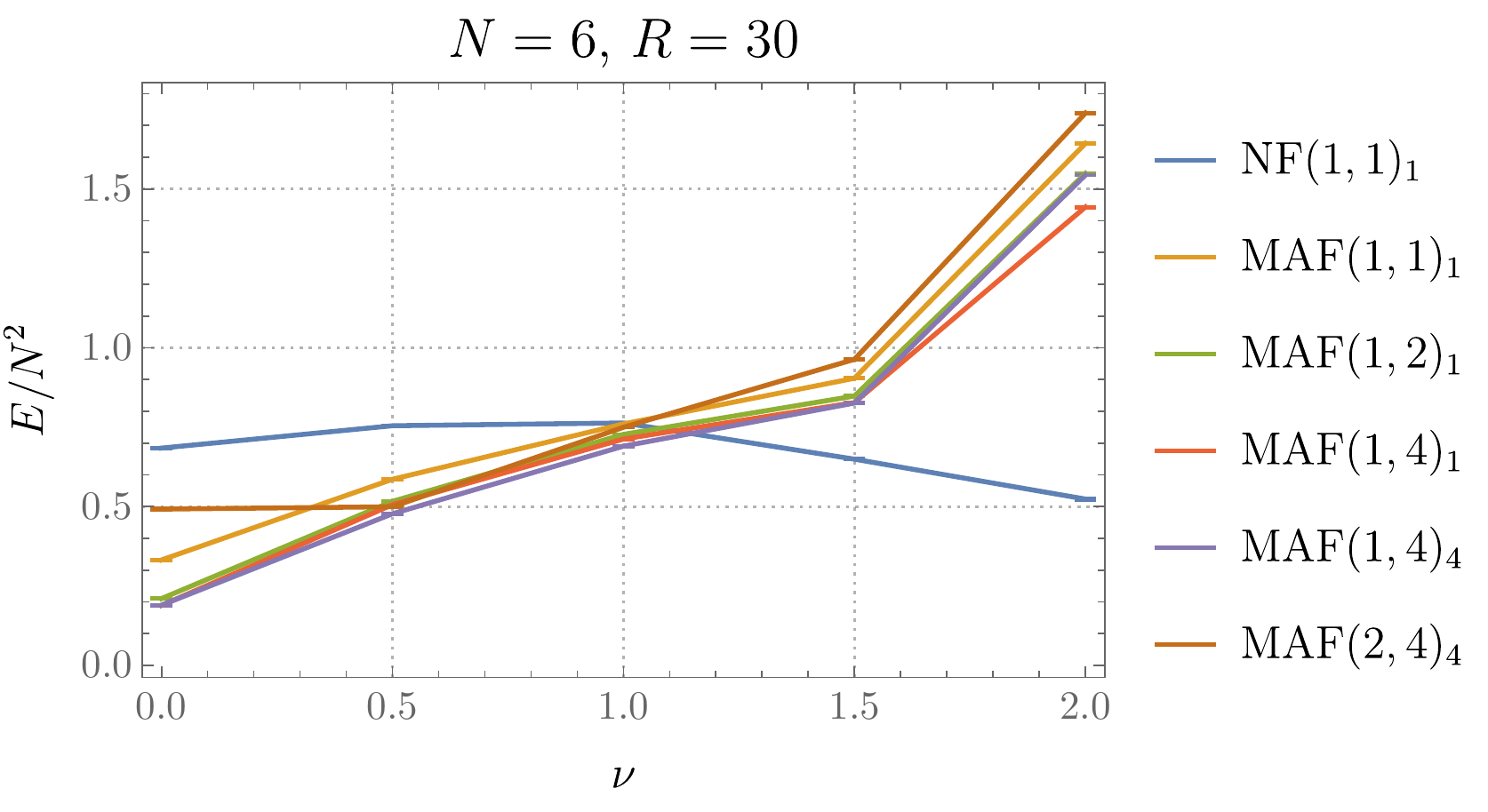}
    \caption{The variational energy for different $N$, $\nu$ and MAF architectures, in the supersymmetric sector. The wavefunctions are initialized near the fuzzy sphere. Error bars (largely invisible) are Monte Carlo uncertainties of the final energy. }
    \label{fig:qmafi}
    \vskip -0.1in
\end{figure}

\begin{figure}[ht!]
    \vskip 0.2in
    \centering
    \includegraphics[width=0.75\textwidth]{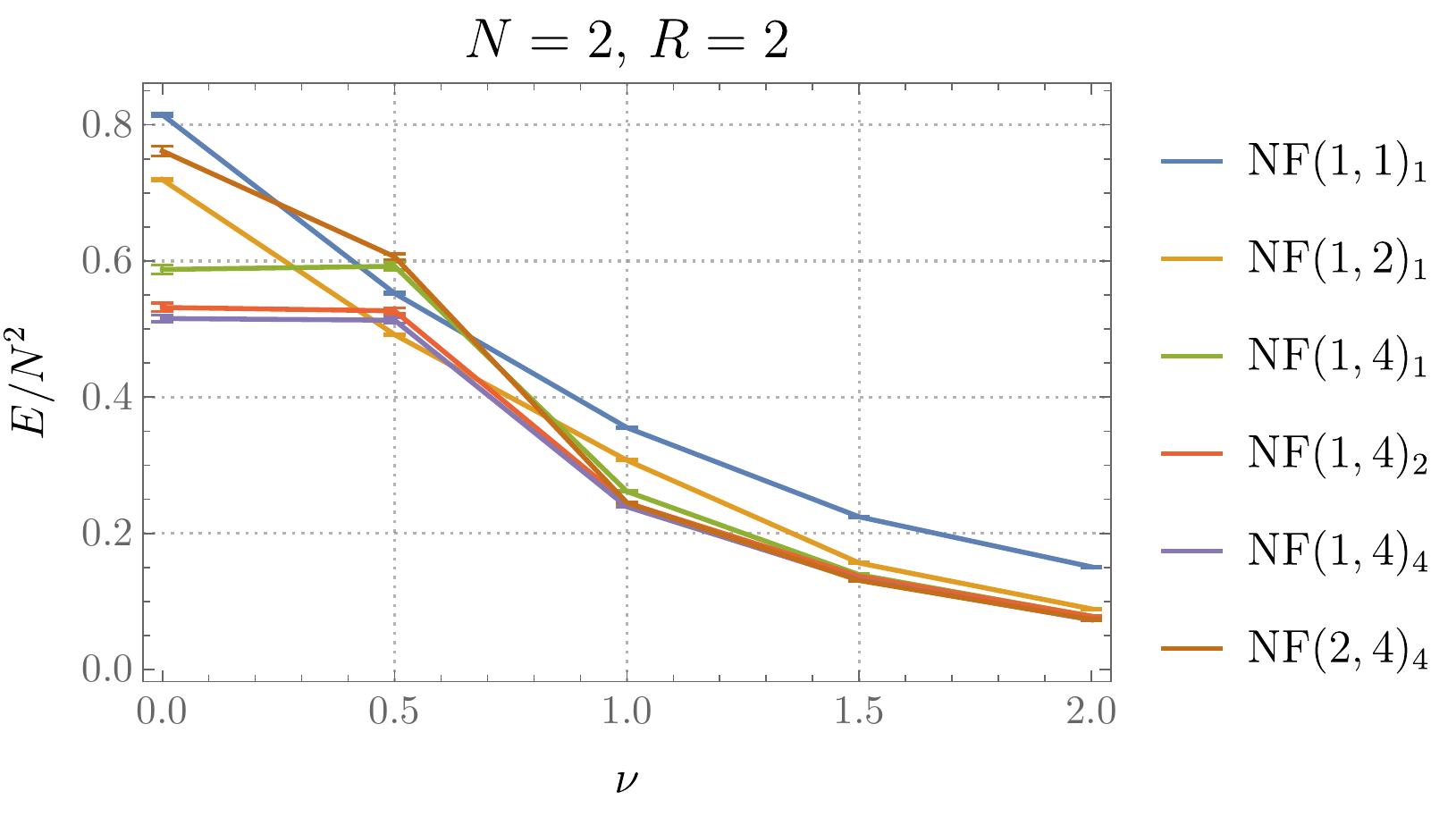}
    \includegraphics[width=0.75\textwidth]{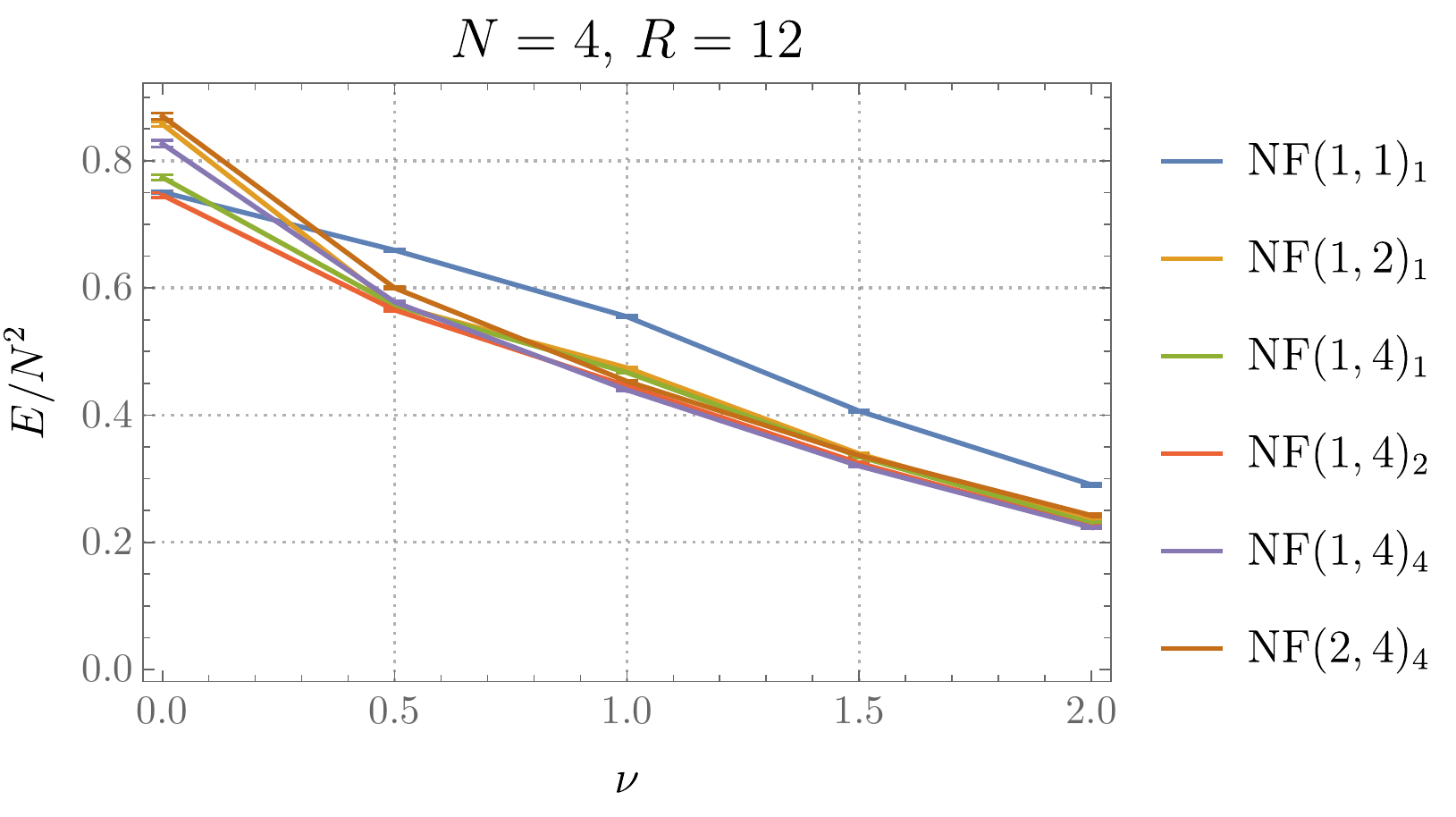}
    \includegraphics[width=0.75\textwidth]{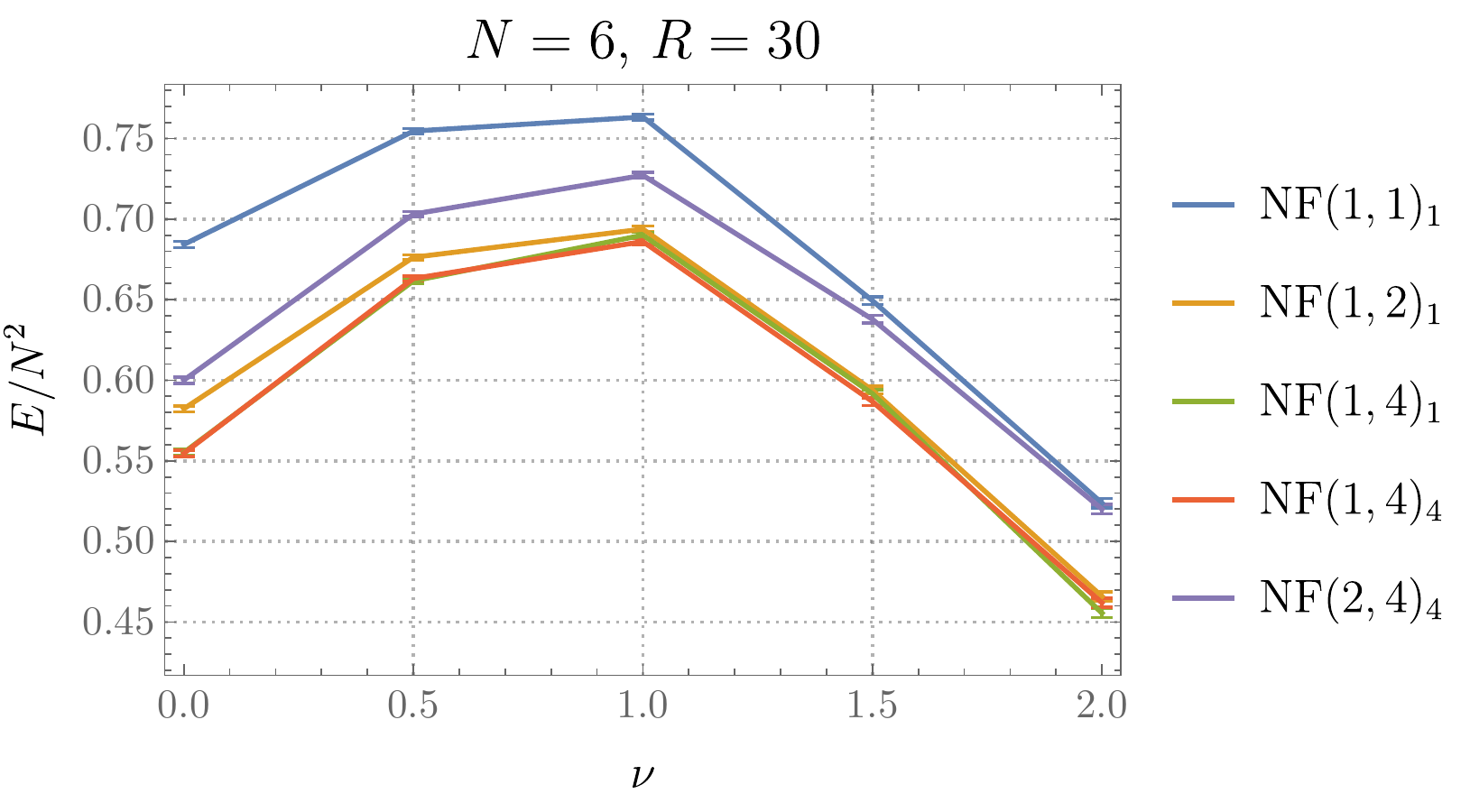}
    \caption{The variational energy for different $N$, $\nu$ and NF architectures, in the supersymmetric sector. The wavefunctions are initialized near zero. Error bars (largely invisible) are Monte Carlo uncertainties of the final energy. }
    \label{fig:qnf}
    \vskip -0.1in
\end{figure}

\begin{figure}[ht!]
    \vskip 0.2in
    \centering
    \includegraphics[width=0.75\textwidth]{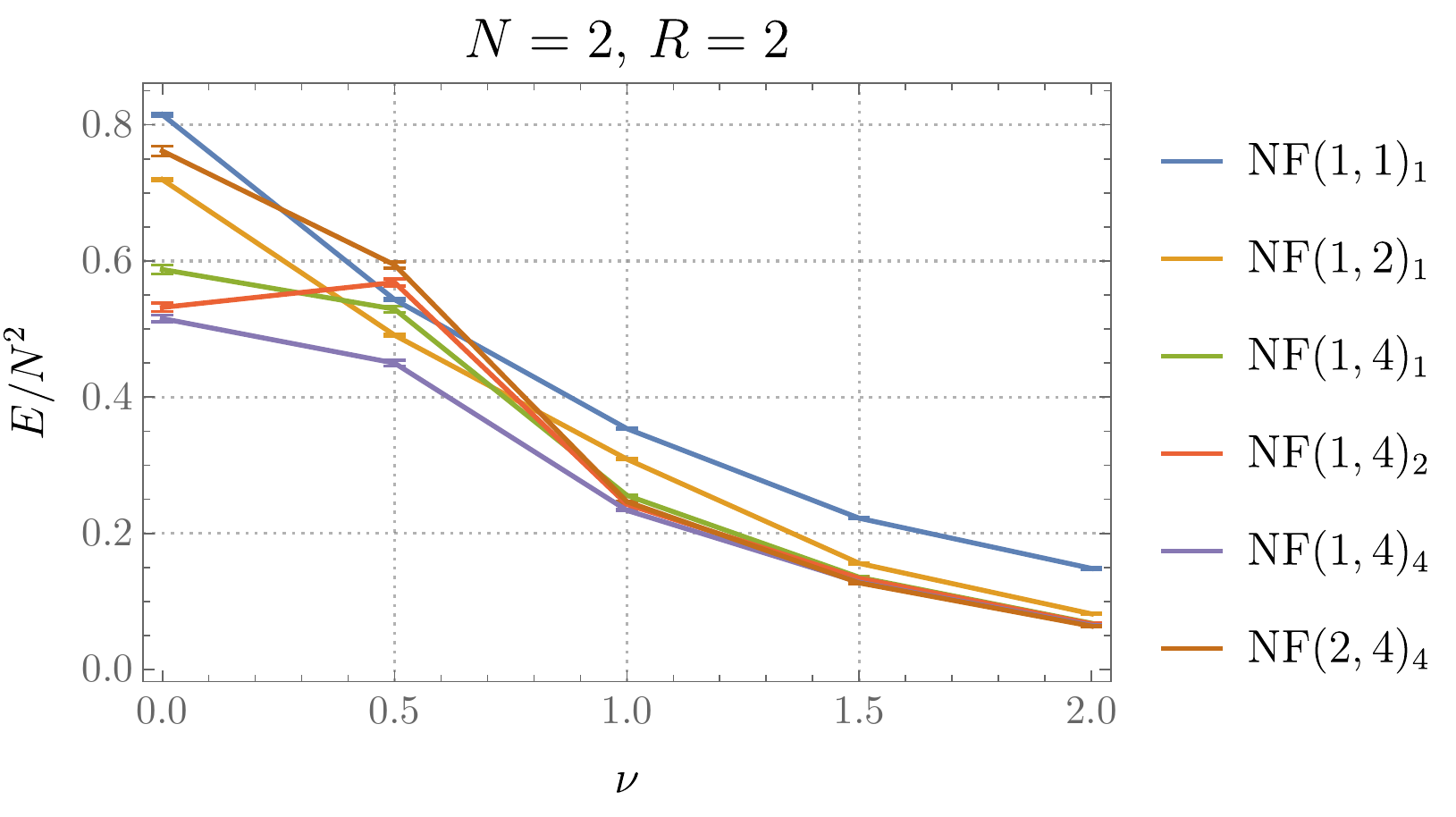}
    \includegraphics[width=0.75\textwidth]{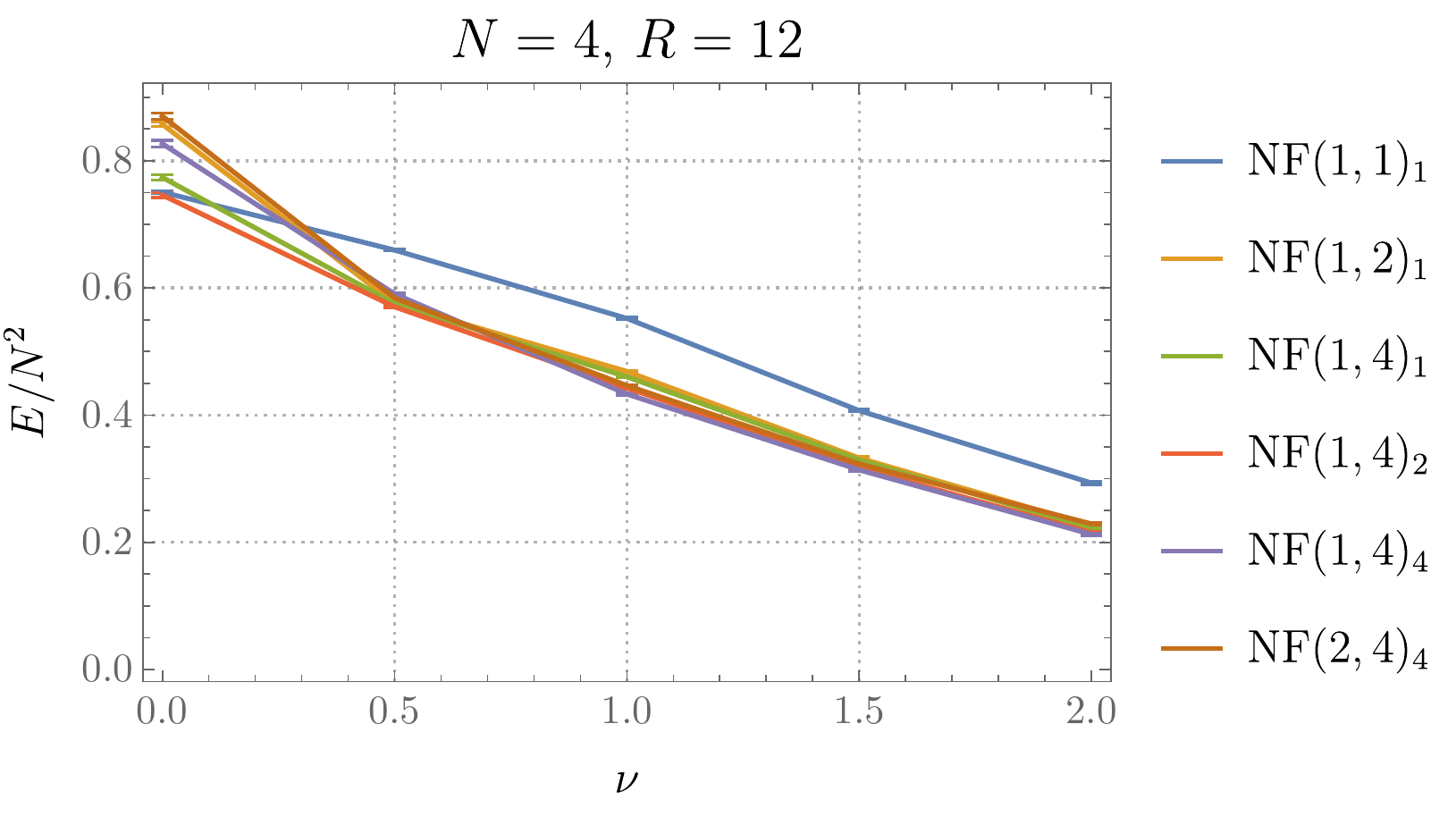}
    \includegraphics[width=0.75\textwidth]{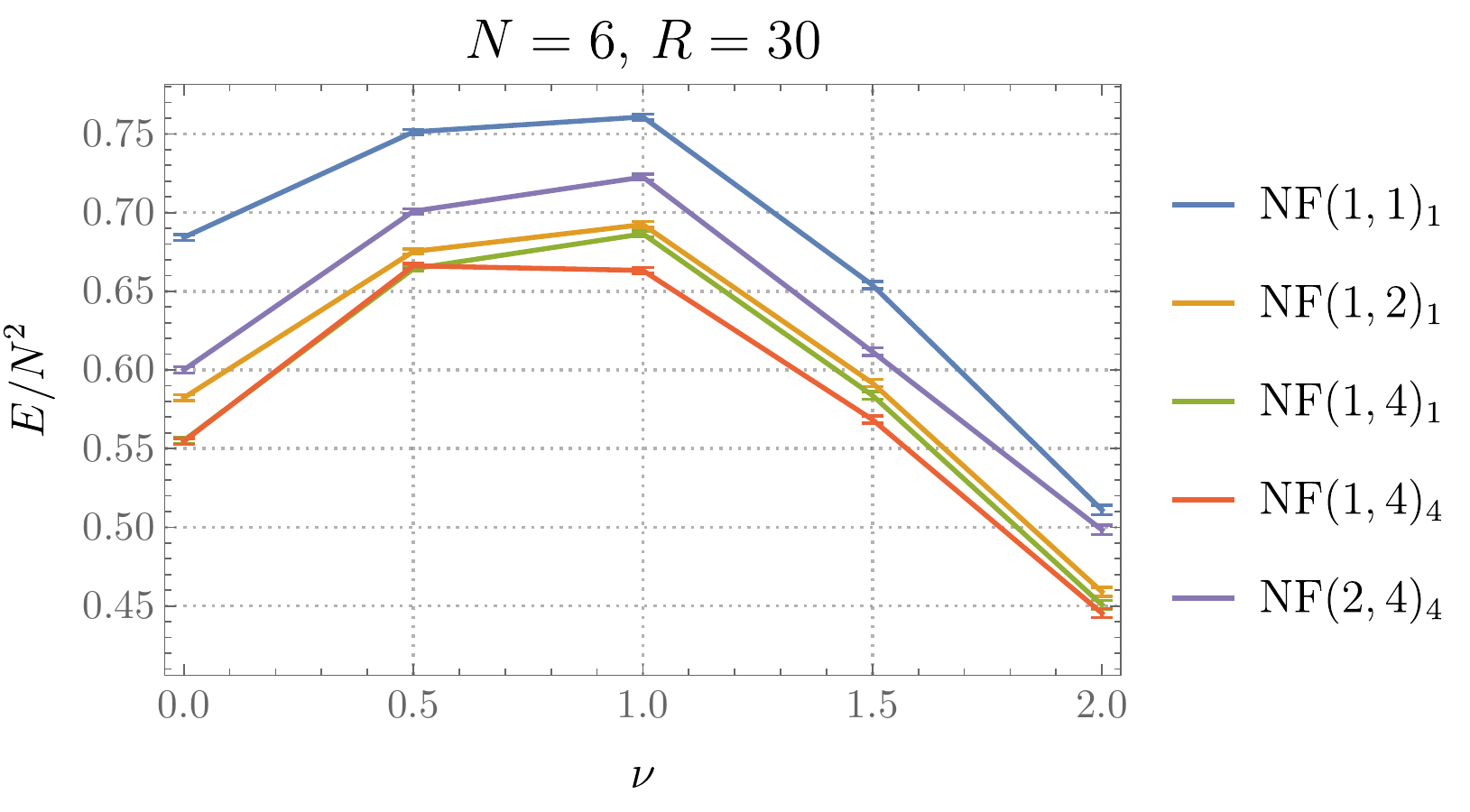}
    \caption{The variational energy for different $N$, $\nu$ and NF architectures, in the supersymmetric sector. The wavefunctions are initialized near the fuzzy sphere. Error bars are Monte Carlo uncertainties of the final energy. }
    \label{fig:qnfi}
    \vskip -0.1in
\end{figure}

The final energy of the trained variational wavefunction is evaluated from 5 million samples, with Monte Carlo uncertainties shown as error bars in Figs.\,\ref{fig:qmaf}, \ref{fig:qmafi}, \ref{fig:qnf} and \ref{fig:qnfi}.  In these figures we compare performance of various architectures and observe that
\begin{itemize}
    \item MAF obtains lower energies for small $\nu$ and NF has lower energies at larger $\nu$.
    \item The result does not significantly depend on the initialization for small $\nu$. 
    \item In the supersymmetric sector the variational energy is close to zero (compared to a typical energy scale, say the bosonic energies). 
    \item Consistent improvement is observed in MAFs if we increase the number of distributions in the mixture or $D$ as in the fermionic wavefunction. However, increasing the number of layers in neural networks does not improve the results.
\end{itemize}


\section{Entanglement of free fields on a sphere} \label{app:ent}

\subsection*{Solution for the projector}

We wish to solve the following optimization problem: find an orthogonal projection operator $P$ such that $\|P - Q\|$ is minimal given another Hermitian operator $Q$. We will now do this in the case that $\|\cdot\|$ is the Frobenius norm. In this case, diagonalize $Q = U Q' U^\dagger$ such that $Q'$ is diagonal with diagonal elements nonincreasing. Then $\|P - Q\|$ is minimized if and only if $\|P' - Q'\|$ is minimized and $P = U P' U^\dagger$. 

Firstly we search for $P'$ that minimizes $\|P' - Q'\|$ in the subspace of projectors with fixed rank $r$. It is equivalent to maximizing $\tr (P' Q')$ by definition of the Frobenius norm. Let $F(V) = \tr (V P' V^\dagger Q')$ for unitary $V$. If $P'$ maximizes $\tr (P' Q')$, $d F = 0$ at $V = I$ for any $d V$ in the Lie algebra of the unitary group:
\begin{equation}
    d F = \tr P' [Q', d V] = 0. \label{eq:df}
\end{equation}
If $Q'$ is diagonal with distinct eigenvalues, (\ref{eq:df}) implies that $P'$ should be diagonal as well. Then the $P'$ that maximizes $\tr (P' Q')$ should be such that $(P')_{i i} = 1$ for $1 \leq i \leq r$ and 0 otherwise, and the minimal value of $\|P - Q\|$ is 
\begin{equation}
    \min_{P^\dagger = P, P^2 = P}^{\tr P = r} \|P - Q\|^2 = \sum_{1 \leq i \leq r} (1 - Q'_{i i})^2 + \sum_{i > r} (Q'_{i i})^2. \label{eq:minimal_proj}
\end{equation}
The projector $P$ that achieves the minimum is unique when $Q'$ has distinct eigenvalues; if $Q'$ is degenerate, there may also be nondiagonal $P'$ matrices that attain the minimal $\|P - Q\|$. 

The second step is to minimize (\ref{eq:minimal_proj}) with respect to the rank $r$. If $Q'_{i i} \neq 1 / 2$, the rank should be the number of eigenvalues of $Q$ that are above $1 / 2$. The minimum is then
\begin{equation}
    \min_{P^\dagger = P, P^2 = P} \|P - Q\|^2 = \sum_{i} \min\{ (1 - Q'_{i i})^2, (Q'_{i i})^2\}.
\end{equation}
When one half is among the eigenvalues, there are multiple $P$'s that minimize $\|P - Q\|$. 

To summarize, let $Q = U Q' U^\dagger$ such that $U$ is unitary and $Q'$ is diagonal. Then the following $P$ minimizes $\|P - Q\|_F$ among orthogonal projectors:
\begin{equation}
    P = U P' U^\dagger, \quad P' \text{ is diagonal with } P'_{i i} = 1 \text{ if } Q'_{i i} > 1 / 2, \text{ and 0 otherwise.}
\end{equation}
And this is the unique minimum if none of the eigenvalues of $Q$ is $1 / 2$.

\subsection*{Evaluation of the second R\'enyi entropy}

As discussed in the main text, in the case where the configuration space $Q$ has a linear structure, an orthogonal decomposition $Q = Q_1 \oplus Q_2$ induces a factorization of the Hilbert space $L^2(Q) = L^2(Q_1) \otimes L^2(Q_2)$. For any pure state $| \psi \rangle \in L^2(Q)$, the entanglement entropy is computed as $S(\rho_1)$, where $\rho_1$ is the reduced density matrix of the subsystem $L^2(Q_1)$. For numerical simplicity, we now focus on the R\'enyi entropy (of order $\alpha \geq 0$):
\begin{equation}
    S_\alpha(\rho) = \frac{1}{1 - \alpha} \ln \tr \rho^\alpha.
\end{equation}
The von Neumann entropy is recovered as the limiting case $\alpha \to 1$. And in the following consider $\alpha = 2$ for concreteness; similar methods and arguments apply to the R\'enyi entropies of integer orders $\alpha \geq 2$.  

The decomposition $Q = Q_1 \oplus Q_2$ can be implicitly specified by an orthogonal projection operator $P : Q \to Q$, such that $Q_1 = \im P$ and $Q_2 = \ker P$. For a pure state $|\psi\rangle \in L^2(Q)$, the reduced density matrix $\rho_1$ is
\begin{equation}
    \rho_1(x, x') = \int d y\, \psi(x + y) \psi^*(x' + y),
\end{equation}
where $x, x' \in Q_1 = \im P$ and the integral is over the subspace $Q_2 = \ker P$. Consequently the second R\'enyi entropy is
\begin{equation}
    S_2(\rho_1) = - \ln \int d x d x' d y d y'\, \psi(x + y) \psi^*(x' + y) \psi(x' + y') \psi^*(x + y'). \label{eq:entropy_complex}
\end{equation}
To further simplify the integral, let $z = x + y \in Q$ and $z' = x' + y' \in Q$, so that
\begin{equation}
    x = P z, \quad x' = P z', \quad y = (I - P) z, \quad y' = (I - P) z'. 
\end{equation}
Thus the integral in (\ref{eq:entropy_complex}) can be done over the full space $Q$ instead:
\begin{equation}
    S_2(\rho_1) = - \ln \int d z d z'\, \psi(z) \psi^*(P z' + (I - P) z) \psi(z') \psi^*(P z + (I - P) z'). \label{eq:entropy_simple}
\end{equation}
Numerically the integral in (\ref{eq:entropy_simple}) can be estimated by Monte Carlo:
\begin{equation}
    S_2(\rho_1) = - \ln \E_{z, z' \sim |\psi|^2} \left[\frac{\psi^*(P z' + (I - P) z) \psi^*(P z + (I - P) z')}{\psi^*(z) \psi^*(z')}\right],
\end{equation}
where in the square bracket, the overall normalization of the wavefunction is unimportant.

The integral in (\ref{eq:entropy_simple}) is analytically tractable for Gaussian states:
\begin{equation}
    \psi(x) = \frac{1}{Z} \exp (- x^\dagger V x), \label{eq:free_wavefunc}
\end{equation}
where $V$ is some positive definite matrix and $Z$ is the normalization factor. Up to numerical factors, for any positive definite matrix $A$,
\begin{equation}
    \int d x \, \exp (- x^\dagger A x) \propto (\det A)^{-1}. \label{eq:gaussian_integral}
\end{equation}
Substituting (\ref{eq:free_wavefunc}) into (\ref{eq:entropy_simple}) and performing the integral using (\ref{eq:gaussian_integral}), for Gaussian pure states, one obtains
\begin{equation}
    S_2(\rho_1) = \ln (\det R / \det S),
\end{equation}
where 
\begin{align}
    R 
    &= \left(
    \begin{array}{cc}
        2 V + 2 P V P - P V - V P & P V + V P - 2 P V P  \\
        V P + P V - 2 P V P & 2 V + 2 P V P - P V  - V P
    \end{array}
    \right), \nonumber \\
    S &= \left(
    \begin{array}{cc}
        2 V & 0  \\
        0 & 2 V
    \end{array}
    \right).
\end{align}
The factor of $\det S$ comes from the normalization $Z$ in (\ref{eq:free_wavefunc}).
It is simpler to write 
\begin{align}
    S_2(\rho_1) &= \ln \det \sqrt{S^{-1}} R \sqrt{S^{-1}} = \ln \det \left(
    \begin{array}{cc}
        I + K & -K  \\
        -K & I + K
    \end{array}
    \right) \nonumber \\
    &=  \ln \det (I + 2 K) = \tr \ln (I + 2 K)
    , \label{eq:entropy_perturb}
\end{align}
where
\begin{equation}
    K =\sqrt{V^{-1}} P V P \sqrt{V^{-1}} - \frac{1}{2} \left(\sqrt{V^{-1}} P \sqrt{V} + \sqrt{V} P \sqrt{V^{-1}}\right). \label{eq:def_K}
\end{equation}
In the next subsection, geometric features of entanglement for free fields are understood analytically from the formulae (\ref{eq:entropy_perturb}) and (\ref{eq:def_K}).

\subsection*{Derivation of the geometric features of entanglement}

Consider a free field on a sphere as in (\ref{eq:free_hamil}) with angular momentum cutoff $j \leq \jm$. The ground state is a Gaussian state (\ref{eq:free_wavefunc}) with $V$ diagonal in the basis of spherical harmonic modes with eigenvalues $\sqrt{j (j + 1) + \mu^2}$ and multiplicities $2 j + 1$. The projector $P$ is the one that minimizes $\|P - \chi_A\|$, with the region $A$ being a spherical cap with polar angle $\theta_A$. We would like to confirm the following numerical findings with analytic computations: as $\jm \to \infty$, \emph{(i)} $S_2 \propto \jm \sin \theta_A \propto \jm |\partial A|$ and \emph{(ii)} $\tr P \propto j_\text{max}^2 \int_0^{\theta_A} \sin \theta d\theta \propto j_\text{max}^2 |A|$.

To start, observe that from (\ref{eq:entropy_perturb}) naively we would expect $S_2 \sim (\jm)^2$ because of the trace, and thus if $S_2 \sim \jm$ it must be the case that the matrix $K$ is small. Hence it is reasonable to make the approximation
\begin{equation}
    S_2 \approx 2 \tr K = 2 \tr P V P V^{-1} - 2 \tr P.
\end{equation}
In terms of matrix elements of the projector, (recall that $P^\dagger = P$ and $P^2 = P$)
\begin{equation}
    S_2 \approx  \sum_{j j' m} |P_{j m, j' m}|^2 \frac{(j - j')^2}{j j'}, \label{eq:entropy2}
\end{equation}
where we have noticed that the projector preserves the $J^z$ quantum number because of the symmetry of region $A$. Also the eigenvalues of $V$ are approximated as $j$. Subleading terms will not modify the scaling as $\jm \to \infty$, where $j$ is typically large. 

For $j, j' \ll \jm$, the projector $P_{j m, j' m}$ should converge to its value at infinite $\jm$, which is the matrix element of multiplication by $\chi_A$:
\begin{equation}
    P_{j m, j' m} \sim \frac{1}{4 \pi} \int_0^{\theta_A} d \theta \sin \theta \int_0^{2 \pi} d \phi \, Y_{j m}^*(\theta, \phi) Y_{j' m}(\theta, \phi),
\end{equation}
where $\chi_A$ restricts the $\theta$ integral to $[0, \theta_A]$. Up to numerical factors,
\begin{equation}
    P_{j m, j' m} \propto \sqrt{\frac{(2 j + 1)(2 j' + 1) (j - m)! (j' - m)!}{(j + m)! (j' + m)!}} \int_{\cos \theta_A}^1 d x\, P_j^m(x) P_{j'}^m(x), \label{eq:projection}
\end{equation}
where $P_j^m(x)$ are associated Legendre polynomials.

The asymptotic form of associated Legendre polynomials $P_j^{-m}(x)$ in the limit $j, m \to \infty$ with $\alpha = m / (j + 1 / 2)$ fixed ($0 < \alpha < 1$) is given by the WKB formulae eqs.\,(3.28) and (3.30) in \cite{Thorne:1957phi}: for $\beta = \sqrt{1 - \alpha^2}$ and $\beta < x \leq 1$,
\begin{equation}
P_j^{-m}(x) \sim \Lambda^{jm} (x^2 - \beta^2)^{-1/4} e^{(j + 1/2) \chi^{jm}_1(x)}, \label{eq:legendre1}
\end{equation}
while for $0 \leq x < \beta$,
\begin{equation}
    P_j^{-m}(x) \sim 2 \Lambda^{jm} (\beta^2 - x^2)^{-1/4} \cos\left(\left(j + \frac{1}{2}\right) \chi^{jm}_2(x) - \frac{\pi}{4}\right), \label{eq:legendre2}
\end{equation}
where
\begin{align}
    \Lambda^{jm} &= \frac{1}{\sqrt{\pi (2 j + 1)}} \sqrt{\frac{(j - m)!}{(j + m)!}}, \nonumber \\
    \chi^{jm}_1(x) &= \cosh^{-1}\left(\frac{x}{\beta}\right) - \alpha \cosh^{-1} \left( \frac{\alpha x}{\beta \sqrt{1- x^2}}\right) < 0, \nonumber \\
    \chi^{jm}_2(x) &= \cos^{-1} \left(\frac{x}{\beta}\right) - \alpha \cos^{-1} \left(\frac{\alpha x}{\beta \sqrt{1- x^2}}\right) > 0.
\end{align}

Let $x = \cos \theta$. At large $j$ the oscillating region of the integral in (\ref{eq:projection}), where (\ref{eq:legendre2}) holds, is $0 < \alpha < \sin \theta$. Outside of this region, the Legendre polynomial is approximately (\ref{eq:legendre1}), and hence exponentially small. We need therefore only consider the region where both Legendre polynomials are oscillating.
In order to get the parametric dependence of observables right, we can furthermore restrict attention to $m \ll j,j'$. In this limit $\beta \to 1$, $\alpha \to 0$ and hence
\be
\chi_2^{jm}(x) = \theta.
\ee
So in this limit the integrand in (\ref{eq:projection}) can be approximated as
\be\label{eq:difff}
dx \, P_j^{-m}(x) P_{j'}^{-m}(x) = d\theta \, 2 \Lambda^{jm} \Lambda^{j'm} \cos\left[(j - j') \theta \right] + \cdots \,.
\ee
The terms $\cdots$ necessarily oscillate strongly at large $j,j'$ and will not contribute to leading order. In the remaining term in (\ref{eq:difff}), in contrast, the oscillations are slower when $j \sim j'$. Performing the integral we obtain
\be\label{eq:Pjjmm}
P_{j (-m), j' (-m)} \propto \frac{\sin \left[(j-j') \theta_A  \right]}{j - j'} \,.
\ee
The lower limit of integration (at $m = [\min(j,j') + 1/2] \sin\theta$) can be ignored so long as $m \ll \min(j,j') \sin\theta_A$. This is stronger than the previous assumption $m \ll j,j'$.
We can now use (\ref{eq:Pjjmm}) to evaluate observables, using the fact that $P_{j (-m), j' (-m)} = P_{jm, j'm}$.

The R\'enyi entropy (\ref{eq:entropy2}) is now (with $j_\text{m} = \min(j,j')$)
\bea
S_2 & \propto & \sum_{jj'}^{|m| \ll  j_\text{m} \sin \theta_A} \frac{\sin^2[(j-j') \theta_A]}{jj'} \\
& \propto & \int^{j_\text{max}} \frac{dj'}{j'} \int^{j'} dj \sin(\theta_A) \sin^2[(j-j') \theta_A] \\
& \propto & j_\text{max} \sin(\theta_A) \,.
\eea
In the second line we used $j_\text{m} \sin\theta_A$ as a cutoff on the sum over $m$, to get an estimate of the scaling with $\sin\theta_A$. This is the boundary law entanglement that was observed numerically in the main text.

To get the rank of the projector one must treat the sum over $m$ a little more carefully. In particular, we refrain from taking $\a \to 0$, $\beta \to 1$. Keeping $\a = m/(j + 1 /2)$,
\bea
\tr P & = & \sum_{jm} P_{jm,jm} \\
& \propto & \sum_{jm} \int^{\theta_A}_{\arcsin{|\a|}} \frac{\sin(\theta) d\theta}{\sqrt{\sin(\theta)^2-\alpha^2}} + \cdots \,.
\eea
Here $\cdots$ again denote terms that oscillate strongly in the large $j$ limit and are therefore subleading. The integrand in the second line is directly the non-oscillating part of (\ref{eq:legendre2}) squared. At large $j_\text{max}$ we therefore have, approximating the sums as integrals and letting $\a = \sin \gamma$,
\bea
\tr P & \propto & j_\text{max}^2 \int_{0}^{\theta_A} d\gamma \int^{\theta_A}_{\gamma} d\theta \frac{\sin(\theta) \cos(\gamma) }{\sqrt{\sin(\theta)^2-\sin(\gamma)^2}} \\
& \propto & j_\text{max}^2 \int_0^{\theta_A} d\theta \sin(\theta) \,. \label{eq:fff}
\eea
The integrals are most easily done by exchanging the order of integration to $\int_0^{\theta_A} d\theta \int_0^\theta d\gamma$. This result shows that the rank of the projector goes like the area of the region on the sphere, as seen numerically in the main text. The prefactor in the final result (\ref{eq:fff}) is easily restored by noting that when $\theta_A = \pi$, corresponding to the whole sphere, $\tr P \sim j_\text{max}^2$ at large $j_\text{max}$.


\begin{thebibliography}{10}

\bibitem{Maldacena:1997re}
J.~M. Maldacena, {{The Large N limit of superconformal field theories and
  supergravity}}, \href{http://dx.doi.org/10.1023/A:1026654312961,
  10.4310/ATMP.1998.v2.n2.a1}{Int. J. Theor. Phys. {\bf 38}, 1113--1133, 1999},
  [\href{http://arxiv.org/abs/arXiv:hep-th/9711200}{{arXiv:hep-th/9711200
  [hep-th]}}].

\bibitem{Polchinski:2010hw}
J.~Polchinski, {{Introduction to Gauge/Gravity Duality}},  in
  \emph{{Proceedings, Theoretical Advanced Study Institute in Elementary
  Particle Physics (TASI 2010). String Theory and Its Applications: From meV to
  the Planck Scale: Boulder, Colorado, USA, June 1-25, 2010}}, pp.~3--46, 2010.
\newblock [\href{http://arxiv.org/abs/arXiv:1010.6134}{{arXiv:1010.6134
  [hep-th]}}].

\bibitem{Anagnostopoulos:2007fw}
K.~N. Anagnostopoulos, M.~Hanada, J.~Nishimura and S.~Takeuchi, {{Monte Carlo
  studies of supersymmetric matrix quantum mechanics with sixteen supercharges
  at finite temperature}},
  \href{http://dx.doi.org/10.1103/PhysRevLett.100.021601}{Phys. Rev. Lett. {\bf
  100}, 021601, 2008},
  [\href{http://arxiv.org/abs/arXiv:0707.4454}{{arXiv:0707.4454 [hep-th]}}].

\bibitem{Catterall:2008yz}
S.~Catterall and T.~Wiseman, {{Black hole thermodynamics from simulations of
  lattice Yang-Mills theory}},
  \href{http://dx.doi.org/10.1103/PhysRevD.78.041502}{Phys. Rev. {\bf D78},
  041502, 2008}, [\href{http://arxiv.org/abs/arXiv:0803.4273}{{arXiv:0803.4273
  [hep-th]}}].

\bibitem{Berkowitz:2016jlq}
E.~Berkowitz, E.~Rinaldi, M.~Hanada, G.~Ishiki, S.~Shimasaki and P.~Vranas,
  {{Precision lattice test of the gauge/gravity duality at large-$N$}},
  \href{http://dx.doi.org/10.1103/PhysRevD.94.094501}{Phys. Rev. {\bf D94},
  094501, 2016},
  [\href{http://arxiv.org/abs/arXiv:1606.04951}{{arXiv:1606.04951 [hep-lat]}}].

\bibitem{Bombelli:1986rw}
L.~Bombelli, R.~K. Koul, J.~Lee and R.~D. Sorkin, {{A Quantum Source of Entropy
  for Black Holes}}, \href{http://dx.doi.org/10.1103/PhysRevD.34.373}{Phys.
  Rev. {\bf D34}, 373--383, 1986}.

\bibitem{Srednicki:1993im}
M.~Srednicki, {{Entropy and area}},
  \href{http://dx.doi.org/10.1103/PhysRevLett.71.666}{Phys. Rev. Lett. {\bf
  71}, 666--669, 1993},
  [\href{http://arxiv.org/abs/arXiv:hep-th/9303048}{{arXiv:hep-th/9303048
  [hep-th]}}].

\bibitem{MPS}
D.~{Perez-Garcia}, F.~{Verstraete}, M.~M. {Wolf} and J.~I. {Cirac}, {{Matrix
  Product State Representations}}, {Quantum Info. Comput. {\bf 7}, 401--430,
  2007},
  [\href{http://arxiv.org/abs/arXiv:quant-ph/0608197}{{arXiv:quant-ph/0608197
  [quant-ph]}}].

\bibitem{ORUS2014117}
R.~Or\'us, {A practical introduction to tensor networks: Matrix product states
  and projected entangled pair states},
  \href{http://dx.doi.org/https://doi.org/10.1016/j.aop.2014.06.013}{Annals of
  Physics {\bf 349}, 117 -- 158, 2014}.

\bibitem{Hinton504}
G.~E. Hinton and R.~R. Salakhutdinov, {Reducing the dimensionality of data with
  neural networks}, \href{http://dx.doi.org/10.1126/science.1127647}{Science
  {\bf 313}, 504--507, 2006}.

\bibitem{LeCun}
Y.~LeCun, Y.~Bengio and G.~Hinton, {Deep learning},
  \href{http://dx.doi.org/10.1038/nature14539}{Nature {\bf 521}, 436, 2015}.

\bibitem{Goodfellow-et-al-2016}
I.~Goodfellow, Y.~Bengio and A.~Courville, \emph{Deep Learning}.
\newblock MIT Press, 2016.

\bibitem{Krizhevsky:2012:ICD:2999134.2999257}
A.~Krizhevsky, I.~Sutskever and G.~E. Hinton, {Imagenet classification with
  deep convolutional neural networks},  in \emph{Proceedings of the 25th
  International Conference on Neural Information Processing Systems - Volume
  1}, NIPS'12, (USA), pp.~1097--1105, Curran Associates Inc., 2012.

\bibitem{Go}
D.~Silver, A.~Huang, C.~J. Maddison, A.~Guez, L.~Sifre, G.~van~den Driessche,
  J.~Schrittwieser, I.~Antonoglou, V.~Panneershelvam, M.~Lanctot, S.~Dieleman,
  D.~Grewe, J.~Nham et~al., {Mastering the game of go with deep neural networks
  and tree search}, \href{http://dx.doi.org/10.1038/nature16961}{Nature {\bf
  529}, 484, 2016}.

\bibitem{doi:10.1063/PT.3.4164}
S.~Das~Sarma, D.-L. Deng and L.-M. Duan, {Machine learning meets quantum
  physics}, \href{http://dx.doi.org/10.1063/PT.3.4164}{Physics Today {\bf 72},
  48--54, 2019}.

\bibitem{Carleo602}
G.~Carleo and M.~Troyer, {Solving the quantum many-body problem with artificial
  neural networks}, \href{http://dx.doi.org/10.1126/science.aag2302}{Science
  {\bf 355}, 602--606, 2017}.

\bibitem{PhysRevX.7.021021}
D.-L. Deng, X.~Li and S.~Das~Sarma, {Quantum entanglement in neural network
  states}, \href{http://dx.doi.org/10.1103/PhysRevX.7.021021}{Phys. Rev. X {\bf
  7}, 021021, 2017}.

\bibitem{Gao2017}
X.~Gao and L.-M. Duan, {Efficient representation of quantum many-body states
  with deep neural networks},
  \href{http://dx.doi.org/10.1038/s41467-017-00705-2}{Nature Communications
  {\bf 8}, 662, 2017}.

\bibitem{PhysRevX.8.011006}
I.~Glasser, N.~Pancotti, M.~August, I.~D. Rodriguez and J.~I. Cirac,
  {Neural-network quantum states, string-bond states, and chiral topological
  states}, \href{http://dx.doi.org/10.1103/PhysRevX.8.011006}{Phys. Rev. X {\bf
  8}, 011006, 2018}.

\bibitem{dinh2014nice}
L.~{Dinh}, D.~{Krueger} and Y.~{Bengio}, {{NICE: Non-linear Independent
  Components Estimation}},  2014,
  [\href{http://arxiv.org/abs/arXiv:1410.8516}{{arXiv:1410.8516 [cs.LG]}}].

\bibitem{2015arXiv150505770J}
D.~{Jimenez Rezende} and S.~{Mohamed}, {{Variational Inference with Normalizing
  Flows}},  2015,
  [\href{http://arxiv.org/abs/arXiv:1505.05770}{{arXiv:1505.05770 [stat.ML]}}].

\bibitem{DBLP:journals/corr/DinhSB16}
L.~Dinh, J.~Sohl{-}Dickstein and S.~Bengio, {Density estimation using real
  {NVP}}, { CoRR, 2016},
  [\href{http://arxiv.org/abs/arXiv:1605.08803}{{arXiv:1605.08803}}].

\bibitem{DBLP:journals/corr/GermainGML15}
M.~Germain, K.~Gregor, I.~Murray and H.~Larochelle, {{{\{}MADE:{\}} Masked
  Autoencoder for Distribution Estimation}}, { CoRR, 2015},
  [\href{http://arxiv.org/abs/arXiv:1502.03509}{{arXiv:1502.03509}}].

\bibitem{DBLP:journals/corr/KingmaSW16}
D.~P. Kingma, T.~Salimans and M.~Welling, {{Improving Variational Inference
  with Inverse Autoregressive Flow}}, { CoRR, 2016},
  [\href{http://arxiv.org/abs/arXiv:1606.04934}{{arXiv:1606.04934}}].

\bibitem{papamakarios2017masked}
G.~Papamakarios, I.~Murray and T.~Pavlakou, {Masked autoregressive flow for
  density estimation},  in \emph{Advances in Neural Information Processing
  Systems}, pp.~2338--2347, 2017.

\bibitem{Carrasquilla2019}
J.~Carrasquilla, G.~Torlai, R.~G. Melko and L.~Aolita, {Reconstructing quantum
  states with generative models},
  \href{http://dx.doi.org/10.1038/s42256-019-0028-1}{Nature Machine
  Intelligence {\bf 1}, 155--161, 2019}.

\bibitem{PhysRevX.8.031012}
Z.-Y. Han, J.~Wang, H.~Fan, L.~Wang and P.~Zhang, {Unsupervised generative
  modeling using matrix product states},
  \href{http://dx.doi.org/10.1103/PhysRevX.8.031012}{Phys. Rev. X {\bf 8},
  031012, 2018}.

\bibitem{PhysRevLett.122.080602}
D.~Wu, L.~Wang and P.~Zhang, {Solving statistical mechanics using variational
  autoregressive networks},
  \href{http://dx.doi.org/10.1103/PhysRevLett.122.080602}{Phys. Rev. Lett. {\bf
  122}, 080602, 2019}.

\bibitem{PhysRevB.97.045153}
Y.-Z. You, Z.~Yang and X.-L. Qi, {Machine learning spatial geometry from
  entanglement features},
  \href{http://dx.doi.org/10.1103/PhysRevB.97.045153}{Phys. Rev. B {\bf 97},
  045153, 2018}.

\bibitem{PhysRevD.98.046019}
K.~Hashimoto, S.~Sugishita, A.~Tanaka and A.~Tomiya, {Deep learning and the
  $\mathrm{AdS}/\mathrm{CFT}$ correspondence},
  \href{http://dx.doi.org/10.1103/PhysRevD.98.046019}{Phys. Rev. D {\bf 98},
  046019, 2018}.

\bibitem{Hu:2019nea}
H.-Y. Hu, S.-H. Li, L.~Wang and Y.-Z. You, {{Machine Learning Holographic
  Mapping by Neural Network Renormalization Group}},  2019,
  [\href{http://arxiv.org/abs/arXiv:1903.00804}{{arXiv:1903.00804
  [cond-mat.dis-nn]}}].

\bibitem{Asplund:2015yda}
C.~T. Asplund, F.~Denef and E.~Dzienkowski, {{Massive quiver matrix models for
  massive charged particles in AdS}},
  \href{http://dx.doi.org/10.1007/JHEP01(2016)055}{JHEP {\bf 01}, 055, 2016},
  [\href{http://arxiv.org/abs/arXiv:1510.04398}{{arXiv:1510.04398 [hep-th]}}].

\bibitem{Berenstein:2002jq}
D.~E. Berenstein, J.~M. Maldacena and H.~S. Nastase, {{Strings in flat space
  and pp waves from N=4 superYang-Mills}},
  \href{http://dx.doi.org/10.1088/1126-6708/2002/04/013}{JHEP {\bf 04}, 013,
  2002},
  [\href{http://arxiv.org/abs/arXiv:hep-th/0202021}{{arXiv:hep-th/0202021
  [hep-th]}}].

\bibitem{Anous:2017mwr}
T.~Anous and C.~Cogburn, {{Mini-BFSS in Silico}},  2017,
  [\href{http://arxiv.org/abs/arXiv:1701.07511}{{arXiv:1701.07511 [hep-th]}}].

\bibitem{Itzhaki:1998dd}
N.~Itzhaki, J.~M. Maldacena, J.~Sonnenschein and S.~Yankielowicz,
  {{Supergravity and the large N limit of theories with sixteen supercharges}},
  \href{http://dx.doi.org/10.1103/PhysRevD.58.046004}{Phys. Rev. {\bf D58},
  046004, 1998},
  [\href{http://arxiv.org/abs/arXiv:hep-th/9802042}{{arXiv:hep-th/9802042
  [hep-th]}}].

\bibitem{Madore:1991bw}
J.~Madore, {{The Fuzzy sphere}},
  \href{http://dx.doi.org/10.1088/0264-9381/9/1/008}{Class. Quant. Grav. {\bf
  9}, 69--88, 1992}.

\bibitem{Hoppe:1988gk}
J.~Hoppe, {{Diffeomorphism Groups, Quantization and SU(infinity)}},
  \href{http://dx.doi.org/10.1142/S0217751X89002235}{Int. J. Mod. Phys. {\bf
  A4}, 5235, 1989}.

\bibitem{DEWIT1988545}
B.~de~Wit, J.~Hoppe and H.~Nicolai, {On the quantum mechanics of
  supermembranes},
  \href{http://dx.doi.org/https://doi.org/10.1016/0550-3213(88)90116-2}{Nuclear
  Physics B {\bf 305}, 545 -- 581, 1988}.

\bibitem{Jatkar:2001uh}
D.~P. Jatkar, G.~Mandal, S.~R. Wadia and K.~P. Yogendran, {{Matrix dynamics of
  fuzzy spheres}}, \href{http://dx.doi.org/10.1088/1126-6708/2002/01/039}{JHEP
  {\bf 01}, 039, 2002},
  [\href{http://arxiv.org/abs/arXiv:hep-th/0110172}{{arXiv:hep-th/0110172
  [hep-th]}}].

\bibitem{Dasgupta:2002hx}
K.~Dasgupta, M.~M. Sheikh-Jabbari and M.~Van~Raamsdonk, {{Matrix perturbation
  theory for M theory on a PP wave}},
  \href{http://dx.doi.org/10.1088/1126-6708/2002/05/056}{JHEP {\bf 05}, 056,
  2002},
  [\href{http://arxiv.org/abs/arXiv:hep-th/0205185}{{arXiv:hep-th/0205185
  [hep-th]}}].

\bibitem{Myers:1999ps}
R.~C. Myers, {{Dielectric branes}},
  \href{http://dx.doi.org/10.1088/1126-6708/1999/12/022}{JHEP {\bf 12}, 022,
  1999},
  [\href{http://arxiv.org/abs/arXiv:hep-th/9910053}{{arXiv:hep-th/9910053
  [hep-th]}}].

\bibitem{Alekseev:2000fd}
A.~{\relax Yu}. Alekseev, A.~Recknagel and V.~Schomerus, {{Brane dynamics in
  background fluxes and noncommutative geometry}},
  \href{http://dx.doi.org/10.1088/1126-6708/2000/05/010}{JHEP {\bf 05}, 010,
  2000},
  [\href{http://arxiv.org/abs/arXiv:hep-th/0003187}{{arXiv:hep-th/0003187
  [hep-th]}}].

\bibitem{McGreevy:2000cw}
J.~McGreevy, L.~Susskind and N.~Toumbas, {{Invasion of the giant gravitons from
  Anti-de Sitter space}},
  \href{http://dx.doi.org/10.1088/1126-6708/2000/06/008}{JHEP {\bf 06}, 008,
  2000},
  [\href{http://arxiv.org/abs/arXiv:hep-th/0003075}{{arXiv:hep-th/0003075
  [hep-th]}}].

\bibitem{Myers:2003bw}
R.~C. Myers, {{NonAbelian phenomena on D branes}},
  \href{http://dx.doi.org/10.1088/0264-9381/20/12/302}{Class. Quant. Grav. {\bf
  20}, S347--S372, 2003},
  [\href{http://arxiv.org/abs/arXiv:hep-th/0303072}{{arXiv:hep-th/0303072
  [hep-th]}}].

\bibitem{NIPS2017_7203}
Z.~Lu, H.~Pu, F.~Wang, Z.~Hu and L.~Wang, {The expressive power of neural
  networks: A view from the width},  in \emph{Advances in Neural Information
  Processing Systems 30} (I.~Guyon, U.~V. Luxburg, S.~Bengio, H.~Wallach,
  R.~Fergus, S.~Vishwanathan and R.~Garnett, eds.), pp.~6231--6239.
\newblock Curran Associates, Inc., 2017.

\bibitem{Azuma:2004zq}
T.~Azuma, S.~Bal, K.~Nagao and J.~Nishimura, {{Nonperturbative studies of fuzzy
  spheres in a matrix model with the Chern-Simons term}},
  \href{http://dx.doi.org/10.1088/1126-6708/2004/05/005}{JHEP {\bf 05}, 005,
  2004},
  [\href{http://arxiv.org/abs/arXiv:hep-th/0401038}{{arXiv:hep-th/0401038
  [hep-th]}}].

\bibitem{CastroVillarreal:2004vh}
P.~Castro-Villarreal, R.~Delgadillo-Blando and B.~Ydri, {{A Gauge-invariant
  UV-IR mixing and the corresponding phase transition for U(1) fields on the
  fuzzy sphere}},
  \href{http://dx.doi.org/10.1016/j.nuclphysb.2004.10.032}{Nucl. Phys. {\bf
  B704}, 111--153, 2005},
  [\href{http://arxiv.org/abs/arXiv:hep-th/0405201}{{arXiv:hep-th/0405201
  [hep-th]}}].

\bibitem{DelgadilloBlando:2007vx}
R.~Delgadillo-Blando, D.~O'Connor and B.~Ydri, {{Geometry in Transition: A
  Model of Emergent Geometry}},
  \href{http://dx.doi.org/10.1103/PhysRevLett.100.201601}{Phys. Rev. Lett. {\bf
  100}, 201601, 2008},
  [\href{http://arxiv.org/abs/arXiv:0712.3011}{{arXiv:0712.3011 [hep-th]}}].

\bibitem{Iso:2001mg}
S.~Iso, Y.~Kimura, K.~Tanaka and K.~Wakatsuki, {{Noncommutative gauge theory on
  fuzzy sphere from matrix model}},
  \href{http://dx.doi.org/10.1016/S0550-3213(01)00173-0}{Nucl. Phys. {\bf
  B604}, 121--147, 2001},
  [\href{http://arxiv.org/abs/arXiv:hep-th/0101102}{{arXiv:hep-th/0101102
  [hep-th]}}].

\bibitem{Paniak:2002fi}
L.~D. Paniak and R.~J. Szabo, {{Instanton expansion of noncommutative gauge
  theory in two dimensions}},
  \href{http://dx.doi.org/10.1007/s00220-003-0964-8}{Commun. Math. Phys. {\bf
  243}, 343--387, 2003},
  [\href{http://arxiv.org/abs/arXiv:hep-th/0203166}{{arXiv:hep-th/0203166
  [hep-th]}}].

\bibitem{Lizzi:2001nd}
F.~Lizzi, R.~J. Szabo and A.~Zampini, {{Geometry of the gauge algebra in
  noncommutative Yang-Mills theory}},
  \href{http://dx.doi.org/10.1088/1126-6708/2001/08/032}{JHEP {\bf 08}, 032,
  2001},
  [\href{http://arxiv.org/abs/arXiv:hep-th/0107115}{{arXiv:hep-th/0107115
  [hep-th]}}].

\bibitem{deWit:1997ab}
B.~de~Wit, {{Supersymmetric quantum mechanics, supermembranes and Dirichlet
  particles}}, \href{http://dx.doi.org/10.1016/S0920-5632(97)00312-5}{Nucl.
  Phys. Proc. Suppl. {\bf 56B}, 76--87, 1997},
  [\href{http://arxiv.org/abs/arXiv:hep-th/9701169}{{arXiv:hep-th/9701169
  [hep-th]}}].

\bibitem{Anagnostopoulos:2005cy}
K.~N. Anagnostopoulos, T.~Azuma, K.~Nagao and J.~Nishimura, {{Impact of
  supersymmetry on the nonperturbative dynamics of fuzzy spheres}},
  \href{http://dx.doi.org/10.1088/1126-6708/2005/09/046}{JHEP {\bf 09}, 046,
  2005},
  [\href{http://arxiv.org/abs/arXiv:hep-th/0506062}{{arXiv:hep-th/0506062
  [hep-th]}}].

\bibitem{Ydri:2012bq}
B.~Ydri, {{Impact of Supersymmetry on Emergent Geometry in Yang-Mills Matrix
  Models II}}, \href{http://dx.doi.org/10.1142/S0217751X12500881}{Int. J. Mod.
  Phys. {\bf A27}, 1250088, 2012},
  [\href{http://arxiv.org/abs/arXiv:1206.6375}{{arXiv:1206.6375 [hep-th]}}].

\bibitem{Casini:2013rba}
H.~Casini, M.~Huerta and J.~A. Rosabal, {{Remarks on entanglement entropy for
  gauge fields}}, \href{http://dx.doi.org/10.1103/PhysRevD.89.085012}{Phys.
  Rev. {\bf D89}, 085012, 2014},
  [\href{http://arxiv.org/abs/arXiv:1312.1183}{{arXiv:1312.1183 [hep-th]}}].

\bibitem{Lin:2018bud}
J.~Lin and D.~Radi\v{c}evi\'c, {{Comments on Defining Entanglement Entropy}},
  2018, [\href{http://arxiv.org/abs/arXiv:1808.05939}{{arXiv:1808.05939
  [hep-th]}}].

\bibitem{Donnelly:2011hn}
W.~Donnelly, {{Decomposition of entanglement entropy in lattice gauge theory}},
  \href{http://dx.doi.org/10.1103/PhysRevD.85.085004}{Phys. Rev. {\bf D85},
  085004, 2012}, [\href{http://arxiv.org/abs/arXiv:1109.0036}{{arXiv:1109.0036
  [hep-th]}}].

\bibitem{Donnelly:2014gva}
W.~Donnelly, {{Entanglement entropy and nonabelian gauge symmetry}},
  \href{http://dx.doi.org/10.1088/0264-9381/31/21/214003}{Class. Quant. Grav.
  {\bf 31}, 214003, 2014},
  [\href{http://arxiv.org/abs/arXiv:1406.7304}{{arXiv:1406.7304 [hep-th]}}].

\bibitem{Das:1995jw}
S.~R. Das, {{Degrees of freedom in two-dimensional string theory}},
  \href{http://dx.doi.org/10.1016/0920-5632(95)00640-0}{Nucl. Phys. Proc.
  Suppl. {\bf 45BC}, 224--233, 1996},
  [\href{http://arxiv.org/abs/arXiv:hep-th/9511214}{{arXiv:hep-th/9511214
  [hep-th]}}].

\bibitem{Hartnoll:2015fca}
S.~A. Hartnoll and E.~Mazenc, {{Entanglement entropy in two dimensional string
  theory}}, \href{http://dx.doi.org/10.1103/PhysRevLett.115.121602}{Phys. Rev.
  Lett. {\bf 115}, 121602, 2015},
  [\href{http://arxiv.org/abs/arXiv:1504.07985}{{arXiv:1504.07985 [hep-th]}}].

\bibitem{Dou:2006ni}
D.~Dou and B.~Ydri, {{Entanglement entropy on fuzzy spaces}},
  \href{http://dx.doi.org/10.1103/PhysRevD.74.044014}{Phys. Rev. {\bf D74},
  044014, 2006},
  [\href{http://arxiv.org/abs/arXiv:gr-qc/0605003}{{arXiv:gr-qc/0605003
  [gr-qc]}}].

\bibitem{Karczmarek:2013jca}
J.~L. Karczmarek and P.~Sabella-Garnier, {{Entanglement entropy on the fuzzy
  sphere}}, \href{http://dx.doi.org/10.1007/JHEP03(2014)129}{JHEP {\bf 03},
  129, 2014}, [\href{http://arxiv.org/abs/arXiv:1310.8345}{{arXiv:1310.8345
  [hep-th]}}].

\bibitem{Okuno:2015kuc}
S.~Okuno, M.~Suzuki and A.~Tsuchiya, {{Entanglement entropy in scalar field
  theory on the fuzzy sphere}},
  \href{http://dx.doi.org/10.1093/ptep/ptv192}{PTEP {\bf 2016}, 023B03, 2016},
  [\href{http://arxiv.org/abs/arXiv:1512.06484}{{arXiv:1512.06484 [hep-th]}}].

\bibitem{Chen:2017kfj}
H.~Z. Chen and J.~L. Karczmarek, {{Entanglement entropy on a fuzzy sphere with
  a UV cutoff}}, \href{http://dx.doi.org/10.1007/JHEP08(2018)154}{JHEP {\bf
  08}, 154, 2018},
  [\href{http://arxiv.org/abs/arXiv:1712.09464}{{arXiv:1712.09464 [hep-th]}}].

\bibitem{Susskind:1994sm}
L.~Susskind and J.~Uglum, {{Black hole entropy in canonical quantum gravity and
  superstring theory}}, \href{http://dx.doi.org/10.1103/PhysRevD.50.2700}{Phys.
  Rev. {\bf D50}, 2700--2711, 1994},
  [\href{http://arxiv.org/abs/arXiv:hep-th/9401070}{{arXiv:hep-th/9401070
  [hep-th]}}].

\bibitem{Fiola:1994ir}
T.~M. Fiola, J.~Preskill, A.~Strominger and S.~P. Trivedi, {{Black hole
  thermodynamics and information loss in two-dimensions}},
  \href{http://dx.doi.org/10.1103/PhysRevD.50.3987}{Phys. Rev. {\bf D50},
  3987--4014, 1994},
  [\href{http://arxiv.org/abs/arXiv:hep-th/9403137}{{arXiv:hep-th/9403137
  [hep-th]}}].

\bibitem{Bianchi:2012ev}
E.~Bianchi and R.~C. Myers, {{On the Architecture of Spacetime Geometry}},
  \href{http://dx.doi.org/10.1088/0264-9381/31/21/214002}{Class. Quant. Grav.
  {\bf 31}, 214002, 2014},
  [\href{http://arxiv.org/abs/arXiv:1212.5183}{{arXiv:1212.5183 [hep-th]}}].

\bibitem{Faulkner:2013ana}
T.~Faulkner, A.~Lewkowycz and J.~Maldacena, {{Quantum corrections to
  holographic entanglement entropy}},
  \href{http://dx.doi.org/10.1007/JHEP11(2013)074}{JHEP {\bf 11}, 074, 2013},
  [\href{http://arxiv.org/abs/arXiv:1307.2892}{{arXiv:1307.2892 [hep-th]}}].

\bibitem{Donnelly:2016auv}
W.~Donnelly and L.~Freidel, {{Local subsystems in gauge theory and gravity}},
  \href{http://dx.doi.org/10.1007/JHEP09(2016)102}{JHEP {\bf 09}, 102, 2016},
  [\href{http://arxiv.org/abs/arXiv:1601.04744}{{arXiv:1601.04744 [hep-th]}}].

\bibitem{Harlow:2016vwg}
D.~Harlow, {{The Ryu-Takayanagi Formula from Quantum Error Correction}},
  \href{http://dx.doi.org/10.1007/s00220-017-2904-z}{Commun. Math. Phys. {\bf
  354}, 865--912, 2017},
  [\href{http://arxiv.org/abs/arXiv:1607.03901}{{arXiv:1607.03901 [hep-th]}}].

\bibitem{Thorne:1957phi}
R.~C. Thorne and H.~Jeffreys, {The asymptotic expansion of legendre function of
  large degree and order},
  \href{http://dx.doi.org/10.1098/rsta.1957.0008}{Phil. Trans. Roy. Soc. Lon.
  Series A, Mathematical and Physical Sciences {\bf 249}, 597--620, 1957}.

\end{thebibliography}
\end{document}